\newcommand{\hide}[1]{}
\newcommand{\revision}[1]{{%
#1}}
\newcommand{\revisiontwo}[1]{{%
#1}}

\documentclass[12pt]{article}

\usepackage{amsmath}
\usepackage{amsthm}

\usepackage{bm}
\usepackage[inline]{enumitem}
\usepackage{amsfonts}
\usepackage{amssymb}
\usepackage{mathtools}
\usepackage{graphicx,psfrag,epsf}
\usepackage{lmodern}
\usepackage[onehalfspacing]{setspace}
\usepackage{xcolor}
\usepackage{appendix} 
\usepackage{wrapfig}
\usepackage{float}
\usepackage{hyperref}
\usepackage[flushleft]{threeparttable}
\usepackage{pdfpages}
\usepackage{pdflscape}
\usepackage{natbib}
\usepackage{array}
\usepackage{graphicx}
\usepackage{float}
\usepackage{hyperref}
\usepackage{caption}
\usepackage{subcaption}
\usepackage{etoolbox}
\appto\appendix{\counterwithin{equation}{section}}
\usepackage{authblk}
\usepackage{xpatch}
\xpatchcmd{\author}{\relax#1\relax}{\relax\detokenize{#1}\relax}{}{}
\xpatchcmd{\affil}{\relax#1\relax}{\relax\detokenize{#1}\relax}{}{}

\newcommand{\blind}{0}

\newtheorem{proposition}{Proposition}[section]
\newtheorem{theorem}{Theorem}[section]

\begin{document}

\def\spacingset#1{\renewcommand{\baselinestretch}%
{#1}\small\normalsize} \spacingset{1}


\if0\blind
{

\title{Forecasting macroeconomic data with Bayesian VARs: Sparse or dense? It depends!}
\author[\empty]{Luis Gruber\thanks{Corresponding author. Adress: Department of Statistics, University of Klagenfurt. Universit\"{a}tsstra\ss e 65-67,
9020 Klagenfurt am W\"{o}rthersee, Austria. E-Mail: \href{mailto:luis.gruber@aau.at}{luis.gruber@aau.at}.

Variants of this paper were presented at the 12th European Seminar on Bayesian Econometrics (ESOBE 2022), the 5th Vienna Workshop on High-Dimensional Time Series in Macroeconomics and Finance 2022, the 2022 ISBA World Meeting, the 16th International Conference on Computational and Financial Econometrics 2022, the Bayesian Young Statisticians Meeting (BAYSM 2023), and the 44th International Symposium on Forecasting (ISF 2024). The authors thank all participants, in particular Sylvia Fr\"uhwirth-Schnatter, Neil Shephard, Massimiliano Marcellino, Florian Huber, Sylvia Kaufmann, Hedibert Lopes, Mike West, as well as anonymous referees and the associate editor for crucially valuable comments and suggestions.

The authors acknowledge funding from the Austrian Science Fund (FWF) for the project ``High-dimensional statistical learning: New methods to advance economic and sustainability policy'' (ZK 35), jointly carried out by the University of Klagenfurt, Paris Lodron University Salzburg, TU Wien, and the Austrian
Institute of Economic Research (WIFO).}}
\author[\empty]{Gregor Kastner}
\affil[\empty]{Department of Statistics, University of Klagenfurt, Austria}
\hide{\title{\bf Forecasting macroeconomic data with Bayesian VARs: Sparse or dense? It depends!}
  \author{Luis Gruber\thanks{The authors acknowledge funding from the Austrian Science Fund (FWF) for the project ``High-dimensional statistical learning: New methods to advance economic and sustainability policy'' (ZK 35), jointly carried out by the University of Klagenfurt, Paris Lodron University Salzburg, TU Wien, and the Austrian
Institute of Economic Research (WIFO).}\hspace{.2cm}\\
    Department of Statistics, University of Klagenfurt, \\9020 Klagenfurt am Wörthersee, Austria \\\href{mailto:luis.gruber@aau.at}{luis.gruber@aau.at}\\
    and \\
    Gregor Kastner \\
    Department of Statistics, University of Klagenfurt, \\9020 Klagenfurt am Wörthersee, Austria\\\href{mailto:gregor.kastner@aau.at}{gregor.kastner@aau.at}}}
  \date{}
  \maketitle
} \fi

\if1\blind
{
  \bigskip
  \bigskip
  \bigskip
  \begin{center}
    {\LARGE\bf Forecasting macroeconomic data with Bayesian VARs: Sparse or dense? It depends!}
\end{center}
  \medskip
} \fi

\begin{center}
This is a preprint of the version published in:\\\emph{International Journal of Forecasting}\\\url{https://doi.org/10.1016/j.ijforecast.2025.02.001}
\end{center}
\vspace{1em}
\begin{abstract}
Vector autogressions (VARs) are widely applied when it comes to modeling and forecasting macroeconomic variables. In high dimensions, however, they are prone to overfitting. Bayesian methods, more concretely shrinkage priors, have shown to be successful in improving prediction performance. In the present paper, we introduce the semi-global framework, in which we replace the traditional global shrinkage parameter with group-specific shrinkage parameters. We show how this framework can be applied to various shrinkage priors, such as global-local priors and stochastic search variable selection priors. We demonstrate the virtues of the proposed framework in an extensive simulation study and in an empirical application forecasting data of the US economy. Further, we shed more light on the ongoing ``Illusion of Sparsity'' debate, finding that forecasting performances under sparse/dense priors vary across evaluated economic variables and across time frames. Dynamic model averaging, however, can combine the merits of both worlds.
\end{abstract}

\noindent%
\vfill

\newpage
\onehalfspacing
\section{Introduction}
The recent literature suggests that predictions of macroeconomic variables benefit from exploiting large data sets. Especially vector autoregressions (VARs), where the number of free parameters is relatively large compared to the limited length of macroeconomic time series, are prone to overfitting. Bayesian methods have shown to be an effective regularization technique in order to reduce estimation uncertainty by imposing additional structure on the model \citep[e.g.,][]{litterman_forecasting_1986, sims_bayesian_1998, koop_forecasting_2013, giannone_prior_2015, chan_minnesota-type_2021, griffin_luo_2022}.

Prior elicitation for VARs is a long-standing issue which goes back at least to the Minnesota prior in \citet{litterman_forecasting_1986}. It stipulates that lagged coefficients of the respective dependent variables are shrunken less than lagged coefficients of other variables, and that coefficients are generally shrunken stronger with an increasing number of lags. The Minnesota prior, therefore, is a classic example of a prior based on domain knowledge. On the other side of the spectrum, hierarchical priors such as stochastic search variable selection (SSVS) priors or global-local (GL) priors -- where the latter most recently attract more and more attention in combination with large VARs \citep[e.g.,][]{follett_achieving_2019, huber_adaptive_2019, kastner_sparse_2020} -- offer off-the-shelf shrinkage, since they are neither problem-specific nor designed for VARs per se. 

The first contribution of this paper is the introduction of the class of semi-global priors, which aims at combining the merits of domain knowledge with off-the-shelf shrinkage. In short, this framework is characterized by shrinkage parameters acting on pre-specified subgroups of the parameter space, i.e., on the semi-global level. It can easily be applied to already existing priors, e.g., it can be applied to any GL prior or to all kinds of SSVS priors. The framework is very flexible  and nests other extension to GL priors, like the extensions to the normal-gamma prior in \citet{huber_adaptive_2019} and \citet{chan_minnesota-type_2021}, as special cases. We propose a grouping of the coefficients which mimics some features of the Minnesota prior, namely the discrimination between lags in general, and within lags the discrimination between own-lag and cross-lag coefficients. 

Second, we provide a concise comparison of shrinkage priors that are commonly used in the VAR literature, but are scattered across different works and consequently lack a systematic comparison. Here, our attempt is to shed more light on the ``prior zoo'' by an in-depth comparison focusing on the two most important features of shrinkage priors: \begin{enumerate*}[label={\alph*)}]
    \item the concentration at zero for shrinking noise, and
    \item the tail robustness for detecting the relevant signals.
\end{enumerate*} This in turn gives guidance for practitioners in meaningfully selecting hyperparameters. The priors under scrutiny are: The Minnesota prior, a semi-hierarchical version of the Minnesota prior (SHM), the stochastic search variable selection (SSVS) prior as in \citet{george_bayesian_2008}, the Horseshoe (HS) prior \citep{carvalhi2010}, the normal-gamma (NG) prior \citep{griffin2010}, the Dirichlet-Laplace (DL) prior \citep{bhattacharya_dirichletlaplace_2015} and the $R^2$-induced Dirichlet decomposition (R2D2) prior \citep{zhang_bayesian_2020}. Concerning the latter and to the best of our knowledge, we are the first who investigate the R2D2 prior in the context of VARs.

Third, we contribute to the ongoing ``Illusion of Sparsity'' (IoS) debate \citep{cross_macroeconomic_2020, fava_illusion_2020, giannone_economic_2021}. Loosely speaking, within the Bayesian framework, there are two contrasting approaches in order to reduce parameter uncertainty in high dimensions. On the one hand, there are sparse modeling techniques that aim at selecting a small set of important predictors. On the other hand, dense modeling techniques support the assumption that all possible explanatory variables might be important; their individual impact for prediction, however, is expected to be small. \citet{giannone_economic_2021} design one specific prior -- which is basically a discrete mixture prior with point mass at zero -- in order to detect whether specific datasets are best summarized by means of many equally important covariates or rather by a small subset of covariates. Analyzing the posterior results, they conclude that macroeconomic data is rather dense. \citet{fava_illusion_2020} show that with little changes to the prior used in \citet{giannone_economic_2021}, i.e., using fatter tailed distributions, the posterior results appear to be sparser.%
\footnote{\revisiontwo{%
We note at this point that sparsity in the VAR coefficients might imply unrealistic constraints from a purely economic point of view. However, as detailed in \citet{simsMacroeconomicsReality1980}, these can nevertheless be useful for forecasting purposes, as restrictions can reduce forecast errors even if they are false.}}
    
In light of these results, it becomes clear that the level of posterior sparsity can be very sensitive to prior assumptions. Thus, we augment the narrative by quantifying the sparseness of our considered priors \emph{before} looking at the data. By applying the Hoyer sparseness measure \citep{hoyer_non-negative_2004} to simulated data from the prior distributions, we demonstrate that they can be ordered from sparse to dense. Sparsity is best expressed by GL priors; denseness is best expressed by Minnesota(-type) priors. In addition, we find that prior sparseness translates to posterior sparseness. In other words, the prior and the posterior sparsity ranking is similar.
    
Other than \citet{fava_illusion_2020} and \citet{giannone_economic_2021}, who investigate a linear model with a single response variable, the flexible framework of VARs allows us to analyze results for different response variables with only one data set. In an extensive simulation study based on various data generating processes (DGPs), we test the validity of our claims. In sparse DGPs, the variants of GL priors have the highest concentration around the true model parameters. In purely dense DGPs, the Minnesota priors seem to be superior to all competing models.	
    
The gold standard of model evaluation in economic and financial applications is comparing forecasting performance: A model is considered good if it performs well in predicting the future. In our empirical application, we adopt a variant of the quarterly data set of the US economy proposed by \citet{stock_watson_2012} and provided by \citet{mccracken_fred-qd_2020}. Overall, the semi-global versions of GL priors, with specific shrinkage for own-lag and cross-lag coefficients, achieve the highest support from the data in terms of one-step-ahead predictive likelihoods. Nevertheless, we find heterogeneity in model performance over time. The heterogeneous results demonstrate that there is no prior, and hence no sparse or dense modeling approach, that performs best during the whole evaluation period and for all variables. To combine the merits of different priors, we also discuss the possibility of dynamic model averaging as proposed in \citet{raftery_online_2010} and the loss discounting framework proposed in \citet{bernaciakLossDiscountingFramework2024}.
    
Overall, this paper provides a broad overview of specification choices with respect to Bayesian VARs in their reduced forms featuring stochastic volatility (SV). Although all of our considered priors have originally been proposed or put forward to the VAR framework using the reduced form, lately, many authors opt for using the structural form \citep[e.g.,][]{cross_macroeconomic_2020, chan_minnesota-type_2021}. Placing priors on the structural form coefficients is alluring because then one can employ a faster Markov chain Monte Carlo (MCMC) algorithm. Nevertheless, placing the prior on the structural coefficients constitutes a different model where posteriors are (potentially substantially) different as well. Beyond that, there is early empirical evidence that modeling reduced-form coefficients leads to better out-of-sample forecasting performance \citep{bernardi2022}. Modeling reduced-form coefficients, however, comes at the cost of a higher computational burden. To render MCMC computation feasible, we apply the corrected triangular algorithm as in \citet{carriero_corrigendum_2021}.\footnote{\citet{carriero_corrigendum_2021} present a correction to the broadly applied algorithm based on equation per equation estimation put forward in \citet{carriero_large_2019}.}

\revision{\citet{cross_macroeconomic_2020} investigate and compare a similar set of prior distributions for VARs. We differ in three important aspects. First, we demonstrate how forecasting performance of GL priors can be improved with the semi-global shrinkage approach. Second, the aforementioned paper places the priors on the structural form VAR coefficients. Third, \citet{cross_macroeconomic_2020} evaluate forecasting performance only on a three-variate subset of considered variables, whereas we consider also the full-system predictive performance in addition to further robustness checks.}
 
The remainder of the paper is organized as follows: Section \ref{sec:econ} describes the econometric framework and lays out the various prior specifications, while Section~\ref{sec:sparseness} compares the shrinkage behavior and sparsity of the priors.
Section \ref{sec:snythetic} presents the results of an extensive simulation study considering different time series length within sparse and dense data-generating scenarios. In Section \ref{sec:application}, we apply the different model setups to data of the US economy. After inspecting the posterior distributions, we perform a forecasting exercise to assess the predictive performance of the considered models. Finally, Section \ref{sec:conclusion} concludes the article.

\section{Econometric framework}
\label{sec:econ}
For $t=1,\dots,T$, let $\bm{y}_t$ denote an $M$-dimensional column vector containing observations on $M$ time series variables. In a $\mathrm{VAR}$ model of order $p$, $\mathrm{VAR}(p)$, $\bm{y}_t$ is determined by\footnote{For simplicity of exposition we omit the intercept in the following (which we nonetheless include in the empirical application).}
\begin{equation}\label{eq:VAR}
	\bm{y}^\prime_t = \sum_{j=1}^{p}\bm{y}^\prime_{t-j}\bm{A}_j + \bm{\varepsilon}^\prime_t, 
\end{equation}
where $\bm{A}_j$ is an $M \times M$ matrix of coefficients, the lag $p$ a positive integer, and $\bm{\varepsilon}_t$ an $M$-dimensional vector of errors. 

 We assume the distribution of the errors to be multivariate Gaussian with time varying variance-covariance matrix, i.e., $\bm{\varepsilon}_t \sim \bm{N(0,\Sigma_t)}$, and follow \citet{cogley_drifts_2005} in applying the decomposition in the form of
\begin{equation}\label{eq:cogley}
	\bm{\Sigma}_t = \bm{U}^{\prime -1} \bm{D}_t \bm{U}^{-1},
\end{equation}
where $\bm{U}^{-1}$ is an upper triangular matrix with ones on the diagonal, and $\bm{D}_t$ is a diagonal matrix. We denote the free off-diagonal elements in $\bm{U}$ as $\bm{u} = (u_{12}, u_{13}, \dots, u_{(M-1)M})^\prime$. The transformed errors $\bm{\xi}_t = \bm{U}^\prime\bm{\varepsilon}_t$ then have a diagonal variance-covariance matrix $\bm{D}_t$. Borrowing from \citet{jacquier_bayesian_1994} and \citet{kim_stochastic_1998}, the orthogonalized errors are assumed to follow univariate SV models: For $i=1,\dots,M$,
\begin{align}\label{eq:usv}
	\xi_{it} &= \exp\left(\frac{h_{it}}{2}\right)\epsilon_{it},\\
	h_{it} &= \mu_{i} + \rho_{i}(h_{i(t-1)} - \mu_i) + \sigma_i \eta_{it},
\end{align}
where $\bm{\epsilon}_t$ and $\bm{\eta}_t$ are assumed to be i.i.d.\ $N(\bm{0, I}_M)$. Hence, the $ii$th element of $\bm{D}_t$ is $d_{ii,t}=\exp(h_{it})$. The log-variance process $\bm{h_{i}}=(h_{i1},\dots,h_{iT})^\prime$ is initialized by $h_{i0} \sim N\left(\mu_i, \frac{\sigma_i^2}{1-\rho_i^2}\right)$. Here, $\mu_i$ is the level of log-variance, $\rho_i \in (-1,1)$ the persistence of log-variance, and $\sigma_i$ the volatility of log-variance.

To facilitate prior implementation, it proves to be convenient to rewrite the model in matrix form. Define a $K \times 1$ vector of predictors $\bm{x}_t = (\bm{y}^\prime_{t-1}, \dots, \bm{y}^\prime_{t-p})^\prime$ and a $K \times M$ matrix of coefficients $\bm{\Phi}=(\bm{A}_1, \dots, \bm{A}_p)^\prime $, where $K = mp$ is the number of vectorautoregressive coefficients per equation. Then the VAR can be written as
\begin{equation}\label{eq:VARshort}
	\bm{Y}= \bm{X}\bm{\Phi} + \bm{E},
\end{equation}
where $\bm{Y} = (\bm{y}, \dots, \bm{y}_T)^\prime$, $\bm{X}=(\bm{x}_1,\dots,\bm{x}_T)^\prime$, and $\bm{E}=(\bm{\varepsilon}_1,\dots,\bm{\varepsilon}_T)^\prime$. $\bm{Y}$ and $\bm{E}$ are $T \times M$ matrices and $\bm{X}$ is a $T \times K$ matrix. Further, let $\bm{\phi} = \mathrm{vec}(\bm{\Phi})=(\phi_1,\dots, \phi_n)^\prime$, where $n=KM$ denotes the number of autoregressive coefficients.

As our approach to inference is Bayesian, we have to specify prior distributions. The generic prior for the coefficient vector is multivariate normal: 
$\bm{\phi} \sim \bm{N(0, \underline{V})}$. For growth rates and/or approximately stationary transformed data it is common to center the prior at zero \citep[e.g.,][]{george_bayesian_2008,koop_bayesian_2010,cross_macroeconomic_2020, kastner_sparse_2020}, whereas for data in levels often the prior mean of own-lag coefficients in the first lag is set to one \citep{litterman_forecasting_1986, sims_bayesian_1998}. Moreover, $\underline{\bm{V}}$ is a diagonal $n \times n$ matrix with diagonal elements $v_1, \dots, v_n$. The shrinkage priors under scrutiny which are to be detailed in Sections~\ref{sec:HM} to \ref{sec:semi} distinguish themselves in their treatment of $\bm{\underline{V}}$. 

In view of the next sections, we want to highlight that the prior for the VAR coefficients does not depend on $\bm{\Sigma}_t$ and hence neither on $\bm{U}$. This implies, that all considerations in Sections~\ref{sec:HM} to \ref{sec:semi} and Sections~\ref{sec:univariate} to \ref{sec:multivariate} also apply to other specifications of the error covariance matrix, e.g.\ the order-invariant factor stochastic volatility specification without restrictions on the factor loadings proposed in \cite{kastner_sparse_2020}, the order invariant stochastic volatility specification in \citet{chan_large_2021} or the order invariant specification based on the eigendecomposition of the error covariance matrix proposed in \citet{WuKoop2022}.

\revision{Instead of modelling the reduced-form coefficients, one could also model the coefficients $\bm B=\bm{\Phi U}$ of the VAR in structural form: $\bm{y}_t^\prime \bm{U} = \bm{x}_t^\prime \bm{B} + \bm{\xi}_t^\prime$, where $\bm B = \bm \Phi U$ and $\bm \xi_t = \bm \epsilon_t \bm U \sim N(\bm 0, \bm D_t)$. Often, this is done to speed up computations, as it allows for embarrassingly parallel ``equation per equation'' estimation in the spirit of \citet{carriero_large_2019}. Discussing whether it is better to shrink the reduced-form or the structural-form coefficients is out of scope of this paper. However, note that prior distributions are not translation invariant per se, and any comparison amongst the different settings must be treated with great care (cf.\ Appendix \ref{sec:reducedstructural} for a demonstration). Performance of state-of-the art shrinkage priors in the structural setting are discussed in \citet{cross_macroeconomic_2020} and \citet{chan_minnesota-type_2021}.}
\subsection{Shrinkage based on domain knowledge: Minnesota priors}
\label{sec:HM}
\paragraph{Traditional Minnesota prior}
The prior proposed in \citet{litterman_forecasting_1986} -- we refer to as traditional Minnesota prior (MP\_LIT) in the following -- is mainly characterized by two assumptions based on domain knowledge: First, own-lags are assumed to account for most of the variation of a given variable. Second, recent lags are assumed to be more important in predicting current values than distant lags. Hence, $\underline{\bm{V}}$ is structured in a way, such that the \revision{diagonal elements of the blocks in $\bm{\Phi}$ corresponding to $\bm A_j$, $j=1,\dots,p$ (the entries of the own-lag coefficients), are shrunken less than the off-diagonal elements.} And, coefficients associated with more recent lags are shrunken less than the ones associated with more distant lags. We follow \citet{koop_bayesian_2010} in setting the notation: Denote $\mathbf{\underline{V}}_i$ the block of $\mathbf{\underline{V}}$ that corresponds to the $K$ coefficients in the $i$th equation, and let $\mathbf{\underline{V}}_{i,jj}$ be its diagonal elements. The diagonal elements are set to
\begin{equation}\label{eq:HMv}
	\mathbf{\underline{V}}_{i,jj}=
	\begin{cases}
		\frac{\lambda_1}{r^2} & \text{for coefficients on own lag $r$ for $r=1, \dots,p$}, \\
		\frac{\lambda_2 \hat{\sigma}^2_{i}}{r^2 \hat{\sigma}^2_{j}} & \text{for coefficients on lag $r$ of variable $j \neq i$}, 
	\end{cases}
\end{equation}
where $\hat{\sigma}^2_{i}$ is the OLS variance of a univariate AR(6) model of the $i$th variable. We set $\lambda_1 > \lambda_2$ to regularize cross-lag coefficients more heavily. The term $r^2$ in the denominator automatically imposes more shrinkage on the coefficients towards their prior mean as lag length increases. The term $\frac{\hat{\sigma}^2_{i}}{\hat{\sigma}^2_{j}}$ adjusts not only for different scales in the data, it is also intended to account for different scales of the responses of one economic variable to another \citep{litterman_forecasting_1986}.

\paragraph{Semi-hierarchical Minnesota prior}
The semi-hierarchical Minnesota (SHM) prior is a generalization of the traditional Minnesota prior. 
Instead of fixing the shrinkage parameters $\lambda_1$ and $\lambda_2$, SHM treats them as unknown quantities and learns them in a data-based fashion. That is, the strong assumption that $\lambda_1 > \lambda_2$ is now replaced with the weaker assumption that own-lags and cross-lags might account for different amounts in the variation of a given variable. We follow \citet{huber_adaptive_2019} in imposing gamma priors on $\lambda_1$ and $\lambda_2$, i.e.,
$
	\lambda_i \sim G(c_i,d_i) \text{ for } i=1,2,
$
where $G(\alpha,\beta)$ denotes the probability density function of the gamma distribution with shape $\alpha$ and rate $\beta$. 

As default choice, we set $c_i=d_i=0.01$ for $i=1,2$. The resulting gamma distribution has expectation 1 and variance 100. Moreover, the density diverges to infinity as $\lambda_j \rightarrow 0$. As most mass of this distribution is at zero, the prior can heavily shrink the parameters if necessary, though as soon as there are some moderate to large nonzero coefficients (in absolute values), the shrinkage will be only moderate to weak. Since the prior variances of cross-lag coefficients still have to be scaled with $\frac{\hat{\sigma}^2_i}{\hat{\sigma}^2_j}$, it is not a fully hierarchical prior, and we refer to it as the semi-hierarchical Minnesota prior.
\subsection{Off-the-shelf shrinkage: Global-local priors} 
Other than the Minnesota priors, global-local (GL) shrinkage priors are not built upon domain knowledge and are neither specifically designed for VARs nor for forecasting macroeconomic data. They are very popular in sparse and high-dimensional applications, and therefore also well suited for large VARs \citep{follett_achieving_2019, huber_adaptive_2019, kastner_sparse_2020}. Following \citet{polson_shrink_2010}, GL priors can be written as continuous scale mixture distributions. Let $K(\delta)$ denote a symmetric unimodal density with variance $\delta$, then a typical GL prior is of the following form:
	\begin{equation}
		\phi_i | \vartheta_{i} \zeta \sim K(\vartheta_{i}\zeta), \quad \vartheta_{i} \sim f, \quad \zeta \sim g \quad \text{for } i=1,\dots,n,
	\end{equation} 
where $\zeta$ represents the global shrinkage and $\vartheta_{i}$ the local shrinkage. While the global parameter determines the overall shrinkage, the local parameters act to detect the relevant signals. Hence, $g$ is usually a density with substantial mass at zero. In case of $\zeta$ being very small, $f$ must be a heavy-tailed density such that $\vartheta_i$ can override the effect of $\zeta$ for a nonzero coefficient. GL priors distinguish themselves in their choices for $f$ and $g$, which will be discussed in the next section.
\subsection{Structured semi-global(-local) shrinkage}
\label{sec:semi}
In this section, we aim at building a framework that allows to put more structure on off-the-shelf shrinkage priors, but with as little restrictions as possible.

Even though many applications have shown that a simple prior like the traditional Minnesota prior -- where all hyperparameters are treated as known and therefore are not estimated from the data in a hierarchical fashion -- can be very competitive, it lacks flexibility. First, due to the fact that own-lag shrinkage and cross-lag shrinkage are fixed a priori, the traditional Minnesota prior is not (automatically) adaptable. Especially in higher dimensions, well-established default values are difficult to determine, which in turn makes selecting them hard to justify. This issue is partly dealt with by considering semi-hierarchical versions of the prior (cf.\ Section \ref{sec:HM}). Second, it relies on tuning the prior variances with OLS estimates in order to adapt for the different scales of the endogenous variables. Third, shrinkage of higher order lags is deterministic, i.e., prior variances are additionally scaled by $r^x$ where $r$ denotes the lag and usually $x=2$. Theoretically, GL priors could remedy those shortcomings with their local scales, which individually adjust the regularization for each coefficient. A problem arises, however, when the noise level and the absolute level of the coefficients are different in different parts of the parameter space. Then, the adjustment is probably not strong enough. In what follows, we introduce a simple framework that tackles this problem. It only needs little input from the researcher and does not require any pre-tuning of hyperparameters.

Let $\mathcal{A}_j$ denote the generic index set that labels the coefficients of the $j$th group in $\bm{\phi}$ (e.g., the first group could be the own-lag coefficients associated with the first lag, the second group could be the cross-lag coefficients associated with the first lag, etc.), and let $n_j$ denote the number of elements in $\mathcal{A}_j$. Then, as a semi-global local prior with $k$ groups, we define the following hierarchical representation:
\begin{equation}\label{eq:sgl_form}
\phi_i \sim K(\vartheta_i \zeta_j), \quad
\vartheta_i \sim f, \quad \zeta_j \sim g, \quad i \in \mathcal{A}_j, \quad j = 1,\dots,k.    
\end{equation}
The only additional input required is the partitioning of $\bm{\Phi}$ into $k$ subgroups. We propose a partitioning that mimics two features of the Minnesota prior, namely the distinction between own-lag and cross-lag coefficients and the distinction between lags in general: In each lag, the diagonal elements (the own-lags) and the off-diagonal elements (the cross-lags) constitute separate groups, which makes $2p$ groups in total. \revision{The vanilla global-local prior, i.e.\ no partitioning of $\bm{\Phi}$, is nested as the special case where $k=1$.}

The semi-global framework nests other modifications to global-local priors within the VAR framework as special cases. E.g., \citet{huber_adaptive_2019} consider both, equation-specific and covariate-specific shrinkage. The first means that the covariates of each equation form $M$ separate groups (individual degree of shrinkage for each equation) and the latter that the specific covariates across all equations form $K$ separate groups (individual degree of shrinkage for each covariate across all equations). The multiplicative lag-wise specification considered in the same paper, however, is not nested in the semi-global framework, since we assume independence across the $k$ groups.
The adaptive Minne\-sota prior in
\citet{chan_minnesota-type_2021} could also be cast in the semi-global framework. It is the unification of the Minnesota prior with a global-local prior, namely the normal-gamma prior \citep{griffin2010}. That is, there are two semi-global shrinkage parameters: One for own-lags and one for cross-lags. Additionally, each coefficient is assigned a local scale. The prior variances are rounded off with the scaling term $\frac{\hat{\sigma}^2_{i}}{r^2\hat{\sigma}^2_j}$ of the Minnesota prior in analogy to Eq.~\ref{eq:HMv}. Besides using quantities estimated from the data to tune the prior variances, the adaptive Minnesota prior differs from our approach as it is a prior for the structural coefficients. \revision{Another class of priors, which could be extended to semi-global priors are the asymmetric conjugate priors for homoskedastic VARs proposed in \cite{chan2022}. Similar as with the adaptive Minnesota prior, the asymmetric conjugate priors exploit the structural form of the VAR.}

In the remainder, we briefly describe several state-of-the art off-the-shelf shrinkage priors and how they can (in one case even have to) be modified, such that the pooling strategy introduced by the semi-global framework materializes.
\paragraph{DL prior}
The Dirichlet-Laplace (DL) prior, introduced in \citet{bhattacharya_dirichletlaplace_2015} and put forward to the VAR framework in \citet{kastner_sparse_2020}, takes the following hierarchical form:
\begin{align}\label{eq:dl0}
	\phi_i| \varrho_{i},\omega_j\sim DE(\varrho_{i}\omega_j), \quad 
 \bm{\varrho} \sim Dir(a^{\gamma},\dots,a^\gamma), \quad
	\omega_j \sim G(n_ja^\gamma, 1/2),
\end{align}
where $DE(\delta)$ denotes the double exponential (sometimes called Laplace) distribution with variance $2\delta^2$, and $Dir(\alpha_1,\dots,\alpha_K)$ denotes the Dirichlet distribution on the open $K-1$ simplex with concentration parameters $\bm{\alpha}=(\alpha_1,\dots,\alpha_K)$. Letting $\sqrt{\vartheta_i/2} = \varrho_i \omega_j$ for $i \in \mathcal{A}_j$ and $j=1,\dots,k$, it can be shown that $\sqrt{\vartheta_i/2} \sim G(a^\gamma,1/2)$ independently for $i \in \mathcal{A}_j$, $j=1,\dots,k$ \citep[cf.\ Lemma IV.3 in][]{zhou_carin_2012}. Using this result, it becomes clear that the proposed semi-global pooling strategy is not compatible with the vanilla DL prior, since there is no information sharing on the (semi-)global level. Therefore, to make things work, we replace the global shape parameter $a^\gamma$ with the semi-global shape parameter $a_j^\gamma$, $j=1,\dots,k$. Applying the Lemma, the semi-global-local DL prior has the following form:
\begin{align}\label{eq:dle}
 \phi_i|\vartheta_i,\zeta_j &\sim DE(\sqrt{\vartheta_i\zeta_j/2}), \quad 
 \sqrt{\vartheta_i/2} \sim G(a_j^\gamma,1/2), \quad \zeta_j=1,
 \end{align}
 where $Var(\phi_i|\vartheta_i,\zeta_j)=\vartheta_i \zeta_j$. Posterior results can be very sensitive to the choice of $a^\gamma_j$, as it is the parameter that controls the strength of regularization (cf.\ Section \ref{sec:sparseness}). Hence, we place i.i.d.\ discrete hyperpriors $Pr(a_j^\gamma = \tilde{a}^\gamma_r) = p^\gamma_r$ on $a^\gamma_j$ for $j=1,\dots,k$, where $\bm{\tilde{a}}^\gamma=(\tilde{a}^\gamma_1,\dots,\tilde{a}^\gamma_R)^\prime$ is the vector of strictly positive support points. Updates of $a_j^\gamma$ are straightforward, as the full conditional posterior is again discrete on the same support points.
 
 Last but not least, the DL prior can be expressed as a Gaussian scale mixture by introducing the auxiliary scaling parameter $\psi_i \sim Exp(1/2)$, s.t.\ $\phi_i|\vartheta_i,\zeta_j \sim DE(\sqrt{\vartheta_i\zeta_j/2}) \Longleftrightarrow \phi_i|\psi_i,\vartheta_i,\zeta_j \sim N(0,\psi_i\vartheta_i\zeta_j/2)$. Consequently, for DL, $v_i = \psi_i\vartheta_i\zeta_j/2$.
\paragraph{NG prior}
The normal-gamma prior proposed by \citet{griffin2010} is nowadays well-established in different kinds of multivariate time-series modelling setups \citep[e.g.,][]{huber_adaptive_2019, kastner_sparse_2019, bitto2019achieving}. It takes the following form:
\begin{align}\label{eq:ng}
    \phi_i|\vartheta_i\zeta_{j} \sim N(0,\vartheta_i \zeta_j), \quad \vartheta_i \sim G(a^\delta_j, a^\delta_j/2), \quad \zeta_j^{-1} \sim G(b^\delta,c),
\end{align} 
i.e., for NG, $v_i=\vartheta_i \zeta_j$.
As with DL, we place on $a^\delta_j$, $j=1,\dots,k$, i.i.d.\ discrete hyperpriors $Pr(a_j^\delta = \tilde{a}^\delta_r) = p^\delta_r$, where $\bm{\tilde{a}}^\delta=(\tilde{a}^\delta_1,\dots,\tilde{a}^\delta_R)^\prime$ is the vector of strictly positive support points.
\paragraph{R2D2 prior}
Only recently, \citet{zhang_bayesian_2020} proposed the $R^2$-induced Dirichlet decomposition (R2D2) prior. It is based on a beta prior on $R^2 \coloneqq \frac{W}{W + 1}$, where $W$ denotes the sum of all prior variances. The induced prior then takes the following form:
\begin{align}\label{eq:r2d20}
    \phi_i|\varrho_{i},\omega_j\sim DE\left(\sqrt{\varrho_i\omega_j/2}\right), \quad \bm{\varrho}_j = (\varrho_i)_{i\in \mathcal{A}_j} \sim Dir(a^{\pi},\dots,a^{\pi}), \quad \omega_j|\xi_j \sim G(n_j a^{\pi},\xi_j), \quad \xi_j\sim G(b^\pi,1).
\end{align}
Apart from a higher level of hierarchy, it has a similar structure to the DL prior. A major difference is that, for R2D2, the prior variance follows a gamma distribution, whereas for DL, the prior scale follows a gamma distribution. Using the same result as before, conditional on $\xi_j$, $\varpi_i=\varrho_i\omega_j \sim G(a_\pi, \xi_j)$ independently. Exploiting the scaling property of the gamma distribution, the R2D2 prior can be cast in the form of Eq.\ \ref{eq:sgl_form}:
\begin{align}\label{eq:r2d2}
    \phi_i|\vartheta_i,\zeta_j \sim DE\left(\sqrt{\vartheta_i\zeta_j/2}\right), \quad \vartheta_i \sim G(a^{\pi},a^{\pi}/2), \quad \zeta_j^{-1} \sim G(b^\pi,a^{\pi}/2),
\end{align}
where $\zeta_j^{-1}=\frac{2\xi_j}{a_\pi}$ and $\vartheta_i = \frac{\varpi_i}{\zeta_j}$. By introducing auxiliary scaling parameters $\psi_i \sim Exp(1/2)$, the R2D2 prior can be written as a Gaussian scale mixture, s.t.\  $\phi_i|\vartheta_i,\zeta_j \sim DE\left(\sqrt{\vartheta_i\zeta_j/2}\right) \Longleftrightarrow \phi_i|\psi_i,\vartheta_i,\zeta_j \sim N\left(0,\psi_i\vartheta_i\zeta_j/2\right)$, i.e.\ for R2D2, $v_i=\psi_i\vartheta_i\zeta_j/2$.

The R2D2 prior and the NG prior are closely related. For R2D2, the variance of a double exponential distribution follows a gamma distribution, whereas for NG, the variance of a normal distribution follows a gamma distribution. The special case of NG with $c=\frac{a^\delta_j}{2}$, hence, could be seen as an $R^2$-induced Dirichlet decomposition prior with a normal kernel.

As with DL and NG, we place on $a^{\pi}_j$, $j=1,\dots,k$, i.i.d.\ discrete hyperpriors $Pr(a_j^\pi = \tilde{a}^\pi_r) = p^\pi_r$, where $\bm{\tilde{a}}^\pi=(\tilde{a}^\pi_1,\dots,\tilde{a}^\pi_R)^\prime$ is the vector of strictly positive support points.
\paragraph{Horseshoe}
The horseshoe (HS) prior proposed in \citet{carvalhi2010} is given by
\begin{align}\label{eq:hs}
    \phi_i | \vartheta_i\zeta_j \sim N(0,\vartheta_i \zeta_j), \quad \sqrt{\vartheta_i} \sim \mathcal{C}^+ (0,1), \quad \sqrt{\zeta_j} \sim \mathcal{C}^+ (0,1), 
\end{align}
where $\mathcal{C}^+(\cdot,\cdot)$ denotes the half-Cauchy distribution. Other than NG and R2D2, HS places the hyperprior distributions on the standard deviation, rather than on the variance. Compared to all other priors discussed so far, it has the advantage that no tuning parameter has to be specified by the user.
\paragraph{Stochastic search variable selection}
\label{sec:SSVS}
Another popular prior for BVARs is the stochastic search variable selection prior \citep[SSVS,][]{george_bayesian_2008}, which is a discrete mixture of two normal distributions. Although this prior is neither a global-local shrinkage prior nor a continuous scale mixture prior, it can be modified in order to comply with the semi-global-local framework. In hierarchical form and adapted to the semi-global-local framework, the prior is given through
\begin{align}
	\phi_i | \gamma_i \sim N\left(0, (1-\gamma_i)\tau_{0i}^2 + \gamma_i\tau_{1i}^2 \right), \quad
	\gamma_i| \underline{p}_j \sim Bernoulli(\underline{p}_j),
\end{align}
where $\gamma_i$ is an auxiliary dummy variable, $\tau_{0i}$ should be small (close to zero) and it must hold that $\tau_{0i} \ll \tau_{1i}$.  Intuitively speaking, if $\phi_i$ is assigned to the spike variance $\tau^2_{0i}$, it is virtually constrained to zero. If $\phi_i$ is assigned to the slab variance $\tau^2_{1i}$, it is shrunken depending on how $\tau^2_{1i}$ is chosen. Further, we assume that the prior inclusion probability $\underline{p}_j$ might differ between the $k$ groups. In order to learn the parameter in a data-based fashion, we impose i.i.d\ beta priors on each $\underline{p}_j$, $j=1,\dots,k$: $\underline{p}_j \sim Beta(s_1,s_2)$. Note, that for prediction purposes and for inference on the joint posterior $p(\phi_i,\gamma_i|\bullet)$, learning $\underline{p}_j$ in a data-based fashion only has an effect if there is information sharing between coefficients. In the case where each $\phi_i$ constitutes a separate group, i.e., each $\phi_i$ is assigned an independent $\underline{p}_i$ \citep[as in, e.g.,][]{cross_macroeconomic_2020}, it is straightforward to show that the full conditional posterior $p(\gamma_i|\bullet)$ only depends on the prior expectation of $\underline{p}_i$.

\subsection{Priors for the variance-covariance matrix}
\label{sec:prior_sigma}
To complete the models, we have to specify the priors on the decomposed variance-covariance matrix $\bm{\Sigma}_t = \bm{U}^{\prime -1}\bm{D}_t\bm{U}^{-1}$. Note that the precision matrix -- the inverse of the variance-covariance matrix -- can be written as $\bm{\Sigma}_t^{-1} = \bm{U D}_t^{-1} \bm{U}^\prime$. It is well-known that the precision matrix should be a sparse matrix, as zero entries on the off-diagonals imply conditional independence among the respective equations \citep{west_bayesian_2020}. Hence, for $\bm{u}$, we choose the HS prior: $u_{ij} \sim N(0,\vartheta_{ij}^u\zeta^u)$, $\sqrt{\vartheta_{ij}^u} \sim \mathcal{C}^+(0,1)$, $\sqrt{\zeta^u} \sim \mathcal{C}^+(0,1)$, $i=1,\dots,M-1$ and $j=2,\dots M$.\footnote{In Section \ref{sec:robustness}, we provide some robustness checks regarding different priors for $\bm{u}$.} Following \citet{kim_stochastic_1998}, for $i=1, \dots, M$, we choose a normal prior for the level of the log variance, $\mu_i \sim N(0, 100^2)$, and for the persistence parameter, $\rho_i$, a beta distribution on $\frac{\rho_i + 1}{2} \sim Beta(20, 1.5)$. Finally, we follow \citet{kastner_ancillarity-sufficiency_2014} by imposing a gamma prior on the variance of the log variance, $\sigma_i^2 \sim G(1/2, 1/2)$. 

\revision{Regarding our prior choices for the SV parameters and related to the finding in \citet{rossiniLossbasedPriorDegrees2024}, that forecasting performance of homoskedastic VARs in conjunction with a Wishart prior on $\bm \Sigma^{-1}$ can be negatively influenced by choosing an unfavorable default value for the degrees of freedom hyperparameter, we acknowledge that choosing different priors for the vector of SV parameters $(\mu_i, \rho_i, \sigma_i^2)$ can have a non-negligible effect on the posterior distribution of those parameters. However, the impact on the estimated log-variances $h_{it}$, which are most relevant for forecasting, is typically small.}
\subsection{Posterior estimation}
Efficient computer implementations of all posterior samplers discussed in this paper are conveniently bundled into 
the R \citep{R_core} package \texttt{bayesianVARs},
interfacing to C/C++ via \texttt{Rcpp} \citep{rcpp} and \texttt{RcppArmadillo} \citep{rcpparmadillo} for increased computational efficiency. The package is available under the General Public License (GPL $\ge$ 3) from the Comprehensive R Archive Network (CRAN) at \url{https://cran.r-project.org/package=bayesianVARs}. 
We thus refrain from describing the various posterior samplers in detail. Instead, we provide their main building blocks, the necessary conditionals used for Gibbs sampling, in Appendix \ref{sec:fcp}.

\revisiontwo{Concerning the scalability of our approach, we present some rough estimates of computation time. The estimates are based on an Intel\textsuperscript{\textregistered} Core\textsuperscript{\texttrademark} i7-10610U processor using one core. For the dimensions of our empirical application, i.e., $M=21$, $T=198$ and $p = 1,\dots,5$, generating 100 draws from the posterior distribution takes approximately 1.9, 5.7, 11.4, 19.1, and 28.6 seconds, respectively. Since the complexity of the underlying correct triangular algorithm \citep{carriero_corrigendum_2021} is $O(M^4)$, the estimation of larger datasets up to $M\approx50$ is surely possible. For even higher dimensions, approximate estimation techniques or a different structure of the error variance-covariance matrix could be considered. Instead of using the correct triangular algorithm, one could use the equation-per-equation algorithm described in \citet{carriero_large_2019} as an approximation. In that paper, a VAR with $p=13$ lags is successfully estimated using a time series with $M=113$ and $T=648$, which is similar in dimensionality to the application in \citet{banbura_large_2010}. The approach of the latter, however, builds on a substantially different model, namely a homoskedastic and conjugate VAR. Variational inference as proposed in \citet{bernardi2022}, another approximate estimation technique, can also reduce estimation time of VARs which are similar in spirit to those presented in this paper. Concerning the error structure, instead of using the Cholesky stochastic volatility specification in \eqref{eq:cogley} and \eqref{eq:usv}, one could use the the factor stochastic volatility (FSV) specification proposed in \citet{kastner_sparse_2020}. Their variant is also compatible with the priors for the VAR coefficients described in Sections~\ref{sec:HM} to \ref{sec:semi} and Sections~\ref{sec:univariate} to \ref{sec:multivariate}. By conditioning on the factors and their loadings, this approach makes conditional equation-per-equation estimation straightforward. In situations where $K \gg T$, equation-per-equation estimation can be further combined with the algorithm for high-dimensional regressions proposed in \citet{bhattacharyaFastSamplingGaussian2016}. This then has complexity $O(pM^2T^2)$ and \citet{kastner_sparse_2020}, e.g., demonstrate that MCMC inference for a VAR with $p=5$ lags using a time series with $M=215$ and $T=228$ becomes feasible.}
\section{Shrinkage behavior and sparsity of various priors}
\label{sec:sparseness}
\revision{In Section \ref{sec:univariate}, we analyse both the tail behavior and the concentration at zero of the various shrinkage priors. In Section \ref{sec:multivariate}, we characterize the sparsity of the priors.}
\subsection{Univariate analysis}
\label{sec:univariate}
The limiting behavior of univariate global-local prior densities is typically analyzed conditionally on the global scale \citep[e.g.,][]{carvalhi2010, griffin2010}, i.e.\ marginalized over the local scales only. Since the global parameters often introduce dependence among all $\phi_i$, $i \in \mathcal{A}_j$, $j=1,\dots,k$, we follow this line of thinking and conduct the analysis in Section~\ref{sec:univariate} in terms of univariate conditionally independent prior densities $p(\phi_i|\cdot)$, where $\cdot$ stands for all global hyperparameters.
\begin{theorem}[Limiting behavior of shrinkage priors]\label{thm:lim}
    \begin{enumerate}[label=\emph{\alph*})]
        \item\label{thm:lim_tails} \textbf{Tail behavior}

        For $|\phi_i| \rightarrow \infty$, the marginal densities satisfy
        \begin{align}
            p^\infty_{DL}(\phi_i|a_j^\gamma)&=O\left(|\phi_i|^{a_j^\gamma/2 - 3/4}\exp\left\lbrace -\sqrt{2|\phi_i|} \right\rbrace\right), \quad \text{for any } a_j^\gamma>0, \label{thm:dl_tail}\\
            p^\infty_{NG}(\phi_i|\zeta_j,a^\delta_j)&=O\left(|\phi_i|^{a^\delta_j-1} \exp\left\lbrace - \sqrt{\frac{a^\delta_j}{\zeta_j}} |\phi_i| \right\rbrace\right), \quad \text{for any } a_j^\delta>0, \label{thm:ng_tail}\\
            p^\infty_{R2D2}(\phi_i|\zeta_j,a^{\pi}_j)&=O\left(|\phi_i|^{\frac{2a-2}{3}}\exp\left\lbrace-3 \left(\frac{\phi_i^2 a^{\pi}_j}{4\zeta_j}\right)^{1/3} \right\rbrace \right), \quad \text{for any } a_j^\pi>0,\label{thm:r2d2_tail}\\ 
            p^\infty_{HS}(\phi_i|\zeta_j)&=O\left( \frac{1}{\phi_i^{2}} \right),\label{thm:hs_tail}\\
            p^\infty_{SHM}(\phi_i|\mathbf{\underline{V}}_{i,jj})&=O\left(\exp\left\lbrace \frac{-\phi_i^2}{2\mathbf{\underline{V}}_{i,jj}} \right\rbrace\right), \quad \text{($\mathbf{\underline{V}}_{i,jj}$, see \eqref{eq:HMv})} \label{thm:shm_tail}\\
            p^\infty_{SSVS}(\phi_i|\underline{p}_j, \tau_{0i}, \tau_{1i})&=O\left( \exp\left\lbrace - \frac{\phi_i^2}{2(\underline{p}_j \tau_{1i}^2 + (1-\underline{p}_j) \tau_{0i}^2)} \right\rbrace \right).\label{thm:ssvs_tail}
        \end{align}
        \item\label{thm:lim_zero} \textbf{Concentration at zero}

        For $|\phi_i| \rightarrow 0$, the marginal densities satisfy
        \begin{align}
            p^0_{DL}(\phi_i|a_j^\gamma)&=
          \begin{cases}
             O\left( \frac{1}{|\phi_i|^{1-a_j^\gamma}} \right) & \text{if } 0<a_j^\gamma<1, \\
             O\left(\log\left( \frac{1}{|\phi_i|} \right)\right) & \text{if } a_j^\gamma=1, \\
             \text{no singularity} & \text{if } a_j^\gamma>1,
          \end{cases} \label{thm:dl_zero}\\
          p^0_{NG}(\phi_i|\zeta_j,a^\delta_j)&=
          \begin{cases}
             O\left( \frac{1}{|\phi_i|^{1-2a^\delta_j}} \right) & \text{if } 0<a^\delta_j<\frac{1}{2}, \\
             O\left(\log\left( \frac{1}{|\phi_i|} \right)\right) & \text{if } a^\delta_j=\frac{1}{2}, \\
             \text{no singularity} & \text{if } a^\delta_j>\frac{1}{2},         
          \end{cases} \label{thm:ng_zero}\\
          p^0_{R2D2}(\phi_i|\zeta_j,a^{\pi}_j)&=
             O\left( \frac{1}{|\phi_i|^{1-2a^{\pi}_j}} \right),\quad \text{for } 0<a^{\pi}_j<\frac{1}{2}, \label{thm:r2d2_zero}\\
          p^0_{HS}(\phi_i|\zeta_j)&=O\left( \log\left(\frac{1}{|\phi_i|}\right)\right)\label{thm:hs_zero}. 
        \end{align}
        The marginal densities of SHM/MP\_LIT and SSVS have no singularity at zero.
    \end{enumerate}
\end{theorem}
For the proofs of Theorem \ref{thm:lim}\ref{thm:lim_tails}(\ref{thm:dl_tail}) and the first part of \ref{thm:lim}\ref{thm:lim_zero}(\ref{thm:dl_zero}), we refer to \citet{zhang_bayesian_2020}, and for \ref{thm:lim}\ref{thm:lim_tails}(\ref{thm:hs_tail}) and \ref{thm:lim}\ref{thm:lim_zero}(\ref{thm:hs_zero}), we refer to \citet{carvalhi2010}. The remaining proofs can be found in Appendix \ref{ap:thm}.

Some noteworthy remarks follow. All global-local priors share a pole at $\phi_i=0$, at least for certain hyperparameter settings, which is important for handling vectors with many zero elements. The Minnesota priors and SSVS do not diverge to infinity at zero, which could be seen as a disadvantage in sparse settings. For $a_j^\gamma<1$, $a^\delta_j<\frac{1}{2}$, and $a^{\pi}_j<\frac{1}{2}$, DL, NG and R2D2, respectively, diverge to infinity with polynomial speed; this is faster than HS. Concerning the tails: HS has the heaviest (Cauchy-like) tails. It is followed by DL and R2D2, respectively, which are still heavy-tailed in the sense that they have heavier tails than the exponential distribution. NG is lighter-tailed than DL and R2D2, but still heavier-tailed than the Minnesota priors and SSVS, which have Gaussian-like tails.\footnote{A marginal note regarding R2D2: \citet{zhang_bayesian_2020} claim that R2D2 is polynomial in both regions at zero and in the tails, which seems contradictory to Theorem \ref{thm:lim}\ref{thm:lim_tails}(\ref{thm:r2d2_tail}). For the avoidance of doubt, we want to clarify that the univariate density of R2D2 is polynomial in the tails only if the global scale $\zeta_j$ is integrated out. This is somewhat misleading, since for all other priors under scrutiny in both papers, i.e. here and in \citet{zhang_bayesian_2020}, the global scale is fixed.}

\begin{figure}[t]
    \centering
    \includegraphics[width=\textwidth]{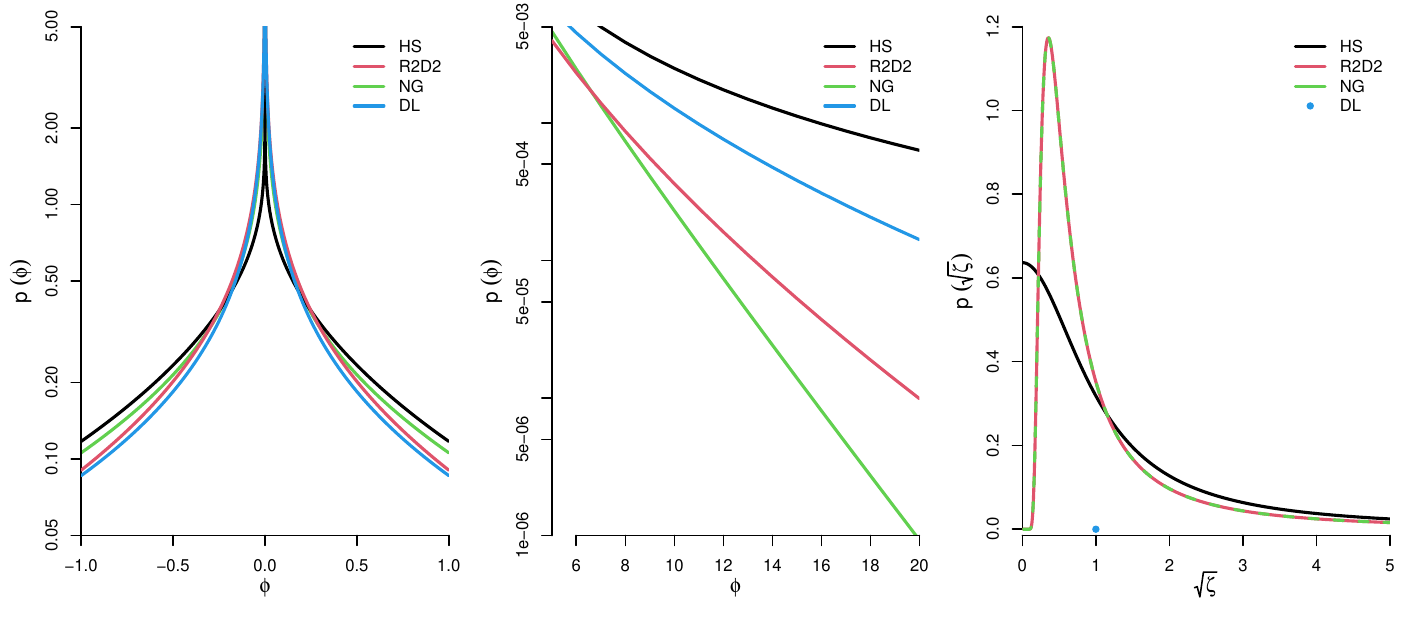}
    \caption{Left and middle panel: Visualization of univariate prior densities conditional on the global scales. Right panel: Visualization of prior densities for the global scales.}
    \label{fig:prior_dens}
\end{figure}

Figure \ref{fig:prior_dens} depicts the univariate densities conditional on the global scales and the prior densities for the global scales separately. For reasons of comparability, we show the densities for $\sqrt{\zeta}_j$ for all priors. It can easily be verified that for NG and R2D2, the gamma hyperpriors on $\zeta_j$ imply generalized gamma hyperpriors on $\sqrt{\zeta}_j$. Clearly, the division of tasks among global and local scales varies among different GL priors. Whereas the global scale of HS has considerable mass at zero, the global scale of NG and R2D2 is bounded away from zero. The global scale of DL is a Dirac-mass at one. That is, for DL, NG, and R2D2, the task of the local scales is twofold: In addition to detecting signals, they must be able to overwhelm the global scale in shrinking the zero coefficients to zero.

\subsection{Multivariate analysis}\label{sec:multivariate}
In order to describe the implied level of sparsity of the different priors, we now analyze their multivariate distributions. We begin by characterizing the sparsity of the priors qualitatively and continue by quantifying it via a sparseness measure.

\begin{figure}[t]
     \centering
     \begin{subfigure}[b]{0.30\textwidth}
         \centering
         \includegraphics[trim = 0 0 0 50, clip,width=\textwidth]{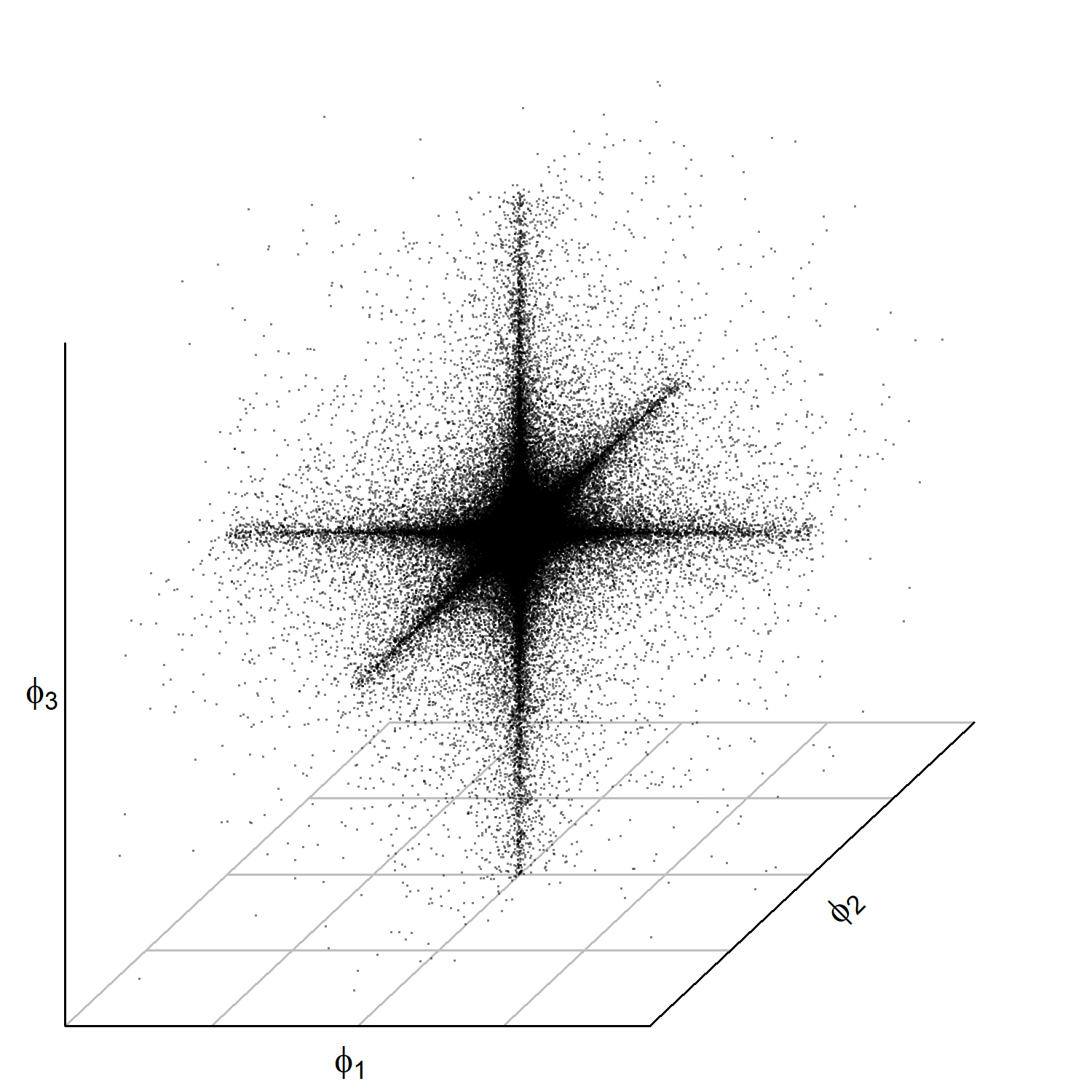}
         \caption{HS}
     \end{subfigure}
     \hfill
     \begin{subfigure}[b]{0.30\textwidth}
         \centering
         \includegraphics[trim = 0 0 0 50, clip,width=\textwidth]{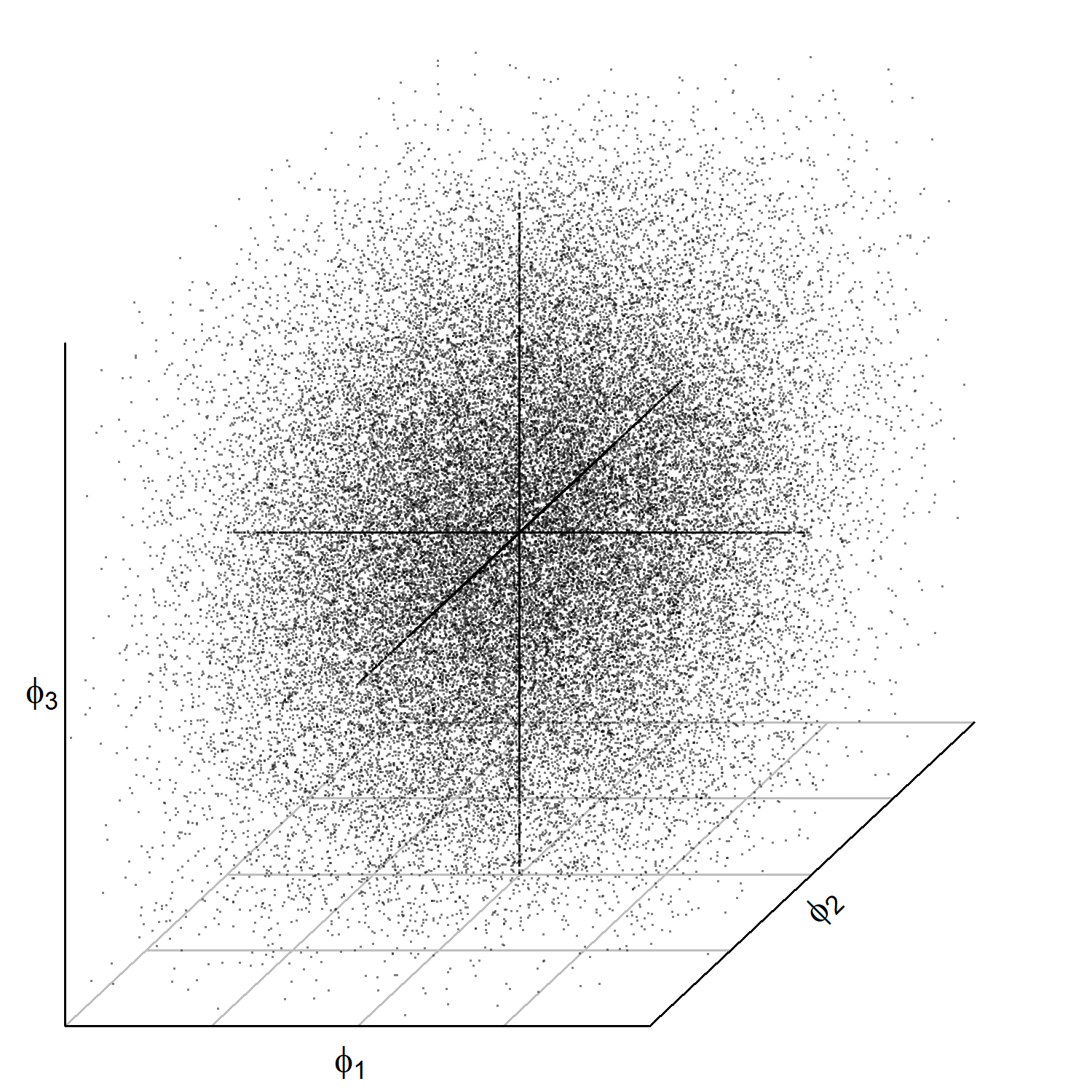}
         \caption{SSVS}
     \end{subfigure}
     \hfill
     \begin{subfigure}[b]{0.30\textwidth}
         \centering
         \includegraphics[trim = 0 0 0 50, clip,width=\textwidth]{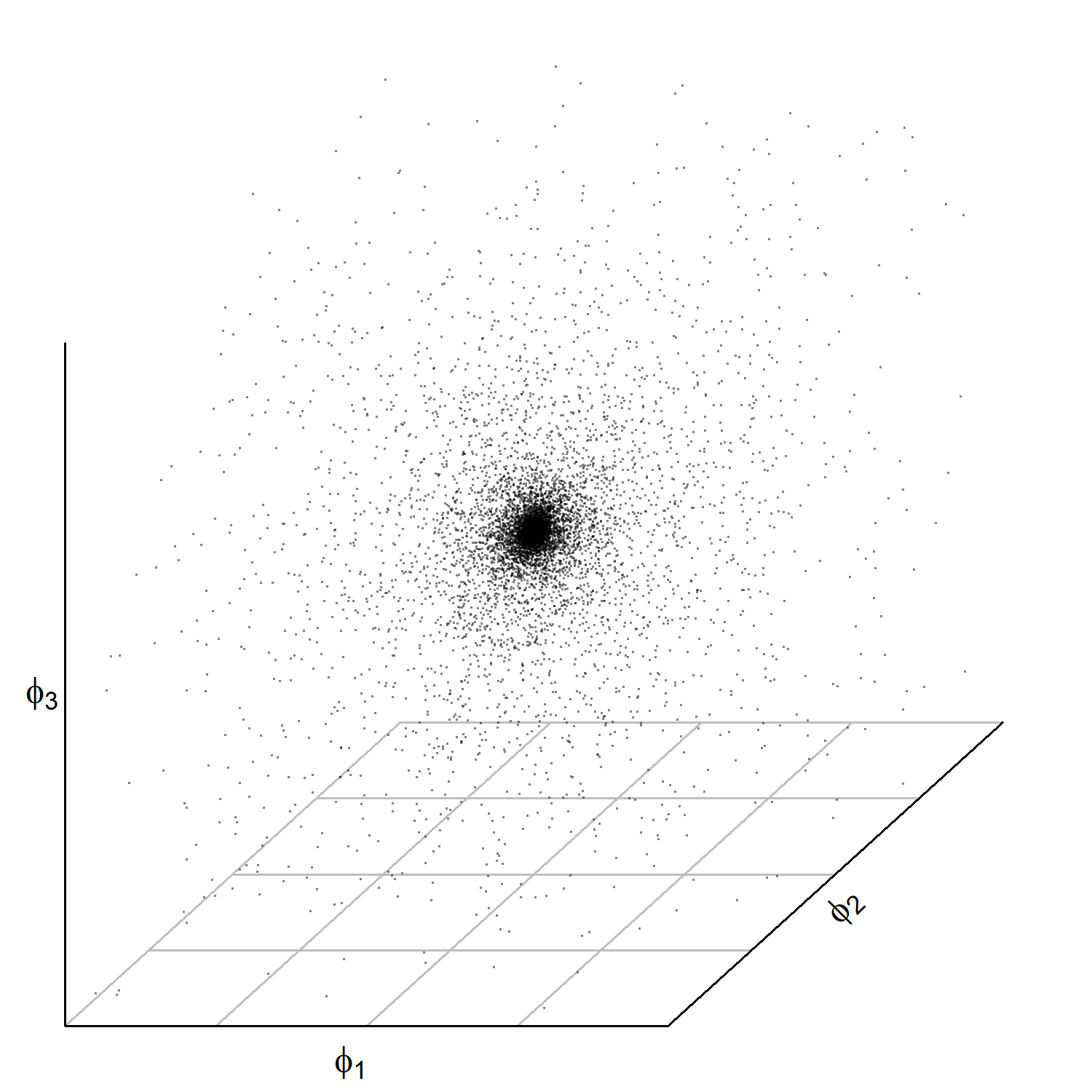}
         \caption{SHM}
     \end{subfigure}
     \hfill
     \begin{subfigure}[b]{0.30\textwidth}
         \centering
         \includegraphics[trim = 0 0 0 50, clip,width=\textwidth]{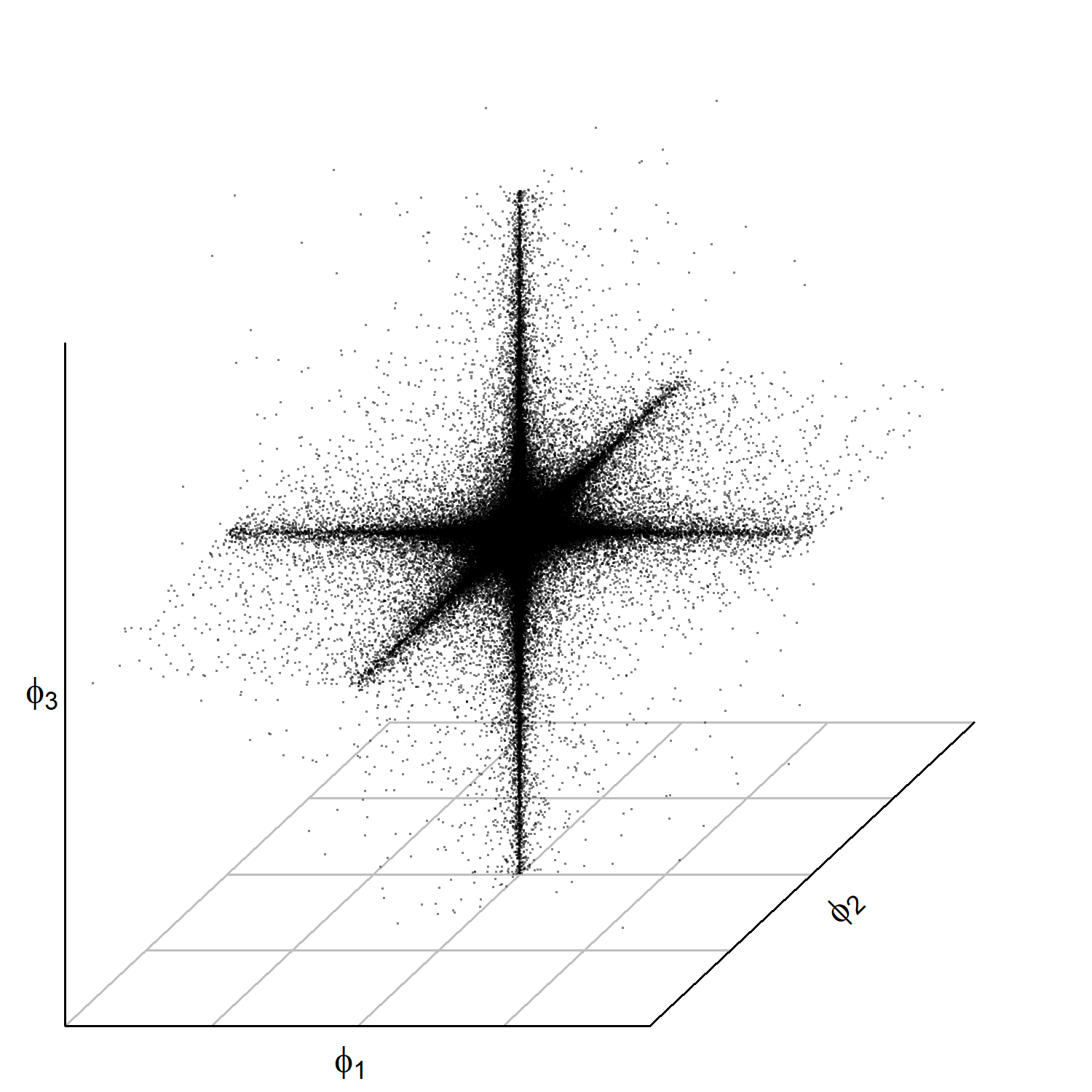}
         \caption{HS*}
     \end{subfigure}
     \hfill
     \begin{subfigure}[b]{0.30\textwidth}
         \centering
         \includegraphics[trim = 0 0 0 50, clip,width=\textwidth]{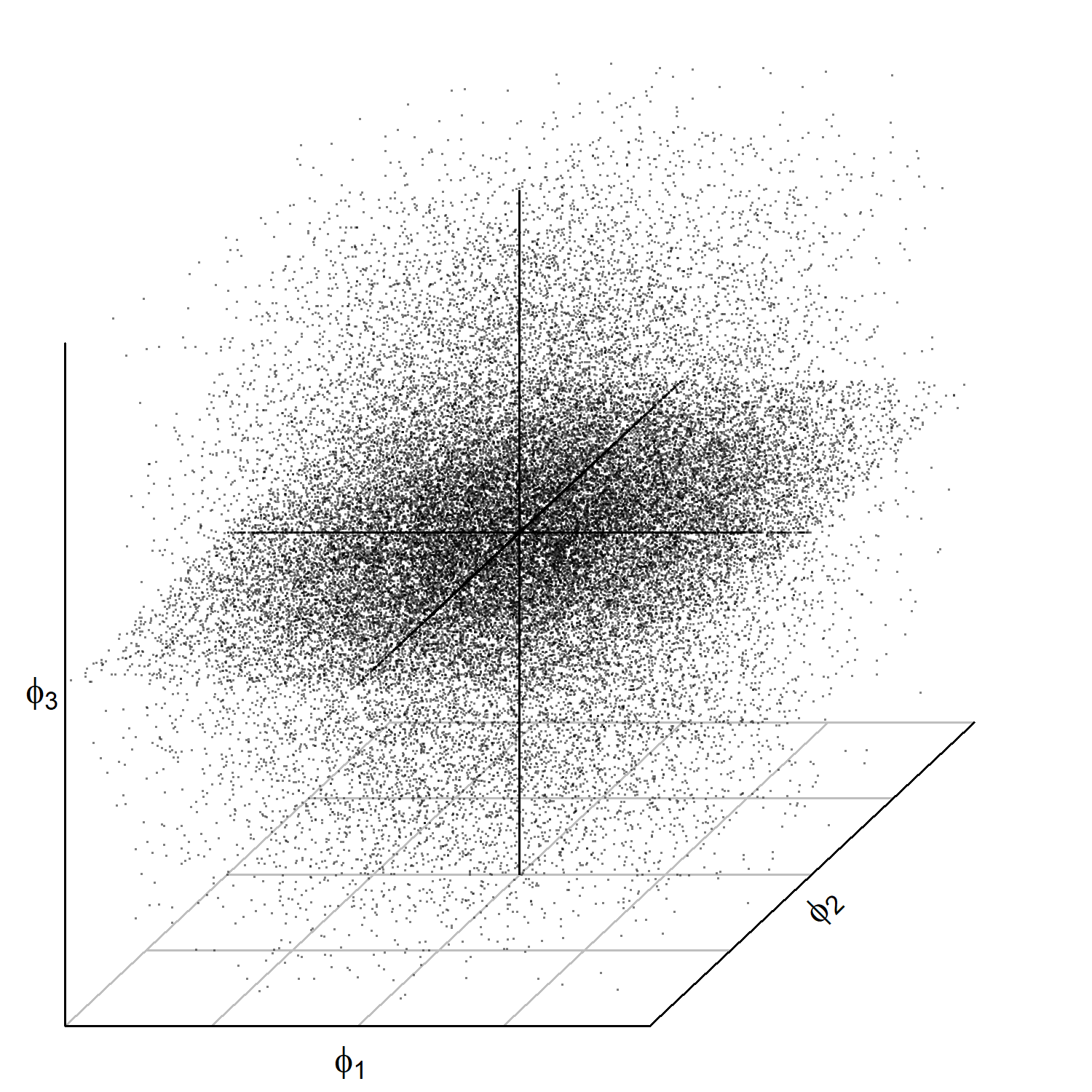}
         \caption{SSVS*}
     \end{subfigure}
     \hfill
     \begin{subfigure}[b]{0.30\textwidth}
         \centering
         \includegraphics[trim = 0 0 0 50, clip,width=\textwidth]{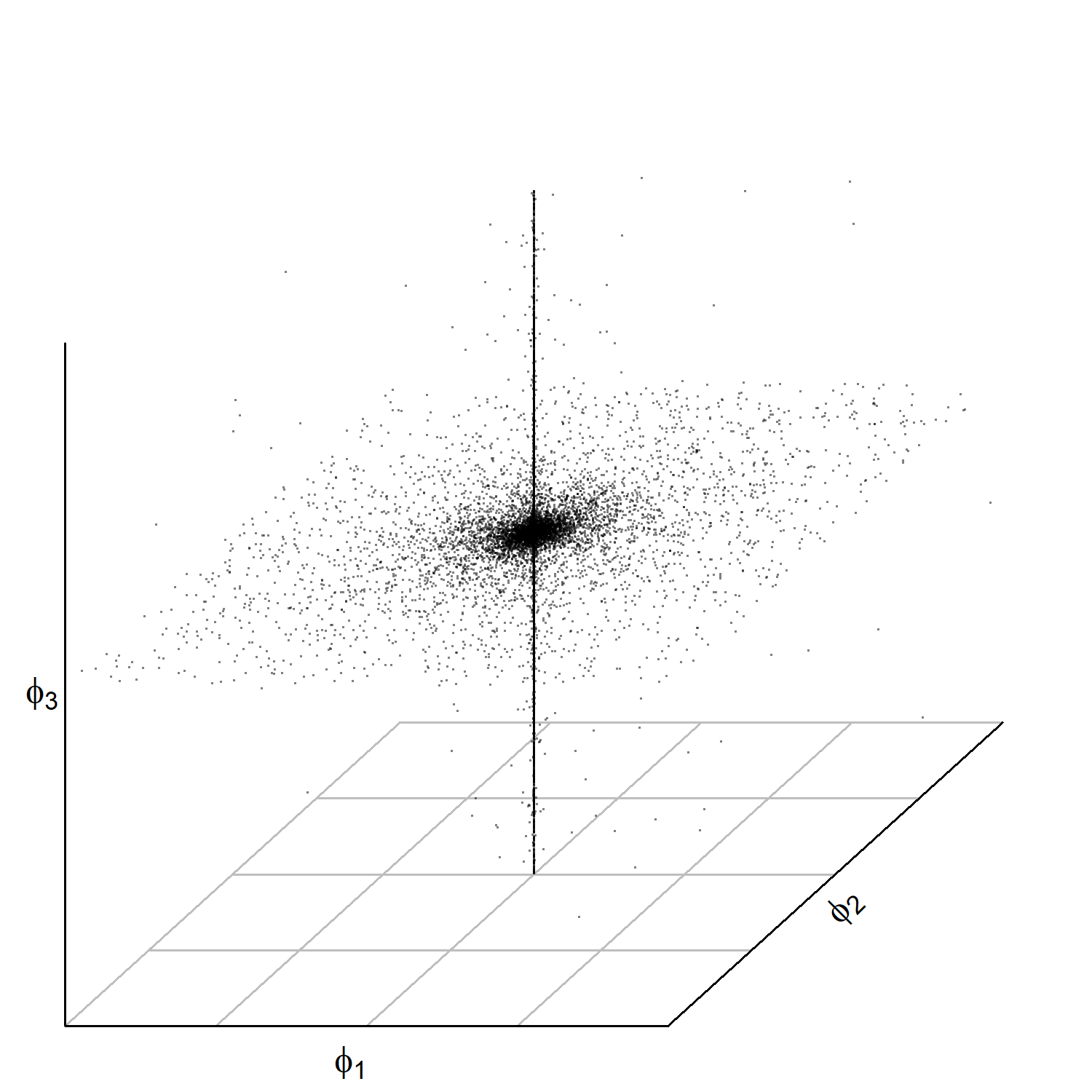}
         \caption{SHM*}
     \end{subfigure}
        \caption{Exemplary visualization of structured shrinkage. 100,000 samples from the HS prior representing the global-local priors, SSVS prior, and SHM prior. In HS, SSVS and SHM, all coefficients belong to the same group, whereas in HS*, SSVS*, and SHM*, $\phi_1$ and $\phi_2$ are from the same, whereas $\phi_3$ is from a different group.}
        \label{fig:3d_prior}
\end{figure}

Figure \ref{fig:3d_prior} depicts three-dimensional scatter plots of prior simulations. The star-like shapes of GL priors (here the HS prior) have most of its mass where all coefficients are almost zero or where at most one is substantial different from zero, indicating sparseness. By contrast, the ball-like shape of SHM, which does not allow for outliers, indicates denseness. SSVS is somewhere in the middle of the contrasting approaches. The figure also demonstrates the effect of structured shrinkage. Assume $\phi_1, \phi_2 \in \mathcal{A}_1$ and $\phi_3 \in \mathcal{A}_2$, then for all priors there is relatively more mass where $\phi_1=\phi_2=0 \And \phi_3 \neq 0$ or else $\phi_3=0 \And \phi_1\neq 0 \And \phi_2\neq 0$, i.e., the structuring introduces discrimination between the groups. For GL priors, this implies discrimination between groups, while being sparse within groups. On the other hand, SHM only discriminates between own-lag and cross-lag coefficients. Within own-lag and within cross-lag coefficients it remains dense, which explains the spinning-top shape.

We quantify the sparseness of each prior by means of the Hoyer sparseness measure \citep{hoyer_non-negative_2004}. The Hoyer measure for the vector $\bm{\phi}_j=(\phi_i)_{i \in \mathcal{A}_j}$,
\begin{align}
	H(\bm{\phi}_j) &= \frac{\sqrt{n_j} - \left( \sum_{i\in\mathcal{A}_j} |\phi_i| \right)/ \sqrt{\sum_{i\in\mathcal{A}_j} \phi_i^2}}{\sqrt{n_j} - 1},  \quad j=1,\dots,k,
\end{align}
is a normalized measure where $H=1$ indicates maximum sparseness, i.e., only one single nonzero value, and $H=0$ indicates maximum denseness, i.e., the absolute values of all components are equal. More importantly, it can express sparseness when all elements of $\bm{\phi}_j$ are nonzero, but when almost all are very close to zero and only very few are large (in absolute values). This is important, as we analyze continuous prior distributions which never strictly exclude single coefficients.

\begin{table}[t]
	\centering
	\begin{tabular}{l c c c c c c}
		& MP\_LIT/SHM & HS & DL & R2D2 & NG  & SSVS\\
		\hline
		 A & 0.21&0.89 &0.51 & 0.45 & 0.37 &0.45\\
		 B & -&-   &0.99 & 0.98 & 0.98 &0.95
	\end{tabular}
    \caption{Hoyer sparseness measure for SHM/MP\_LIT, HS, DL, R2D2, NG, and SSVS. The reported values are the means stemming from 10,000 simulations of vectors of length 1000. In scenario A, for DL $a^\gamma=1$, for R2D2 $a^\pi=a^\gamma/2$, for NG $a^\delta=a^\gamma/2$, for SSVS $\tau_0=0.01$, $\tau_1=100$ and $\underline{p}=0.5$. In scenario B, for DL $a^\gamma=0.001$, for R2D2 $a^\pi=a^\gamma/2$, for NG $a^\delta=a^\gamma/2$, for SSVS $\tau_0=0.01$, $\tau_1=100$ and $\underline{p}=0.01$. For SHM/MP\_LIT and HS, results do not depend on hyperparameter choices.}
	\label{tab:prior_hoyer}
\end{table}

We estimate the Hoyer measure through Monte Carlo simulation. Table \ref{tab:prior_hoyer} reports the arithmetic means stemming from a Monte Carlo simulation with $10,000$ iterations and simulated vectors of length $n=1000$. Interestingly, the simulation suggests that for MP\_LIT/SHM, i.e., all priors of the form $\bm{\phi} \sim N(\bm{0},\bm{I} c)$, $c \sim f$, regardless the specific choice of $f$,
the expectation of the Hoyer measure converges to $0.21$ for $n=1000$. This behavior of the measure is convenient, since sparsity should not depend on the overall noise level. For the remaining priors, results can be sensitive to the choice of hyperparameters; thus, for each prior, we consider two different scenarios. In scenario A, the hyperparameters for DL, R2D2, and NG are selected to ensure that the concentration at zero is comparable to the one of HS. In scenario B, we select the hyperparameters such that the concentration at zero is higher compared to scenario A: For DL we set $a^\gamma=0.001$, for R2D2 we set $a^{\pi}=0.0005$ and for NG we set $a^{\delta}=0.0005$, which results in the same behavior around the origin for those three priors. Our numerical experiments suggest, that the sparseness of DL, R2D2, and NG does not depend upon choices for the global hyperparameters. For SSVS, we set $\tau_0=0.01$ and $\tau_1=100$ in both scenarios. We set $\underline{p}=0.5$ in scenario A and $\underline{p}=0.01$ in scenario B. 

The exercise corroborates our preceding considerations. From sparse to dense, the priors can be ranked in the following order: DL, R2D2, NG, SSVS, HS, and at some distance SHM/MP\_LIT. Moreover, the sparseness of DL, R2D2, NG, and SSVS is very sensitive to the choices of the parameters that control the concentration at zero, namely $a^\gamma, a^\pi$, $a^\delta$, and $\underline{p}$, respectively.

\section{Synthetic data exercises}
\label{sec:snythetic}
This section aims at comparing the performances of the various priors on simulated data. Before that, the merits of the semi-global framework are illustrated on synthetic data. 
\subsection{Structured/informed shrinkage: An illustration}
\label{sec:snythetic_illustration}
\begin{figure}[t]
    \centering
    \begin{subfigure}{0.18\textwidth}
        \includegraphics[width=\textwidth]{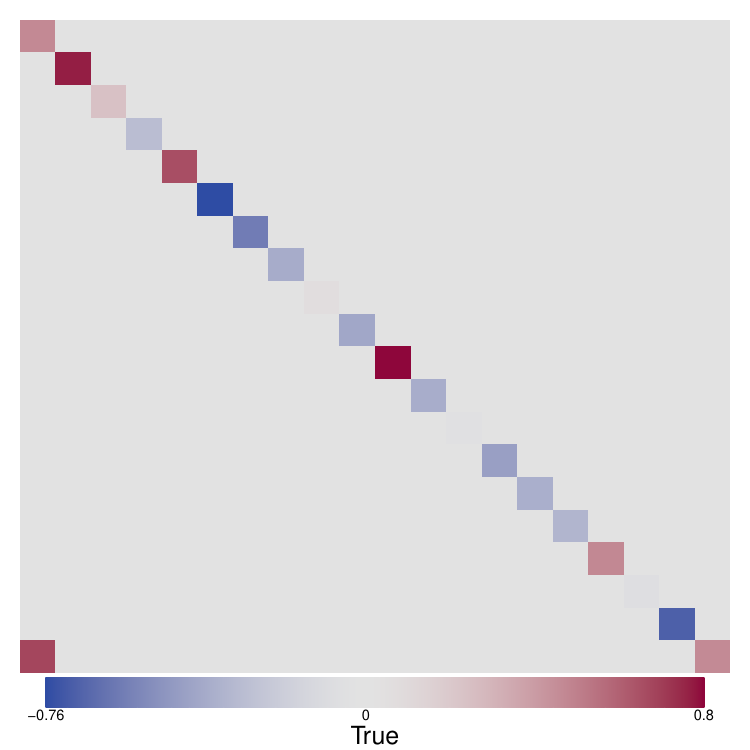}
    \end{subfigure}
    \begin{subfigure}{0.18\textwidth}
        \includegraphics[width=\textwidth]{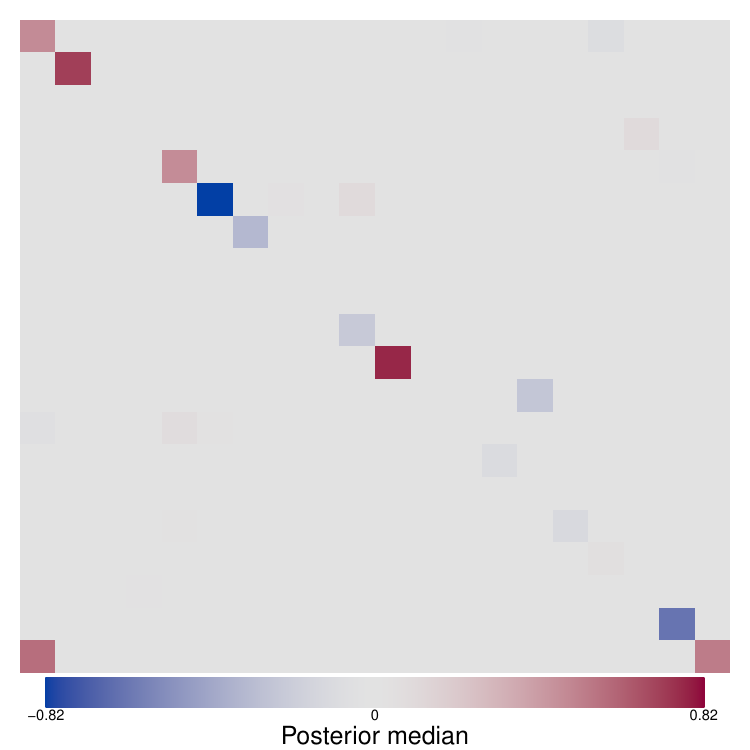}
    \end{subfigure}
    \begin{subfigure}{0.18\textwidth}
        \includegraphics[width=\textwidth]{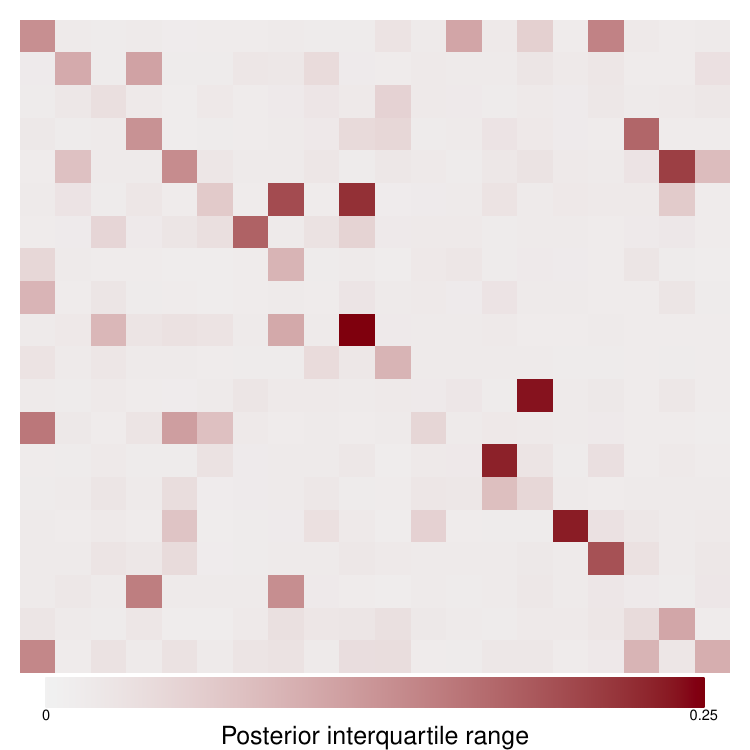}
    \end{subfigure}
    \begin{subfigure}{0.18\textwidth}
        \includegraphics[width=\textwidth]{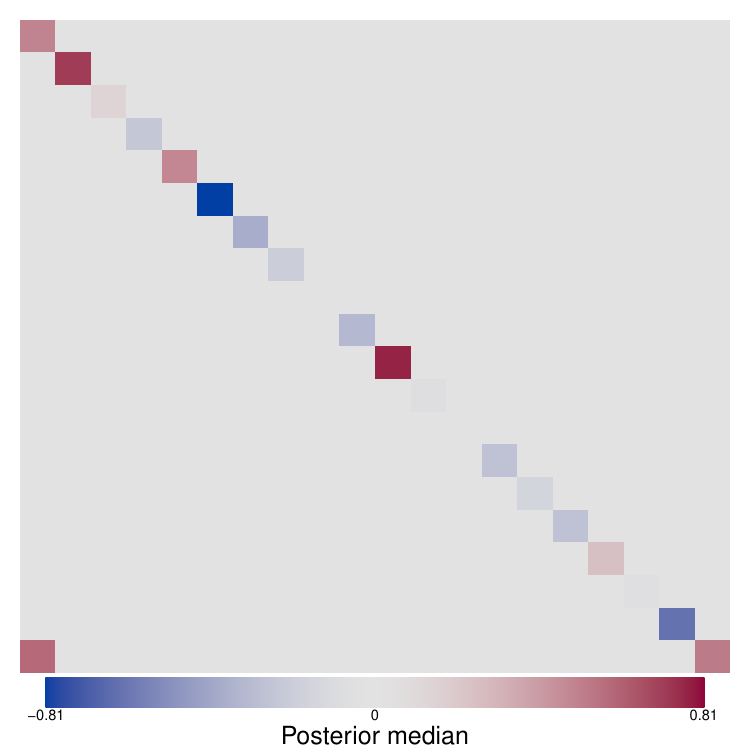}
    \end{subfigure}
    \begin{subfigure}{0.18\textwidth}
        \includegraphics[width=\textwidth]{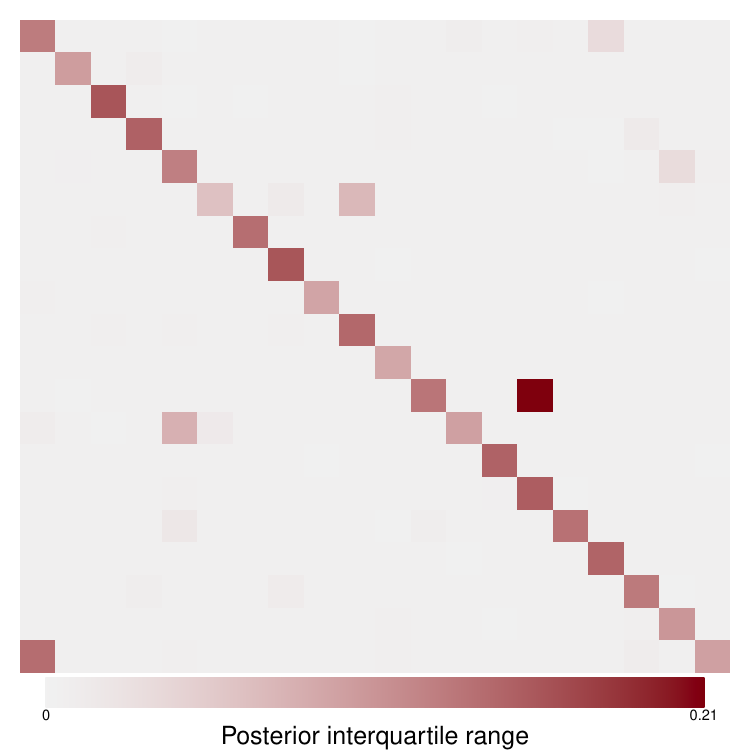}
    \end{subfigure}
    \caption{Exemplary visualization of true and estimated VAR coefficients. From the left to the right: true values, posterior median HS, interquartile range HS, posterior median HS semi-global-local, posterior interquartile range HS semi-global-local.}
    \label{fig:sgl_demo}
\end{figure}

Consider the following example of a VAR(1): Assume that all diagonal elements of $\bm{\Phi}$ are nonzero, whereas there is only one nonzero off-diagonal element. That is, taken as a whole, $\bm{\Phi}$ is not sparse. Figure \ref{fig:sgl_demo} demonstrates that a vanilla global-local prior -- here the HS prior -- overshrinks the diagonal elements and does not shrink the off-diagonal elements strong enough. The structured semi-global local version, however, adapts the degree of shrinkage in the two different subspaces of $\bm{\Phi}$.
\subsection{Simulation study}
\label{sec:sim}
In this section, we compare the performances of the different priors on simulated data. The structure of the simulation study is borrowed from \citet{kastner_sparse_2020}. We consider dense and sparse data generating processes (DGPs), as well as different dimensions in $T \in \lbrace 50, 100, 200 \rbrace$. Dimensionality is $M=20$ throughout. In each setting, we sample 20 replicates of the DGP. In all scenarios, we distinguish between own- and cross-lags. The nonzero coefficients of the DGPs are drawn from Gaussians with mean $\mu_i$ and standard deviation $\sigma_i$ for $i \in \{ol,cl\}$, where $ol$ and $cl$ denote own-lag and cross-lag, respectively. In both scenarios, the probability of own-lag coefficients to be nonzero is 0.8, i.e., own-lags are always considered to be dense. The probability of cross-lag coefficients to be nonzero is 0.01 in the sparse, 0.1 in the medium, and 0.8 in the dense scenario. Further, the scenarios differ regarding the signal-to-noise ratio. In the sparse scenario, we set $\mu_{ol} = \sigma_{ol}=\mu_{cl} = \sigma_{cl}=0.3$. In the medium scenario, we set $\mu_{ol} = \sigma_{ol}=0.15$ and $\mu_{cl} = \sigma_{cl}=0.1$. In the dense scenario, we set $\mu_{ol} = \sigma_{ol}=0.15$ and $\mu_{cl} =\sigma_{cl}=0.01$. To simulate stable VAR processes, we control that none of the eigenvalues of the companion form have modulus greater than one \citep{lutkepohl_new_2005}.

Elements in $\bm{u}$ are nonzero with probability 0.1. Furthermore, we set $\mu_{u} = \sigma_{u} = 0.001$. The AR(1) processes driving the log-variances of the orthogonalized errors have mean $\mu_{\sigma i} = -10$, persistencies $\rho_{\sigma i}$ in the range $[0.85,0.98]$ and standard deviations $\sigma_{\sigma i}$ in the range $[0.1,0.3]$.

The root mean squared error (RMSE) between the posterior mean of the coefficients and the true parameter values is a common measure to evaluate the performances:
\begin{equation}
	RMSE = \sqrt{\frac{1}{n} \sum_{i=1}^n \left(\frac{1}{R} \sum_{r=1}^R \phi_i^{(r)} - \phi_i\right)^2} = \sqrt{\frac{1}{n} \sum_{i=1}^n (\hat{\phi}_i - \phi_i)^2},
\end{equation}
where $\phi_i^{(r)}$ denotes the $r$-th posterior sample and $\hat{\phi}_i = \frac{1}{R} \sum_{r=1}^R \phi_i^{(r)}$ is the posterior mean of the $i$-th VAR coefficient.

The following prior specifications are considered: DL, NG, and R2D2 denote the vanilla global-local implementations with the concentration parameters being fixed. In DL, $a^\gamma=\frac{1}{K}$, the value proposed in \citet{kastner_sparse_2020}. In NG and R2D2, $a^\delta=a^\pi=\frac{1}{2K}$, i.e.\ DL, NG, and R2D2 are specified to have the same concentration at zero. Further, in R2D2, $b^\pi=0.5$, the value proposed in \citet{zhang_bayesian_2020}. In NG, $b^\delta=0.5$ and $c=\frac{a^\delta}{2}$, such that NG and R2D2 differ only in their respective kernels: Whereas NG defines hyperpriors for the variance of a normal distribution, R2D2 defines hyperpriors for the variance of a double-exponential distribution. \revision{In DL$_a$, NG$_a$, and R2D2$_a$, the subscript ``a'' indicates that the following discrete prior is placed on the corresponding concentration parameter:} The support points of the discrete priors $\tilde{\bm{a}}^\gamma=\tilde{\bm{a}}^\delta=\tilde{\bm{a}}^\pi=\tilde{\bm{a}}=(\tilde{a}_1=\frac{1}{1000},\dots,\tilde{a}_{1000}=1)^\prime$ are equally spaced, and the probabilities are proportional to the density of an exponential distribution with rate $1$. In MP\_LIT, $\lambda_1=0.16$ and $\lambda_2=0.004$. In SHM, $c_i=d_i=0.01$ for $i=1,2$. For all SSVS priors, we closely follow the semi-automatic approach in \citet{george_bayesian_2008} by setting $\tau_{0i} = \frac{1}{100} \sqrt{\widehat{var}(\phi_i)}$ and $\tau_{1i}=100 \sqrt{\widehat{var}(\phi_i)}$. Here, $\widehat{var}(\phi_i)$ denotes the variance of the posterior distribution resulting from a conjugate flat normal-Wishart prior, which is available in closed form. \citet{george_bayesian_2008} use the variance of the ordinary least squares (OLS) estimator. However, in situations where the number of covariates per equation exceeds the number of observations, the OLS estimator might not exist. \revision{For SSVS, the prior inclusion probability is fixed at $\underline{p}=0.5$, whereas for SSVS$_p$, the subscript ``$p$'' indicates that a $\text{Beta}(1,1)$ is placed on $\underline{p}$}. A superscript ``$^*$'' following a prior indicates a semi-global modification, e.g., the semi-global-local HS prior is denoted by HS$^*$. \revision{
Sub- and superscripts also can be combined: For instance, NG$_a^*$ stands for the semi-global-local NG prior where the concentration parameters $a^\delta_j$, $j=1,\dots,2p$, are treated hierarchically, whereas NG$^*$ stands for the semi-global-local NG prior where the concentration parameters $a^\delta_j=\frac{1}{2K}$, $j=1,\dots,2p$, are fixed}.

\begin{table}[t]\centering
\resizebox*{\textwidth}{!}{
\begin{tabular}[t]{l!{\extracolsep{4pt}} r !{\extracolsep{0pt}}rr!{\extracolsep{4pt}} r !{\extracolsep{0pt}}rr!{\extracolsep{4pt}} r !{\extracolsep{0pt}}rr}
\hline\hline
&\multicolumn{3}{c}{sparse}&\multicolumn{3}{c}{medium}&\multicolumn{3}{c}{dense}\\
\cline{2-4}\cline{5-7}\cline{8-10}
model  & $T=50$ & $T=100$ & $T=200$ & $T=50$ & $T=100$ & $T=200$ & $T=50$& $T=100$ & $T=200$\\
\hline
MP\_LIT & 5.63 & 4.68 & 4.03 & \textbf{5.09} & 4.49 & 4.08 & 3.18 & 2.88 & 2.60\\
SHM & 5.51 & 4.70 & 4.04 & 5.35 & 4.56 & 4.07 & 3.50 & \textbf{2.50} & \textbf{2.01}\\
DL & 5.38 & 3.27 & 2.22 & 6.02 & 4.63 & 3.57 & 4.50 & 3.77 & 2.82\\
DL$_a$ & 5.83 & 3.49 & 2.37 & 6.17 & 4.69 & 3.56 & 5.07 & 4.02 & 2.94\\
DL$^*_a$ & 5.29 & 3.22 & 2.19 & 6.07 & 4.58 & 3.40 & 4.79 & 3.85 & 2.79\\
HS & 5.22 & 3.31 & 2.25 & 5.75 & 4.63 & 3.55 & 4.05 & 3.52 & 2.62\\
HS$^*$ & \textbf{3.89} & \textbf{2.52} & \textbf{1.77} & 5.26 & \textbf{4.18} & \textbf{3.30} & \textbf{3.14} & 2.63 & 2.10\\
NG & 5.25 & 3.24 & 2.19 & 5.73 & 4.62 & 3.55 & 4.16 & 3.67 & 2.78\\
NG$_a$ & 5.52 & 3.45 & 2.36 & 5.74 & 4.62 & 3.54 & 4.44 & 3.82 & 2.85\\
NG$^*$ & 4.83 & 3.04 & 2.11 & 5.56 & 4.63 & 3.49 & 3.99 & 3.51 & 2.72\\
NG$^*_a$ & 4.65 & 2.94 & 2.15 & 5.47 & 4.36 & \textbf{3.30} & 4.08 & 3.44 & 2.57\\
R2D2 & 5.31 & 3.19 & 2.19 & 5.72 & 4.67 & 3.57 & 4.09 & 3.59 & 2.73\\
R2D2$_a$ & 5.49 & 3.39 & 2.34 & 5.73 & 4.60 & 3.54 & 4.35 & 3.80 & 2.83\\
R2D2$^*$ & 4.82 & 2.98 & 2.09 & 5.57 & 4.65 & 3.49 & 3.96 & 3.48 & 2.60\\
R2D2$^*_a$ & 4.56 & 2.91 & 2.10 & 5.47 & 4.36 & 3.31 & 4.05 & 3.36 & 2.50\\
SSVS & 5.48 & 3.67 & 2.68 & 5.96 & 5.17 & 4.29 & 4.11 & 3.65 & 2.90\\
SSVS$_p$ & 5.93 & 4.38 & 2.60 & 6.11 & 5.51 & 4.85 & 4.18 & 3.91 & 3.42\\
SSVS$^*_p$ & 5.57 & 3.56 & 2.25 & 6.15 & 5.45 & 4.69 & 4.18 & 3.81 & 3.08\\
\hline
\end{tabular}
}
\caption{100$\times$RMSE for simulated data. The reported values are the medians stemming from 20 simulations per setting. The smallest value in each column is indicated in bold font.}
\label{tab:sim}
\end{table}

Table \ref{tab:sim} reveals the results of the simulation study. HS$^*$ has the lowest RMSE in six out of nine scenarios: It is best in all sparse scenarios, in the medium scenarios with at least 100 observations and in the dense scenarios with only 50 observations. In the dense scenarios with at least 100 observations, SHM has the lowest scores. Concerning GL priors, the semi-global versions perform better than the vanilla versions in all scenarios. In the sparse and the medium scenarios, GL priors in general, i.e., the vanilla and the semi-global versions, tend to perform better than the Minnesota priors (with one notable exception). In the dense scenarios, SHM is overall best, closely followed by HS$^*$. SHM is considerably better than MP\_LIT in the dense scenarios with at least 100 observations, whereas in the remaining scenarios their performances in terms of RMSEs is comparable. Turning to the various specifications of SSVS, we tend to observe competitive performance only in the sparse scenarios.

\section{Empirical application to data of the US economy}
\label{sec:application}
In Section \ref{sec:specs}, we shortly summarize the data set, present the model specifications and the forecasting design. In Section \ref{sec:empirical_posteriors}, we inspect the posterior distributions of VAR coefficients arising from different prior distributions. In Section \ref{sec:model_evidence}, we assess out-of-sample forecasting accuracy of the different models. In Section \ref{sec:DMA}, we demonstrate the merits of combining forecasts of different models in a dynamic fashion via dynamic model averaging (DMA).

\subsection{Data overview, model specification and forecasting design}
\label{sec:specs}
The aim of the empirical application is to forecast economically relevant US time series. We use the quarterly data set provided by \citet{mccracken_fred-qd_2020}, which is based on the well-known \citet{stock_watson_2012} data set. The sample period ranges from 1959:Q4 to 2020:Q1. All in all, we include $M=21$ quarterly time series with the intention to cover the most important segments of the US economy. The complete list of all variables used in the following illustration is provided in Table \ref{tab:data} (Appendix \ref{sec:Data}). 

\revision{Similar to \citet{huber_adaptive_2019}, \citet{cross_macroeconomic_2020}, and \citet{chan_minnesota-type_2021}, most variables enter the models in log differences to be interpreted as growth rates, except interest rates, which are already defined in rates and hence taken in levels. In our experience, this choice results in more stable predictive densities compared to data taken in (log-)levels. Another stream of literature deals explicitly with the problem of potentially non-stationary data by eliciting priors accordingly \citep[cf., e.g.,][]{sims_bayesian_1998,villaniSteadystatePriorsVector2009, giannonePriorsLongRun2019}. It has to be acknowledged that results about the performances of the priors might be different when considering different transformations. A preliminary analysis available upon request, however, suggests that many of the results are qualitatively similar when the data is taken in log-levels.} 

A very important and often underemphasized choice when modeling macroeconomic time series with VARs is the number of lags of endogenous variables included as predictors. In the Bayesian framework, fixing $p$ at a certain number implies prior information that higher lags do not carry any important information. Hence, \citet{litterman_forecasting_1986} suggests estimating as many lags as is computationally feasible, in which the prior has to be designed such that irrelevant coefficients get shrunken to zero. However, the more lags we include, the severer the problem of overfitting becomes, and the harder it will be to regularize the parameter space. Not to mention the increase of computational burden. Typical choices for quarterly data in similar dimensions are either four or five lags \citep{chan_minnesota-type_2021,cross_macroeconomic_2020, huber_adaptive_2019, giannone_prior_2015}. We estimate VAR($p$)s for $p\in\lbrace1,2,3,4,5\rbrace$ to investigate to what degree longer lags improve/worsen forecasting performance.

The forecasting performance is evaluated by means of a recursive pseudo out-of-sample forecasting exercise. Based on the initial estimation window ranging from 1959:Q4 to 1980:Q1, one-step-ahead and \revision{four-steps-ahead} predictive densities are evaluated. After that, the estimation window is expanded by one quarter and the model is re-estimated. For each estimation step, we run ten independent MCMC chains. Per chain, we keep 15,000 draws after a burn-in of 5,000 iterations. This procedure is consequently repeated until the end of the sample 2020:Q1 is reached.

The prior specifications we consider in the empirical application are the same as in Section \ref{sec:sim}. \revision{In what is to follow, we also include an intercept term. The prior for the intercept is independent normal centered at zero with variance 1000 for all models.}
\subsection{Inspecting the posterior distributions}
\label{sec:empirical_posteriors}

\begin{figure}[t]
    \centering
    \includegraphics[width=\textwidth]{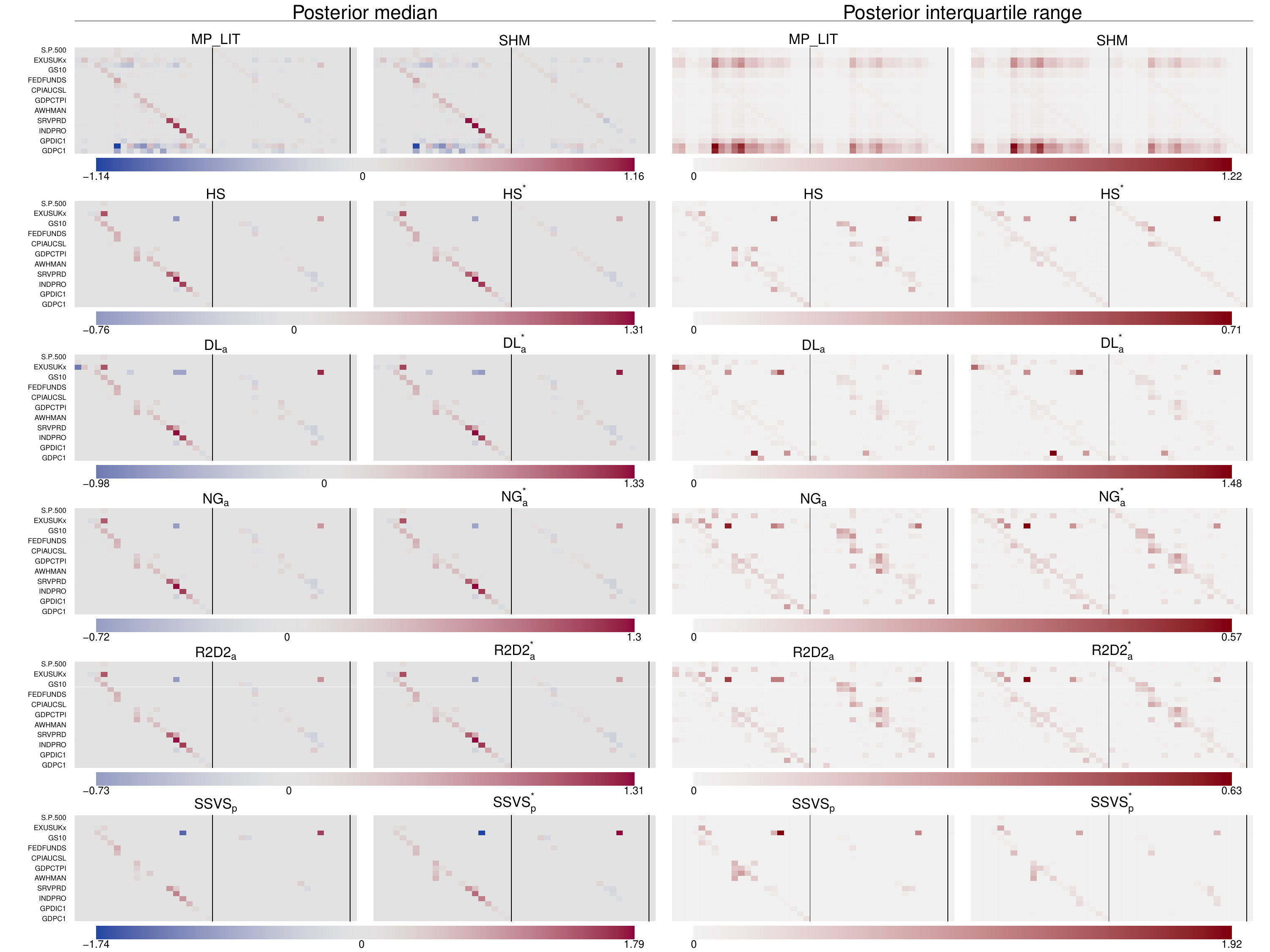}
    \caption{Posterior summaries for VAR(2) coefficients arising from different prior distributions.}
    \label{fig:post}
\end{figure}

Before presenting the results of the forecasting exercise, it is worth analyzing the posterior distributions arising from the different priors. Figure \ref{fig:post} depicts posterior medians and posterior interquartile ranges of different VAR(2) models.\footnote{Although we find signals with higher lag orders, the out-of-sample results indicate that two lags usually are enough (cf.\ Section \ref{sec:model_evidence}).} All models, even the vanilla SSVS and GL priors, detect many signals associated with the first own-lag and little signals in the reaming part of the coefficient space. Concerning the cross-lag coefficients, the posteriors arising from the Minnesota priors show relatively high uncertainty, whereas the remaining priors only find sparse signals. The differences between the vanilla and the semi-global versions of SSVS and GL priors are hardly detectable by visual exploration only.

\begin{table}[t]
\centering\tiny
\begin{subtable}[t]{0.07\textwidth}
\centering
\begin{tabular}[t]{rc}
\hline\hline
p & ol/cl \\
\hline
1&ol  \\
&cl \\
2&ol  \\
&cl \\
3&ol  \\
&cl \\
4&ol  \\
&cl \\
5&ol  \\
&cl \\
\hline
\end{tabular}
\label{test}
\end{subtable}
\hfill
\begin{subtable}[t]{0.28\textwidth}
\centering
\begin{tabular}[t]{rrrrr}
\hline\hline
 $l=1$ & $l=2$ & $l=3$ & $l=4$ & $l=5$\\
\hline
0.30 & - & - & - & -\\
0.78 & - & - & - & -\\
0.37 & 0.48 & - & - & -\\
0.82 & 0.84 & - & - & -\\
0.37 & 0.53 & 0.68 & - & -\\
0.82 & 0.84 & 0.85 & - & -\\
0.37 & 0.56 & 0.71 & 0.75 & -\\
0.82 & 0.85 & 0.85 & 0.88 & -\\
0.38 & 0.56 & 0.74 & 0.76 & 0.78\\
0.82 & 0.85 & 0.85 & 0.86 & 0.84\\
\hline
\end{tabular}
\caption{DL$_a$}
\end{subtable}
\hfill
\begin{subtable}[t]{0.28\textwidth}
\centering
\begin{tabular}[t]{rrrrr}
\hline\hline
$l=1$ & $l=2$ & $l=3$ & $l=4$ & $l=5$\\
\hline
0.26 & - & - & - & -\\
0.78 & - & - & - & -\\
0.33 & 0.40 & - & - & -\\
0.82 & 0.84 & - & - & -\\
0.33 & 0.44 & 0.64 & - & -\\
0.83 & 0.84 & 0.86 & - & -\\
0.32 & 0.45 & 0.65 & 0.72 & -\\
0.83 & 0.85 & 0.85 & 0.89 & -\\
0.33 & 0.45 & 0.70 & 0.74 & 0.76\\
0.82 & 0.85 & 0.85 & 0.87 & 0.85\\
\hline
\end{tabular}
\caption{DL$_a^*$}
\end{subtable}
\hfill
\begin{subtable}[t]{0.28\textwidth}
\centering
\begin{tabular}[t]{rrrrr}
\hline\hline
 $l=1$ & $l=2$ & $l=3$ & $l=4$ & $l=5$\\
\hline
 0.23 & - & - & - & -\\
0.60 & - & - & - & -\\
0.30 & 0.19 & - & - & -\\
0.61 & 0.61 & - & - & -\\
0.31 & 0.22 & 0.22 & - & -\\
0.61 & 0.61 & 0.61 & - & -\\
0.31 & 0.21 & 0.22 & 0.25 & -\\
0.61 & 0.61 & 0.61 & 0.61 & -\\
0.31 & 0.22 & 0.22 & 0.25 & 0.26\\
0.61 & 0.61 & 0.61 & 0.61 & 0.61\\
\hline
\end{tabular}
\caption{SHM}
\end{subtable}
\\[1em]
\begin{subtable}[t]{0.07\textwidth}
    \centering
    \begin{tabular}[t]{rc}
    \hline\hline
    p & ol/cl \\
    \hline
         1&ol  \\
         &cl \\
         2&ol  \\
         &cl \\
         3&ol  \\
         &cl \\
         4&ol  \\
         &cl \\
         5&ol  \\
         &cl \\
         \hline
    \end{tabular}
\end{subtable}
\hfill
\begin{subtable}[t]{0.28\textwidth}
\centering
\begin{tabular}[t]{rrrrr}
\hline\hline
$l=1$ & $l=2$ & $l=3$ & $l=4$ & $l=5$\\
\hline
0.31 & - & - & - & -\\
0.75 & - & - & - & -\\
0.38 & 0.54 & - & - & -\\
0.83 & 0.82 & - & - & -\\
0.38 & 0.65 & 0.76 & - & -\\
0.85 & 0.81 & 0.86 & - & -\\
0.38 & 0.77 & 0.81 & 0.77 & -\\
0.86 & 0.83 & 0.84 & 0.85 & -\\
0.40 & 0.77 & 0.81 & 0.76 & 0.85\\
0.87 & 0.84 & 0.84 & 0.86 & 0.81\\
\hline
\end{tabular}
\caption{HS}
\end{subtable}
\hfill
\begin{subtable}[t]{0.28\textwidth}
\centering
\begin{tabular}[t]{rrrrr}
\hline\hline
$l=1$ & $l=2$ & $l=3$ & $l=4$ & $l=5$\\
\hline
0.25 & - & - & - & -\\
0.80 & - & - & - & -\\
0.32 & 0.27 & - & - & -\\
0.84 & 0.89 & - & - & -\\
0.32 & 0.30 & 0.37 & - & -\\
0.85 & 0.91 & 0.93 & - & -\\
0.32 & 0.29 & 0.40 & 0.46 & -\\
0.85 & 0.91 & 0.92 & 0.88 & -\\
0.32 & 0.29 & 0.41 & 0.48 & 0.58\\
0.85 & 0.89 & 0.91 & 0.87 & 0.85\\
\hline
\end{tabular}
\caption{HS$^*$}
\end{subtable}
\hfill
\begin{subtable}[t]{0.28\textwidth}
\centering
\begin{tabular}[t]{rrrrr}
\hline\hline
$l=1$ & $l=2$ & $l=3$ & $l=4$ & $l=5$\\
\hline
0.24 & - & - & - & -\\
0.60 & - & - & - & -\\
0.29 & 0.21 & - & - & -\\
0.60 & 0.61 & - & - & -\\
0.29 & 0.22 & 0.22 & - & -\\
0.60 & 0.61 & 0.61 & - & -\\
0.29 & 0.22 & 0.23 & 0.26 & -\\
0.60 & 0.61 & 0.61 & 0.61 & -\\
0.29 & 0.23 & 0.23 & 0.26 & 0.25\\
0.60 & 0.61 & 0.61 & 0.61 & 0.61\\
\hline
\end{tabular}
\caption{MP\_LIT}
\end{subtable}
\\[1em]
\begin{subtable}[b]{0.07\textwidth}
    \centering
    \begin{tabular}[t]{rc}
    \hline\hline
    p & ol/cl \\
    \hline
         1&ol  \\
         &cl \\
         2&ol  \\
         &cl \\
         3&ol  \\
         &cl \\
         4&ol  \\
         &cl \\
         5&ol  \\
         &cl \\
         \hline
    \end{tabular}
\end{subtable}
\hfill
\begin{subtable}[t]{0.28\textwidth}
\centering
\begin{tabular}[t]{rrrrr}
\hline\hline
$l=1$ & $l=2$ & $l=3$ & $l=4$ & $l=5$\\
\hline
0.29 & - & - & - & -\\
0.75 & - & - & - & -\\
0.37 & 0.47 & - & - & -\\
0.79 & 0.78 & - & - & -\\
0.37 & 0.55 & 0.66 & - & -\\
0.80 & 0.77 & 0.81 & - & -\\
0.37 & 0.63 & 0.68 & 0.71 & -\\
0.80 & 0.76 & 0.79 & 0.79 & -\\
0.38 & 0.63 & 0.71 & 0.73 & 0.75\\
0.80 & 0.76 & 0.78 & 0.78 & 0.78\\
\hline
\end{tabular}
\caption{NG$_a$}
\end{subtable}
\hfill
\begin{subtable}[t]{0.28\textwidth}
\centering
\begin{tabular}[t]{rrrrr}
\hline\hline
$l=1$ & $l=2$ & $l=3$ & $l=4$ & $l=5$\\
\hline
0.25 & - & - & - & -\\
0.77 & - & - & - & -\\
0.32 & 0.31 & - & - & -\\
0.80 & 0.79 & - & - & -\\
0.31 & 0.36 & 0.57 & - & -\\
0.80 & 0.78 & 0.82 & - & -\\
0.31 & 0.37 & 0.57 & 0.68 & -\\
0.80 & 0.78 & 0.81 & 0.81 & -\\
0.32 & 0.37 & 0.63 & 0.69 & 0.73\\
0.80 & 0.77 & 0.80 & 0.80 & 0.80\\
\hline
\end{tabular}
\caption{NG$_a^*$}
\end{subtable}
\hfill
\begin{subtable}[t]{0.28\textwidth}
\centering
\begin{tabular}[t]{rrrrr}
\hline\hline
$l=1$ & $l=2$ & $l=3$ & $l=4$ & $l=5$\\
\hline
0.29 & - & - & - & -\\
0.76 & - & - & - & -\\
0.37 & 0.48 & - & - & -\\
0.80 & 0.79 & - & - & -\\
0.37 & 0.56 & 0.66 & - & -\\
0.81 & 0.78 & 0.82 & - & -\\
0.37 & 0.63 & 0.69 & 0.72 & -\\
0.82 & 0.77 & 0.80 & 0.80 & -\\
0.38 & 0.64 & 0.73 & 0.74 & 0.76\\
0.82 & 0.77 & 0.79 & 0.79 & 0.79\\
\hline
\end{tabular}
\caption{R2D2$_a$}
\end{subtable}
\\[1em]
\begin{subtable}[t]{0.07\textwidth}
    \centering
    \begin{tabular}[t]{rc}
    \hline\hline
    p & ol/cl \\
    \hline
         1&ol  \\
         &cl \\
         2&ol  \\
         &cl \\
         3&ol  \\
         &cl \\
         4&ol  \\
         &cl \\
         5&ol  \\
         &cl \\
         \hline
    \end{tabular}
\end{subtable}
\hfill
\begin{subtable}[t]{0.28\textwidth}
\centering
\begin{tabular}[t]{rrrrr}
\hline\hline
$l=1$ & $l=2$ & $l=3$ & $l=4$ & $l=5$\\
\hline
0.25 & - & - & - & -\\
0.77 & - & - & - & -\\
0.32 & 0.31 & - & - & -\\
0.81 & 0.80 & - & - & -\\
0.32 & 0.35 & 0.55 & - & -\\
0.81 & 0.78 & 0.83 & - & -\\
0.32 & 0.36 & 0.55 & 0.66 & -\\
0.81 & 0.79 & 0.82 & 0.82 & -\\
0.32 & 0.37 & 0.62 & 0.68 & 0.72\\
0.81 & 0.78 & 0.82 & 0.81 & 0.81\\
\hline
\end{tabular}
\caption{R2D2$_a^*$}
\end{subtable}
\hfill
\begin{subtable}[t]{0.28\textwidth}
\centering
\begin{tabular}[t]{rrrrr}
\hline\hline
$l=1$ & $l=2$ & $l=3$ & $l=4$ & $l=5$\\
\hline
0.33 & - & - & - & -\\
0.82 & - & - & - & -\\
0.40 & 0.64 & - & - & -\\
0.84 & 0.83 & - & - & -\\
0.38 & 0.73 & 0.60 & - & -\\
0.85 & 0.81 & 0.82 & - & -\\
0.36 & 0.75 & 0.42 & 0.57 & -\\
0.84 & 0.74 & 0.64 & 0.87 & -\\
0.38 & 0.73 & 0.47 & 0.45 & 0.67\\
0.84 & 0.70 & 0.64 & 0.81 & 0.58\\
\hline
\end{tabular}
\caption{SSVS}
\end{subtable}\hfill
\begin{subtable}[t]{0.28\textwidth}
\centering
\begin{tabular}[t]{rrrrr}
\hline\hline
$l=1$ & $l=2$ & $l=3$ & $l=4$ & $l=5$\\
\hline
0.32 & - & - & - & -\\
0.81 & - & - & - & -\\
0.36 & 0.62 & - & - & -\\
0.87 & 0.89 & - & - & -\\
0.34 & 0.72 & 0.55 & - & -\\
0.84 & 0.58 & 0.73 & - & -\\
0.34 & 0.70 & 0.52 & 0.41 & -\\
0.85 & 0.60 & 0.70 & 0.41 & -\\
0.36 & 0.70 & 0.41 & 0.37 & 0.43\\
0.85 & 0.68 & 0.68 & 0.46 & 0.31\\
\hline
\end{tabular}
\caption{SSVS$_p^*$}
\end{subtable}\hfill
\caption{Posterior means of the Hoyer measure, separately for own-lag (ol) and cross-lag (cl) coefficients for VARs of order $p\in \{1,\dots,5\}$.}
\label{tab:hoyer_post}
\end{table}

Table \ref{tab:hoyer_post} displays the posterior mean of the Hoyer measure.\footnote{We compute the Hoyer measure for every single coefficient posterior draw, which then provides us with the posterior distribution of the measure.} In the first lag, the posteriors under all priors are dense, whereas for the Minnesota priors own-lags in general are dense. Moreover, the Minnesota priors are always denser than the remaining priors for all considered lag-lengths with respect to both own-lag and cross-lag coefficients. Last but not least, especially on the off-diagonals, the posteriors under all priors except SHM and MP\_LIT can get extremely sparse. The heterogeneous results regarding sparsity show that posterior inference is strongly affected by the prior assumptions, which stresses the importance of careful prior elicitation. \revision{This corroborates findings in \citet{jarocinski2017}: In order to answer the question whether one variable is relevant for another variable, the authors derive a closed-form expression for the posterior probability of Granger noncausality and a generalization thereof in a homoskedastic VAR with a conjugate prior. They find that the posterior probability for zero restrictions increases as the prior gets looser. Since prior choices matter,} 
we think it can be delusive to draw general conclusions about sparseness mainly by analyzing posterior quantities. \citet{fava_illusion_2020} elaborate on how such posterior results often just spread illusions. 

\begin{figure}[t]
\centering
\begin{subfigure}{.45\textwidth}
\centering
    \includegraphics[width=.9\textwidth]{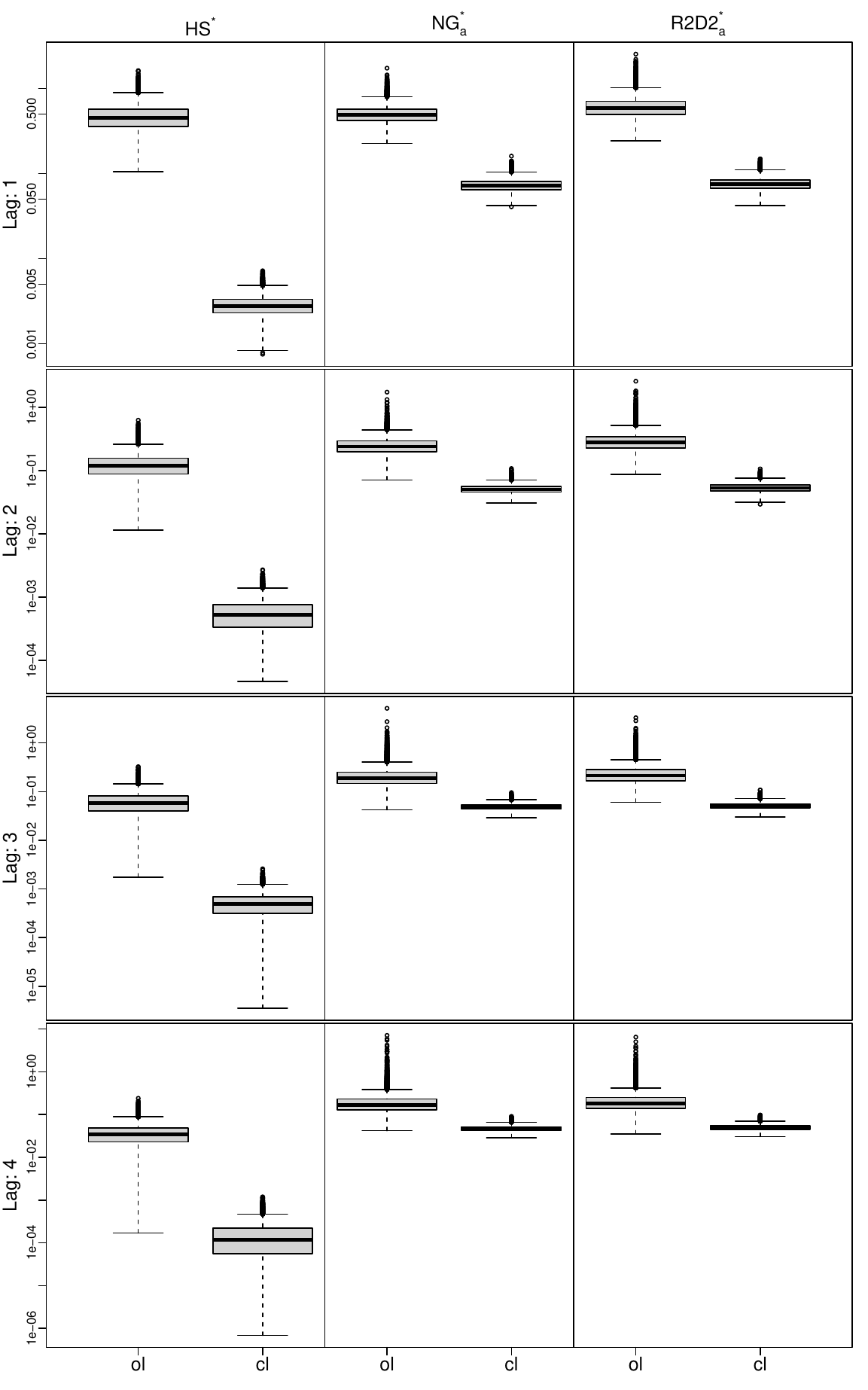}
    \caption{Posteriors of $\sqrt{\zeta}_j$ for HS$^*$ (left), NG$^*_a$ (middle), R2D2$^*_a$ (right), $j \in \{\text{lag: } 1,\dots,4\} \times \{\text{own-lag, cross-lag}\}$.}
\end{subfigure}
\hfill
\begin{subfigure}{.45\textwidth}
\centering
    \includegraphics[width=.9\textwidth]{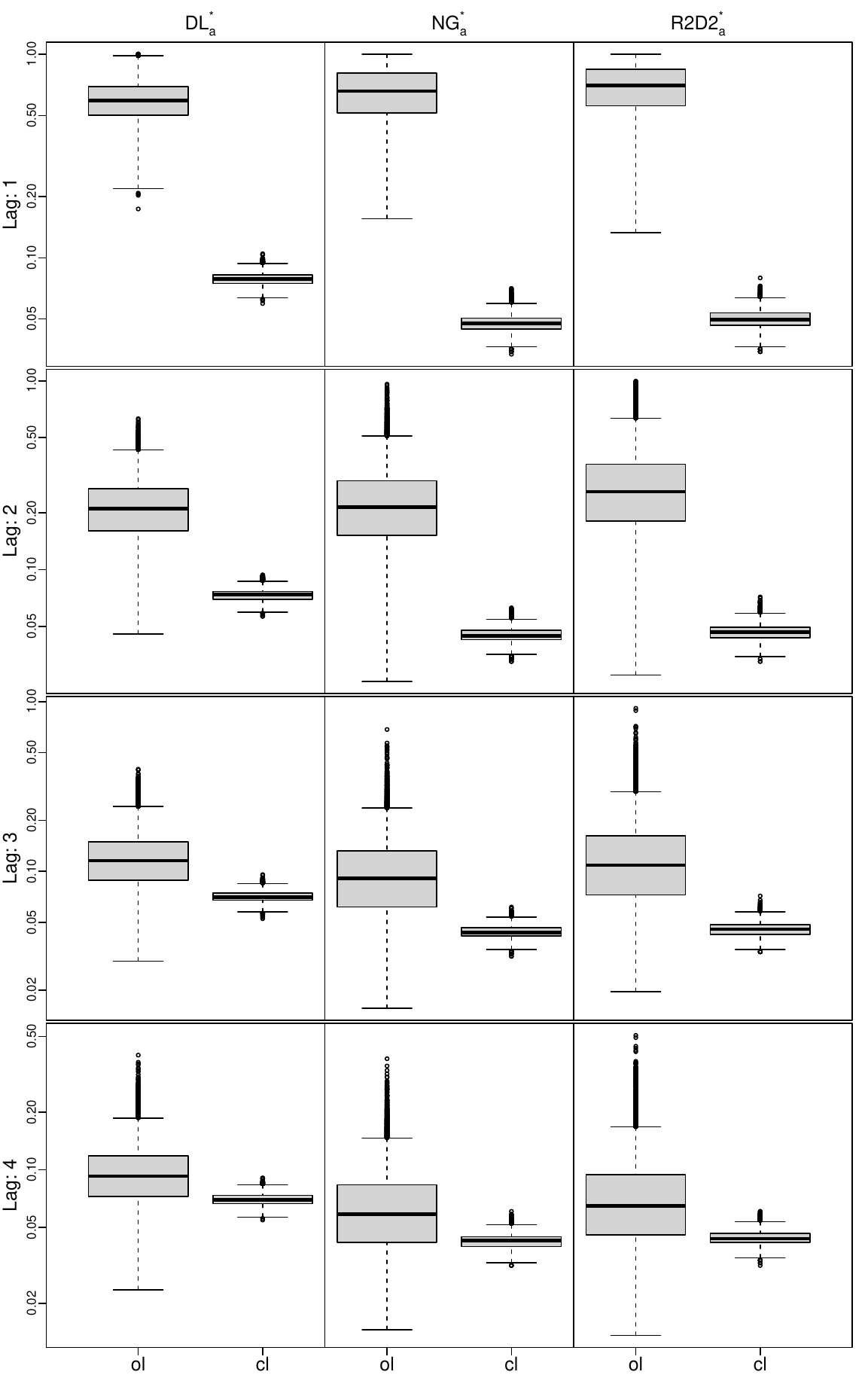}
    \caption{Posteriors of $a_j^\gamma$ for DL$^*_a$ (left), $a_j^\delta$ for NG$^*_a$ (middle), $a_j^\pi$ for R2D2$^*_a$ (right), $j \in \{\text{lag: } 1,\dots,4\} \times \{\text{own-lag, cross-lag}\}$.}
\end{subfigure}
	\caption{Box plots of the posterior distributions of the semi-global hyperparameters for a VAR(4).}
	\label{fig:post_box}
\end{figure}
Comparing the sparseness between GL priors, we observe that the own-lag coefficients of the semi-global modifications are denser compared to their vanilla GL counterparts. HS$^*$ discriminates the strongest between own-lags and cross-lags in terms of sparsity, especially in more distant lags. In this context, it is worth analyzing the posterior distributions of the (semi-)global hyperparameters. Figure \ref{fig:post_box} depicts, exemplary for a VAR(4), box plots of the posterior distribution of $\sqrt{\zeta}_j$ for HS$^*$, NG$^*_a$, and R2D2$^*_a$, and the posterior distribution of $a_j^\gamma$ for DL$^*_a$, $a_j^\delta$ for NG$^*_a$, $a_j^\pi$ for R2D2$^*_a$, $j \in \{\text{lag}: 1,\dots,4\} \times \{\text{own-lag, cross-lag}\}$. For all models there is a clear discrimination in all lags between own-lag and cross-lag, i.e., shrinkage is always stronger for cross-lag coefficients than for own-lag coefficients in all lags. The strength of discrimination between own-lag and cross-lag decreases with higher lags, and is the strongest for HS. Last but not least, the posterior distributions of the semi-global hyperparameters concentrate around smaller values for more distant lags than for closer lags, i.e., shrinkage increases with increasing lags. 

\subsection{Model evidence: Log predictive likelihoods}
\label{sec:model_evidence}
\revision{Following \citet{geweke_comparing_2010} we evaluate forecasting performance in terms of log predictive likelihoods, treating the period ranging from 1959:Q4 to 1980:Q1 as training sample (initial estimation window). Evaluation of the predictive densities begins in 1980:Q2 or 1981:Q1 for one-step-ahead forecasts and four-steps-ahead forecasts, respectively. The estimation window is recursively expanded by one observation until the end of the evaluation period 2020:Q1 is reached.}

\begin{table}[t]
\centering
\resizebox{\textwidth}{!}{
\begin{tabular}[t]{l !{\extracolsep{10pt}}r!{\extracolsep{0pt}}rrrr !{\extracolsep{10pt}}r!{\extracolsep{0pt}}rrrr}
\hline\hline
&\multicolumn{5}{c}{One-step-ahead}&\multicolumn{5}{c}{Four-steps-ahead}\\
\cline{2-6}\cline{7-11}
model & $p=1$ & $p=2$ & $p=3$ & $p=4$ & $p=5$ & $p=1$ & $p=2$ & $p=3$ & $p=4$ & $p=5$\\
\hline
MP\_LIT & -67 & -17 & -8 & 0 & -13 & -64 & -10 & -1 & 0 & -1\\
SHM & -55 & 0 & 10 & 7 & 4 & -61 & -10 & -4 & 2 & 5\\
HS & -58 & -27 & -23 & -36 & -43 & -39 & 17 & 14 & 4 & 10\\
DL & -44 & 10 & 11 & -6 & -23 & -25 & 45 & 31 & 29 & 22\\
NG & -46 & 11 & 9 & -2 & -14 & -30 & 40 & 41 & 30 & 27\\
R2D2 & -40 & 8 & 9 & 3 & -13 & -30 & 44 & 35 & 30 & 25\\
DL$_a$ & -41 & 16 & 6 & -12 & -53 & -21 & 39 & 36 & 12 & -13\\
NG$_a$ & -44 & 9 & -2 & -28 & -53 & -27 & 42 & 40 & 16 & 8\\
R2D2$_a$ & -45 & 8 & -6 & -20 & -43 & -25 & 42 & 37 & 16 & -1\\
NG$^*$ & -47 & 17 & 16 & 4 & -14 & -30 & 40 & 43 & 33 & 33\\
R2D2$^*$ & -39 & 13 & 18 & 6 & -11 & -27 & 37 & 39 & 41 & 22\\
HS$^*$ & -33 & 24 & \textbf{41} & \textbf{50} & \textbf{38} & -39 & 21 & 43 & \textbf{45} & \textbf{53}\\
DL$^*_a$ & -30 & 32 & 32 & 13 & -29 & \textbf{-20} & \textbf{52} & 43 & 21 & 1\\
NG$^*_a$ & \textbf{-28} & \textbf{33} & 34 & 19 & -16 & -27 & \textbf{52} & \textbf{56} & 36 & 20\\
R2D2$^*_a$ & \textbf{-28} & \textbf{33} & 40 & 17 & -10 & -23 & 50 & 53 & 36 & 21\\
SSVS & -63 & -29 & -29 & -49 & -90 & -45 & -3 & -11 & -32 & -42\\
SSVS$_p$ & -87 & -61 & -62 & -88 & -106 & -53 & -23 & -14 & -36 & -38\\
SSVS$^*_p$ & -84 & -39 & -36 & -63 & -84 & -69 & -35 & -15 & -39 & -43\\
\hline
\end{tabular}
}
\caption{\revision{Sum of one and four-steps-ahead log predictive likelihoods relative to MP\_LIT(4), which itself has scores of $12602$ (one step) and $11522$ (four steps). The highest value in each column is indicated in bold font.}}
\label{tab:lbfs}
\end{table}

Table \ref{tab:lbfs} depicts sums of one-step- \revision{and four-steps}-ahead log predictive likelihoods relative to the traditional Minnesota prior with four lags (MP\_LIT(4)). \revision{We start by inspecting the one-step-ahead scores.} The most important findings are the following:
The overall best performance is achieved by the semi-global modification of HS with 4 lags (HS$^*$(4)). In almost all cases, the semi-global modifications clearly improve forecasting performance compared to their vanilla GL counterparts. Moreover, the semi-global modifications usually perform better than the Minnesota priors, whereas the vanilla GL priors do not. \revision{SHM priors perform well in general.} Comparing the scores of SHM and MP\_LIT reveals that learning the shrinkage parameters of the Minnesota prior from the data always improves forecasting performance, which corroborates findings in \citet{giannone_prior_2015} and \citet{cross_macroeconomic_2020}. Among all considered priors, SSVS priors achieve the lowest scores.

Concerning the performance of GL priors compared to Minnesota priors: The vanilla GL priors do not produce more accurate forecasts than SHM. This contradicts results in \citet{huber_adaptive_2019} at first sight. The weaker performance of SHM in \citet{huber_adaptive_2019} can be explained (at least to some extent) by the vague prior imposed on $\bm{u}$, a normal prior with zero mean and variance 10. The fact that the prior choice for $\bm{u}$ indeed plays a crucial role in terms of forecasting performance is demonstrated in Section \ref{sec:robustness}. Interestingly, among the vanilla GL priors, HS has the lowest scores: For all orders, it has even lower scores than MP\_LIT(4) \revision{concering one-step-ahead forecasts}. The story changes radically, though, when looking at the semi-global modifications. For $p \in \{3,4,5\}$, HS$^*$ performs best. DL$_a^*$, R2D2$_a^*$, and NG$_a^*$ with $p \in \{2,3,4\}$ also perform considerably better than the best Minnesota prior, SHM(3).

\revision{Regarding four-steps-ahead predictions, we again observe that the semi-global modifications usually perform better than the vanilla implementations. Here, the overall best performance is achieved by NG$^*_a$(3). What stands out compared to one-step-ahead forecasts is that even the vanilla GL priors are doing better than the Minnesota priors. In other words, at longer forecasting horizons there is stronger evidence in favor of sparsity inducing priors.}

Turning the focus to lag-length, we find that for the data set at hand, competitive forecasts require at least two lags of the endogenous variables as predictors. \revision{Most priors perform best with $p=2$ or $p=3$ lags. HS$^*$ and MP\_LIT perform best with $p=4$ lags.} 
Interestingly, \revision{concerning one-step-ahead forecasts, the performances of HS, DL, DL$_a$, DL$_a^*$, NG, NG$_a$, NG$_a^*$, R2D2, R2D2$_a$, and R2D2$_a^*$ deteriorate with $p>3$. In this respect, HS$^*$ and SHM -- the only priors without a clear drop in scores for $p=5$ -- are the most robust priors}. 

The forecasting exercise also reveals both potentials and limitations of the hierarchical treatment of hyperparameters. The Minnesota prior clearly benefits from treating both shrinkage parameters as random variables. For GL priors, the story is more complex. On the one hand, DL, NG, and R2D2 in their vanilla implementations do not improve by placing priors on the concentration parameters $a^\gamma$, $a^\delta$, and $a^\pi$, respectively. On the other hand, the semi-global modifications of both NG and R2D2 unfold their full potential only if both $a^\delta_j$ and $a^\pi_j$, $j=1,\dots,2p$, are equipped with hyperpriors. This finding confirms our conjecture that GL priors have limitations when the noise level and the sparsity differs in different regions of the parameter space.

\begin{figure}[t]
    \centering
    \begin{subfigure}{\textwidth}
    \centering
        \includegraphics[width=\textwidth]{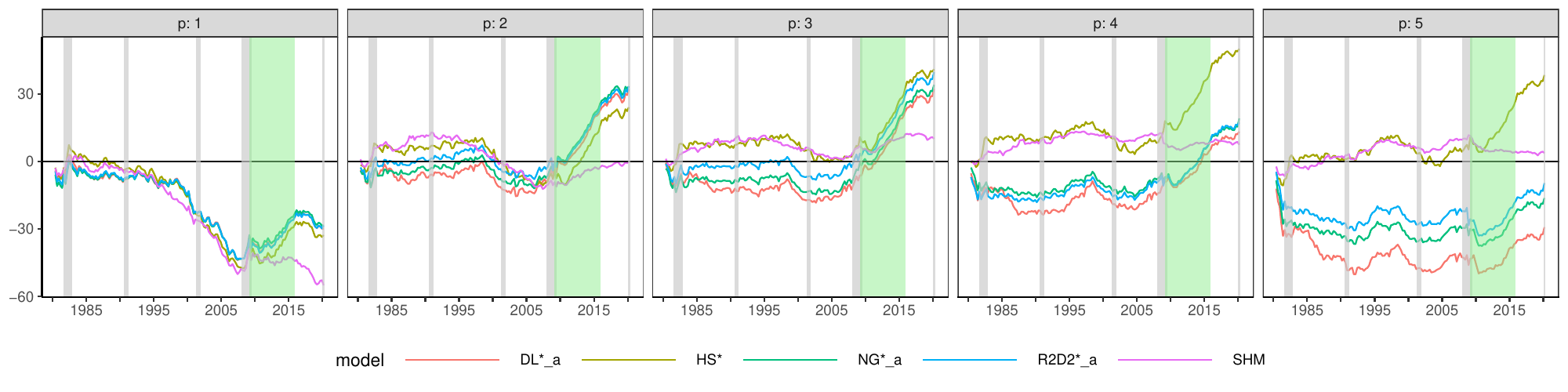}
    \caption{One-step-ahead.}
    \end{subfigure}
    \\[1em]
    \begin{subfigure}{\textwidth}
    \centering
        \includegraphics[width=\textwidth]{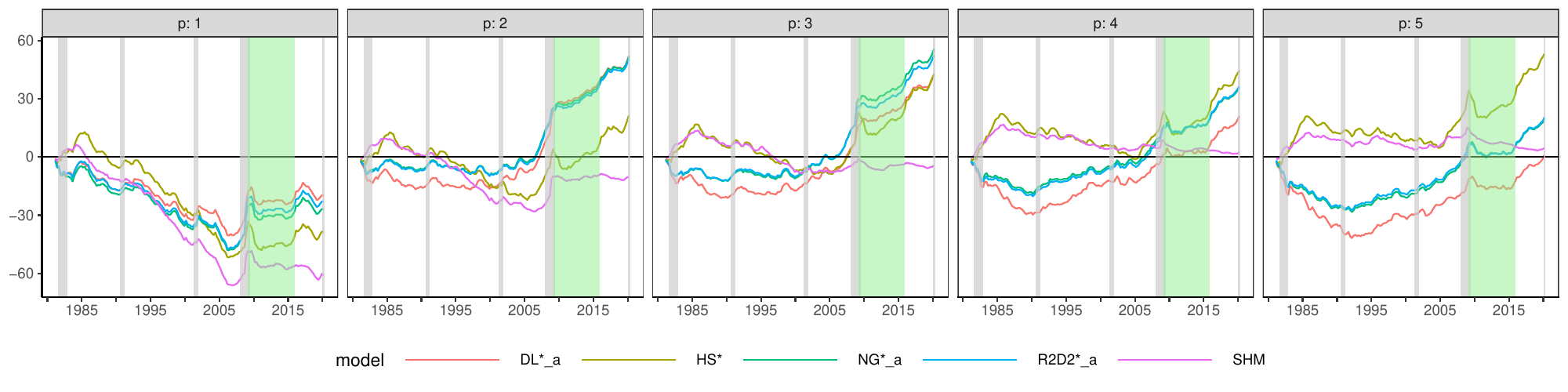}
        \caption{Four-steps-ahead.}
    \end{subfigure}
    \caption{Cumulative log predictive likelihoods relative to MP\_LIT with $p=4$ considering the full 21-dimensional predictive density. \revision{Gray shaded areas are recessions dated by the National Bureau of Economic Research. The green shaded area denotes the period of the zero lower bound.}}
    \label{fig:clbf}
\end{figure}

Figure \ref{fig:clbf} depicts how the forecasting scores evolve over time for an illustrative selection of different priors. It reveals that the \revision{time between the early 2000s recession and the financial crisis 2007/2008 marks a turning point} 
in the evolution of the scores. \revision{Especially during the time of the zero lower bound (ZLB), which we define as the period after the financial crisis when the FFR is below 20 basis points \citep[cf.][]{mavroeidisIdentificationZeroLower2021}, } 
the sparse models -- the semi-global modifications of the GL priors -- accumulate much higher scores than the denser SHM. \revision{The sparsity inducing priors seem to capture the ZLB considerably well, by virtually zeroing almost all cross-lag coefficients in the equations of the interest rates.} Before the crisis, however, SHM is very competitive. The only sparse prior en par with SHM before the crisis is HS*. Overall, there is no single model that performs best over the whole evaluation period. \revision{The findings suggest that there may be no general answer to the question whether sparse or dense modeling techniques perform better in forecasting economic data.}

\begin{figure}[t]
    \centering
    \begin{subfigure}{\textwidth}
    \centering
        \includegraphics[width=\textwidth]{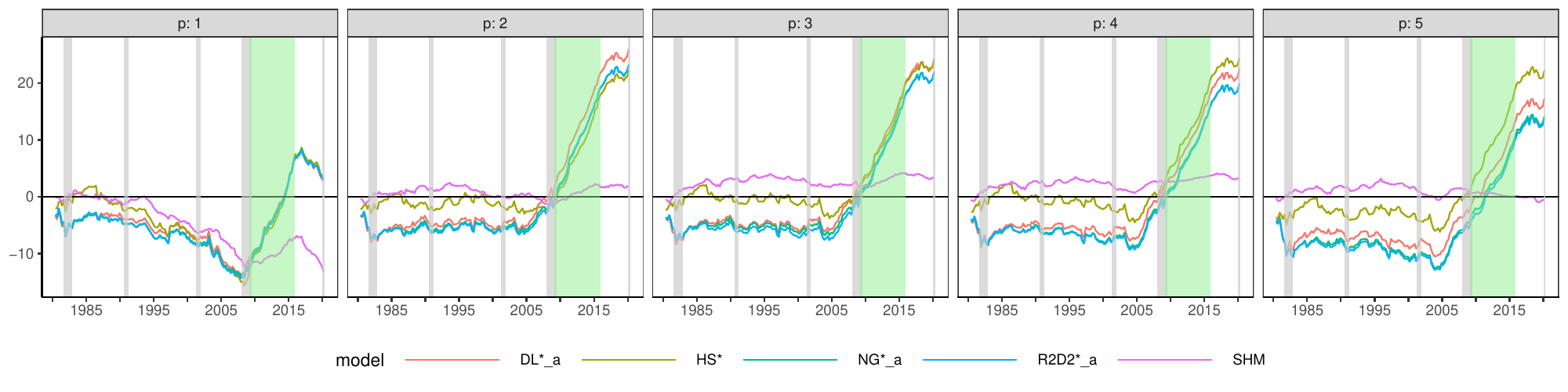}
    \caption{One-step-ahead.}
    \end{subfigure}
    \\[1em]
    \begin{subfigure}{\textwidth}
    \centering
        \includegraphics[width=\textwidth]{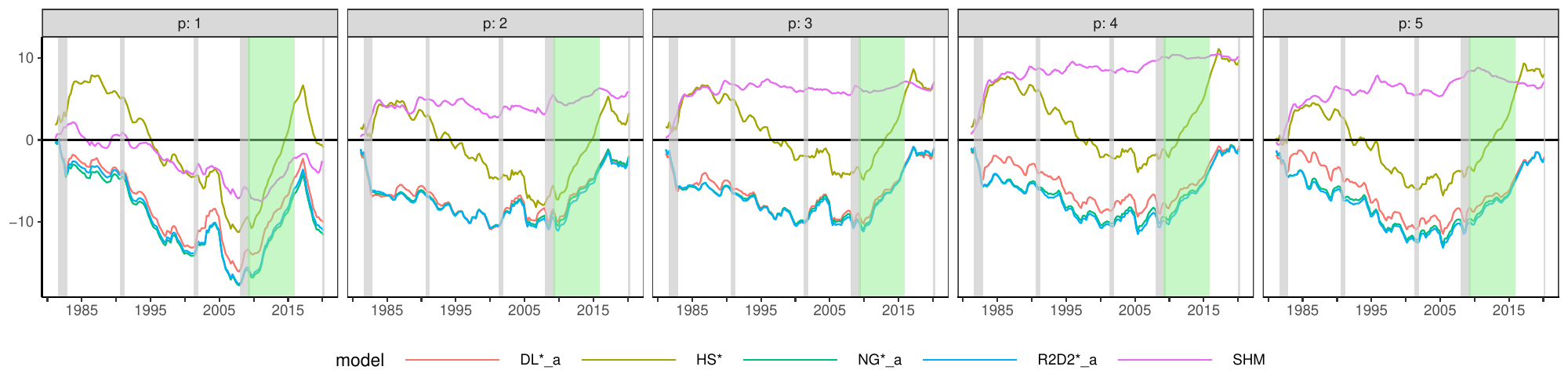}
        \caption{Four-steps-ahead.}
    \end{subfigure}
    \caption{\revision{Cumulative log predictive likelihoods relative to MP\_LIT with $p=4$ considering the joint predictive density of GDP, CPI and FFR. Gray shaded areas are recessions dated by the National Bureau of Economic Research. The green shaded area denotes the period of the zero lower bound.}}
    \label{fig:clbf_VoI}
\end{figure}

\revision{In real applications, there might be interest in forecasts of only a small set of variables \citep[cf., e.g.,][]{banbura_large_2010, feldkircherSophisticatedSmallSimple2024}. Figure \ref{fig:clbf_VoI} depicts cumulative log predictive likelihoods where instead of the full 21-dimensional predictive density the joint predictive density of three focal variables, namely GDP, CPI and FFR, is evaluated. Similar to before, we observe that SHM dominates until the mid 2000s, and the semi-global-local shrinkage priors since the mid 2000s.}

\subsection{Dynamic model averaging}
\label{sec:DMA}
The previous sections highlighted the heterogeneity of model performance over time. The question arises whether combining weighted forecasts can improve forecasting performance. 

Dynamic model averaging \citep[DMA,][]{raftery_online_2010,koop_forecasting_2012,onorante_dynamic_2016} is a straightforwardly implementable averaging scheme, where the weights are sequentially updated as new observations become available, depending on the forecasting performance up to this point in time. A model with strong support from the data for a given period will receive a higher weight for the following period. By contrast, a model performing relatively worse will receive a lower weight.

In a nutshell, DMA works as follows. Denote $PL_{t|t-1,i}$ the one-step-ahead PL in $t$ for model $i$ within model space $M$. The predicted weight $\omega_{t|t-1,i}$ associated with model $i$ is computed as follows,
\begin{equation}\label{eq:weights_predict}
	\omega_{t|t-1,i} = \frac{\omega_{t-1|t-1,i}^\alpha}{\sum_{i\in \mathcal{M}} \omega_{t-1|t-1,i}^\alpha},
\end{equation}
where $0 \leq \alpha \leq 1$ denotes a discount factor. The model updating equation then is
\begin{equation}\label{eq:dma}
	\omega_{t|t,i} = \frac{\omega_{t|t-1,i} PL_{t|t-1,i}}{\sum_{i\in \mathcal{M}} \omega_{t|t-1,i} PL_{t|t-1,i}}.
\end{equation}
To shed more light on the discount factor: Standard (static) Bayesian model averaging is achieved by setting $\alpha=1$, as $\omega_{t|t-1,i}$ would be proportional to the (training-sample) marginal likelihood of model $i$ using data through the time $t-1$. We follow \citet{raftery_online_2010} in specifying $\alpha = 0.99$, indicating that $PL_{t-1|t-2}$ receives $99\%$ as much weight as $PL_{t|t-1}$. 

The sparse models that we consider for DMA are DL$_a^*$, HS$^*$, NG$_a^*$, and R2D2$_a^*$. MP\_LIT and SHM, by contrast, are representing the dense models. Concerning lag-length, we average over VARs of order $p \in \{1,2,3,4,5\}$. Last but not least, we initialize the weights to be equally distributed over all considered models for the first prediction.

\begin{figure}[tp]
     \centering
     \begin{subfigure}{\textwidth}
         \centering
         \includegraphics[width=\textwidth]{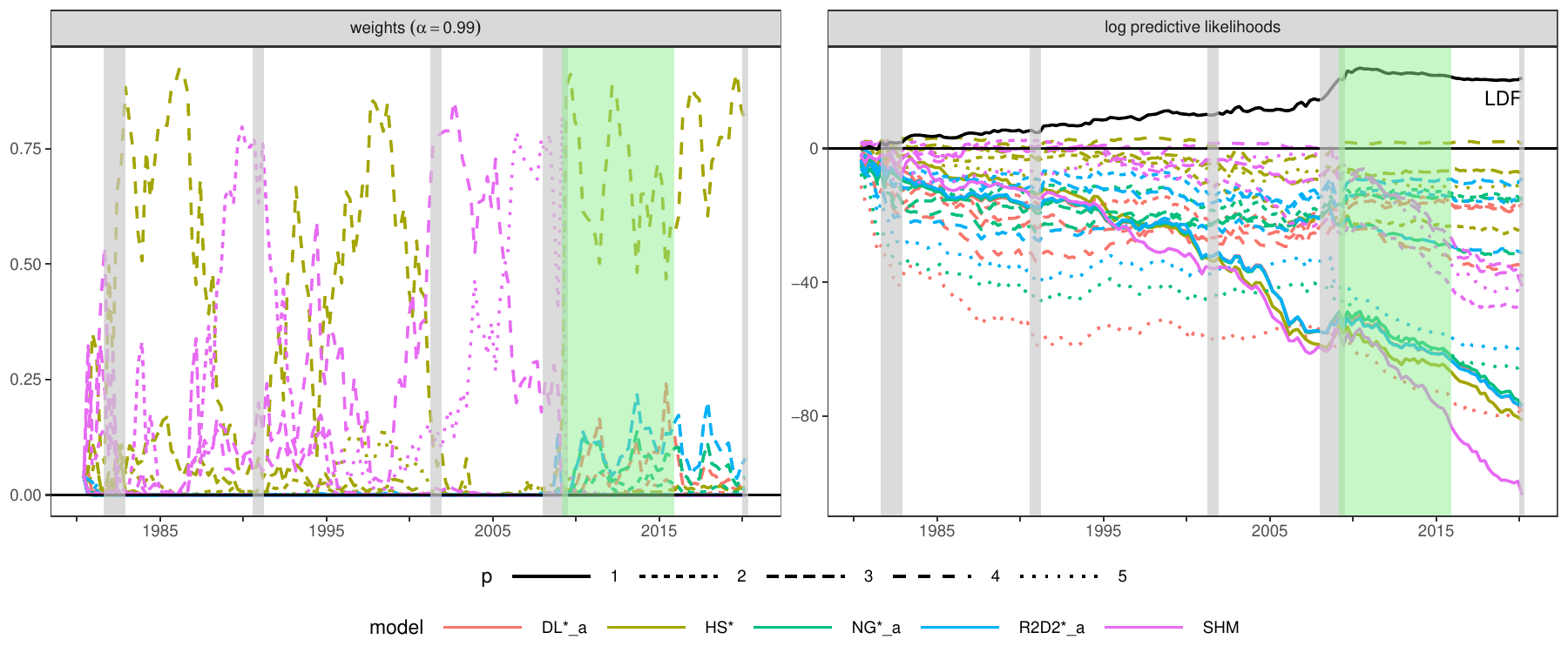}
         \caption{Weights determined by full 21-dimensional predictive density.}
         \label{fig:dma_full}
     \end{subfigure}
     \\[1em]
     \begin{subfigure}{\textwidth}
         \centering
         \includegraphics[width=\textwidth]{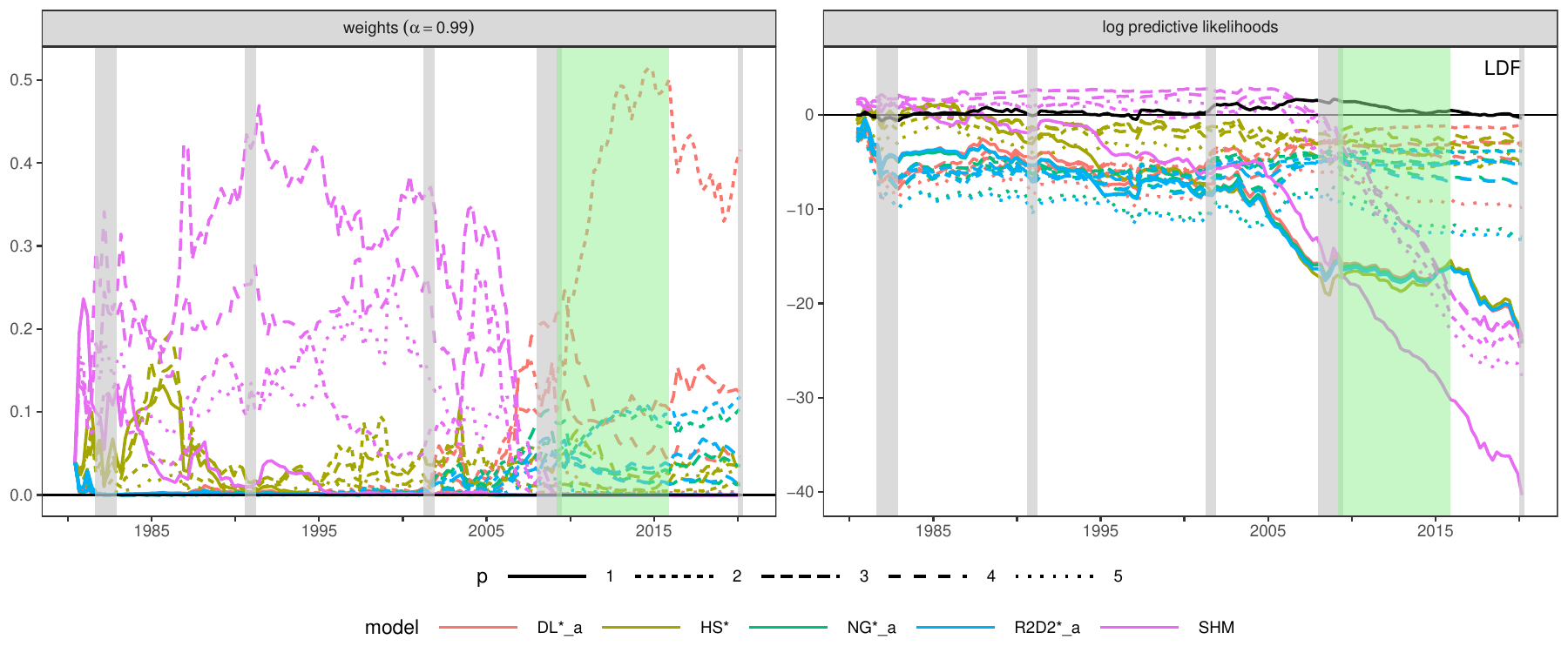}
         \caption{Weights determined by joint predictive density of GDP, CPI, and FFR.}
         \label{fig:dma_tri}
     \end{subfigure}
        \caption{Dynamic model averaging (DMA) for one-step-ahead predictions: Model weights (left) and cumulative log predictive likelihoods relative to DMA (right) with discount factor $\alpha=0.99$. \revision{The black solid line stands for the cumulative log predictive likelihood of the loss discounting framework (LDF) relative to DMA. The gray shaded areas are recessions dated by the National Bureau of Economic Research. The green shaded area denotes the period of the zero lower bound.}}
        \label{fig:dma}
\end{figure}

Turning towards the results, we first consider the one-step-ahead predictive density of all 21 variables. Figure \ref{fig:dma_full} shows the evolution of model weights and cumulative log predictive likelihoods of the considered models relative to DMA over the whole evaluation period. Until the end of the 2000s, SHM with $p \in \{2,3,4,5\}$ and HS$^*$ with $p=4$ are dominating alternately. From \revision{2009 until the end of the evaluation period, HS$^*$ with $p=4$ receives the majority of the weights, the remaining semi-global-local priors have some weights, and SHM has quasi zero weights.} 
Analyzing the cumulative log predictive likelihoods, it turns out that DMA is practically never worse than any single model specification.

DMA can also improve forecasts if only a subset of variables is considered.
Figure \ref{fig:dma_tri} depicts DMA model weights and cumulative log predictive likelihoods relative to DMA when the predictive density of GDP, CPI and FFR is taken into account, only. By and large, the evolution of the predictive scores is similar to when all variables are considered. However, the analysis of model weights reveals that the distribution of weights over the models is now more dispersed. DL$_a^*$(2), which is practically not apparent  when evaluating the full 21-dimensional predictive density, now receives the largest weights after 2010.

\revision{The result of DMA, of course, depends on the choice of $\alpha$. To address this issue, \citet{bernaciakLossDiscountingFramework2024} introduce the loss discounting framework (LDF) which extends DMA to general loss functions and more general discounting dynamics. 
We consider the following variant of LDF: Let $S_\alpha=\{\alpha_1,\dots,\alpha_s\}$ denote a set of different discount factors. In a first layer, we apply DMA with discount factor $\alpha_i$ for $i=1,\dots,s$ to our model pool. This yields $s$ dynamically averaged meta-models. In the second layer, we apply DMA with discount factor $\alpha_i$ for $i=1,\dots,s$ to the $s$ meta-models. This process can be repeated arbitrarily. In the final meta-model layer a single discount factor needs to be chosen. Adding more layers has a diminishing effect on the sequence of predictive distributions and therefore on the final result. It can be shown that the weights converge as the number of layers approaches infinity. In this respect, the final result is least affected by the choice of the discount factor in the last layer when a large number of layers $L$ is chosen. Concretely, we set $S_\alpha=\{0.1,0.2,0.3,0.4,0.5,0.6,0.7,0.8,0.9,0.95,0.99,1\}$ and $l=4$, which appears to be sufficient in our case in the sense that the exact choice of the discount factor in the final layer shows negligible impact. According to Figure \ref{fig:dma}, LDF outperforms all single model specifications and DMA throughout the whole evaluation period with respect to the full 21-dimensional predictive density. Focusing on the joint predictive distribution of the three focal variables, LDF is en par with DMA and not worse than any single model in the long run.
}

To sum up, model performance is not only heterogeneous over time, but also across evaluated variables. Model averaging is a very effective tool in combining the merits of various modeling techniques, in contrast to the apparently unrewarding search for the single model that can do them all. 

\subsection{Sensitivity of results with respect to the prior for \texorpdfstring{$\bm \Sigma_t$}{Sigma}}
\label{sec:robustness}
\revision{In this section, we investigate the sensitivity of results with respect to different prior choices for $\bm \Sigma_t$.  The main focus is on 
different prior choices for $\bm u$. Additionally, we briefly discuss a homoskedastic specification of the variance-covariance matrix.}

The elements in $\bm u$ are often interpreted as contemporaneous coefficients. 
Hence, an obvious choice for $\bm u$ would be to use a prior of the same family as the prior on $\bm \phi$, e.g., DL/DL$_a^*$ on $\bm \phi$ is combined with DL on $\bm u$. For DL on $\bm u$, \revision{for simplicity we specify $a^\gamma=\frac{1}{n_u}$, although this parameter could be learned from the data as well. Here, $n_u$ is the number of elements in $\bm u$.} For NG and R2D2 on $\bm u$ we set $a^\delta = a^\pi = \frac{1}{2 n_u}$, such that for DL, NG, and R2D2, the concentration at zero is matched. \revision{For SSVS on $\bm u$ we specify $\tau_{0i} = \frac{1}{1000}$ and $\tau_{1i} = 1.$} As a second alternative, we investigate a relatively uninformative normal prior with zero mean and variance 10 for each element in $\bm{u}$, independently, which we refer to as FLAT prior. As a third option, we investigate a global(-but-not-local) shrinkage (GS) prior. More specifically, the prior is conditionally normal with zero mean and a global gamma hyperprior on the variance parameter: $u_{ij} | \lambda_g \sim N(0,\lambda_g)$, $\lambda_g \sim G(0.01,0.01)$. GS is similar in spirit to SHM. In combination with SHM on $\bm \phi$, then the contemporaneous coefficients $\bm u$ represents the third distinct group besides own-lag and cross-lag coefficients. 

\begin{table}[tp]
\centering
\resizebox{\textwidth}{!}{
\begin{tabular}[t]{l !{\extracolsep{4pt}}r!{\extracolsep{0pt}}rrr !{\extracolsep{4pt}}r!{\extracolsep{0pt}}rrr !{\extracolsep{4pt}}r!{\extracolsep{0pt}}rrr}
\hline\hline
& \multicolumn{4}{c}{$p=2$}& \multicolumn{4}{c}{$p=3$}& \multicolumn{4}{c}{$p=4$}\\
& \multicolumn{4}{c}{Prior on $\bm{u}$}& \multicolumn{4}{c}{Prior on $\bm{u}$}& \multicolumn{4}{c}{Prior on $\bm{u}$}\\
\cline{2-5}\cline{6-9}\cline{10-13}
Prior on $\bm{\phi}$ & HS & GS & FLAT &*/*& HS & GS & FLAT &*/*& HS & GS & FLAT &*/*\\
\hline 
MP\_LIT & \textbf{-17} & -96 & -64 & - & \textbf{-8} & -87 & -73 & - & \textbf{0} & -89 & -63 & -\\
SHM & \textbf{0} & -83 & -44 & - & \textbf{10} & -75 & -59 & - & \textbf{7} & -73 & -46 & -\\
HS & \textbf{-27} & -106 & -78 & \textbf{-27} & \textbf{-23} & -103 & -95 & \textbf{-23} & \textbf{-36} & -107 & -92 & \textbf{-36}\\
DL & \textbf{10} & -59 & -35 & 0 & \textbf{11} & -71 & -66 & 3 & -6 & -83 & -74 & \textbf{-3}\\
NG & \textbf{11} & -72 & -51 & 9 & 9 & -70 & -64 & \textbf{15} & -2 & -86 & -74 & \textbf{1}\\
R2D2 & \textbf{8} & -75 & -46 & 7 & 9 & -74 & -64 & \textbf{13} & 3 & -80 & -71 & \textbf{9}\\
SSVS & -29 & -105 & -86 & \textbf{-25} & -29 & -101 & -101 & \textbf{-28} & \textbf{-49} & -125 & -108 & -51\\
HS$^*$ & \textbf{24} & -63 & -27 & \textbf{24} & \textbf{41} & -49 & -35 & \textbf{41} & \textbf{50} & -46 & -19 & \textbf{50}\\
DL$^*_a$ & \textbf{32} & -57 & -26 & 29 & \textbf{32} & -69 & -61 & 22 & \textbf{13} & -82 & -95 & 5\\
NG$^*_a$ & \textbf{33} & -44 & -18 & 32 & \textbf{34} & -59 & -44 & \textbf{34} & \textbf{19} & -77 & -59 & 14\\
R2D2$^*_a$ & 33 & -57 & -21 & \textbf{36} & \textbf{40} & -54 & -36 & 32 & \textbf{17} & -66 & -70 & 16\\
\hline
\end{tabular}
}
\caption{Sum of relative one-step-ahead log predictive likelihoods for several VARs equipped with different priors on $\bm u$. The benchmark model is MP\_LIT on $\bm \phi$ with $p=4$ lags in combination with HS on $\bm u$. The flag */* indicates that the prior on $\bm u$ is of the same family as the prior on $\bm \phi$. \revision{The highest value in
each row of the columns corresponding to $p\in \{2,3,4\}$ is indicated in bold font.}}
\label{tab:rob}
\end{table}

Table \ref{tab:rob} shows sums of one-step-ahead log predictive likelihoods relative to a VAR with $p=4$ lags with MP\_LIT on $\bm \phi$ and HS on $\bm u$ for an illustrative selection of VARs. GS or FLAT on $\bm{u}$ clearly deteriorates forecasting performance of all models compared to HS used in the main analysis. Interestingly, GS is even worse than FLAT, which indicates that there are few strong signals that GS simply overshrinks. GL priors and SSVS on $\bm u$ are serious contenders to HS. 
Overall, the results highlight the importance of a sparsity-inducing prior for $\bm u$. The sensitivity of forecasting performance with respect to the prior choice for $\bm u$ strengthen our view to use the same prior on $\bm u$ for all models in the main analysis and also could explain to some extent the relatively weak performance of SHM in \citet{huber_adaptive_2019}, where it is combined with the FLAT prior on $\bm{u}$.

\revision{Concerning the role of SV in forecasting macroeconomic data, we also consider homoskedastic VARs in a preliminary exercise. We assume that the diagonal matrix $\bm D_t=\bm D$ is constant for $t=1,\dots,T$. In that case, we specify independent inverse gamma priors with both shape and scale equalling 0.01 for the diagonal elements in $\bm D$. The prior on $\bm u$ is HS as described in Section \ref{sec:prior_sigma}. In terms of forecasting performance, the results (available upon request) are overwhelmingly clear in favor of stochastic volatility. Therefore, we do not continue this route of research any further. More comprehensive treatments of this matter with similar findings can be found in \citet{koopLargeTimevaryingParameter2013,clark2015, carriero_large_2019}, among others.}

\section{Conclusion}
\label{sec:conclusion}
Careful prior elicitation in Bayesian vector autoregressions goes back at least to \citet{litterman_forecasting_1986} and has been refined ever since. Our main contribution is a discussion of modern variants of prior choices, both established and novel, and a side-by-side comparison in terms of their structural (sparsity) behavior. In doing so, we borrow both from well-established (Minnesota-type) structural knowledge as well as from more automatic (shrinkage-type) approaches, thereby aiming at combining the best of both worlds. As a result of this aim, we propose what we term semi-global-local priors. We compare the various specifications in terms of their in- and out-of-sample performance on simulated data, as well as on a practically interesting (and notoriously hard) macroeconomic prediction problem. \revision{Overall, we find that the semi-global-local priors outperform the vanilla global-local priors and are strong competitors to the semi-hierarchical Minnesota prior, which itself already constitutes an improvement to the traditional Minnesota prior.} 
Finally, we investigate how dynamic model averaging can help to shed light on the temporal patterns representing historical regime shifts. It becomes clear that no single model dominates the others; much rather, we find that some models are sometimes better than others, and dynamically combining them is fruitful.

While trying to comprehensively and integratively shed light on the Illusion of Sparsity debate, we also want to point out that many questions remain unanswered. Issues left for future research include an in-depth analysis where forecasting performance between modeling reduced-form and modeling structural-form coefficients is compared side by side. Moreover, a systematic analysis of the various different approaches to modeling reduced-form VARs with time-varying variance-covariance matrices (e.g.\ Cholesky-SV vs.\ Factor-SV) is still missing. \revision{A first notable attempt in this direction is undertaken by \citet{chanComparingStochasticVolatility2023}. As mentioned by an anonymous referee, the semi-global approach could be useful for putting structure on the variance-covariance matrix as well, e.g., by imposing group-specific shrinkage according to industry and economic sectors \citep[cf.][]{kastner_sparse_2019}.}
\revisiontwo{%
Furthermore, while our forecasting exercise takes into account the entire predictive distribution, we do not specifically focus on the assessment of tail risk.
Threshold- and quantile-weighted scoring rules as discussed in \citet{gneiting2011} or the asymmetric continuous probability score (ACPS) proposed in \citet{iacopiniProperScoringRules2023} could serve as starting points in this regard.
}%
Lastly, we believe that this work could be expanded to cater for VARs with time-varying regression parameters \citep[e.g.][]{bitto2019achieving, huber2019should, knaussfs, feldkircherSophisticatedSmallSimple2024} where the issue of choosing good shrinkage priors is exacerbated in the sense that in addition to shrinkage towards zero, shrinkage towards constancy needs to be dealt with.

\bibliographystyle{apalike_modified.bst}
\bibliography{BVAR}

\newcommand{\noop}[1]{}
\begin{thebibliography}{}

\bibitem[Bateman, 1953]{bateman1953higher}
Bateman, H. (1953).
\newblock {\em Higher transcendental functions}, volume~1.
\newblock McGraw-Hill Book Company, New York.

\bibitem[Bańbura et~al., 2010]{banbura_large_2010}
Bańbura, M., Giannone, D., and Reichlin, L. (2010).
\newblock Large {Bayesian} vector auto regressions.
\newblock {\em Journal of Applied Econometrics}, 25(1):71--92.

\bibitem[Bernaciak and Griffin, 2024]{bernaciakLossDiscountingFramework2024}
Bernaciak, D. and Griffin, J.~E. (2024).
\newblock A loss discounting framework for model averaging and selection in
  time series models.
\newblock {\em International Journal of Forecasting}, 40(4):1721--1733.

\bibitem[Bernardi et~al., 2024]{bernardi2022}
Bernardi, M., Bianchi, D., and Bianco, N. (2024).
\newblock Variational inference for large {Bayesian} vector autoregressions.
\newblock {\em Journal of Business \& Economic Statistics}, 42(3):1066--1082.

\bibitem[Bhattacharya et~al., 2016]{bhattacharyaFastSamplingGaussian2016}
Bhattacharya, A., Chakraborty, A., and Mallick, B.~K. (2016).
\newblock Fast sampling with {G}aussian scale mixture priors in
  high-dimensional regression.
\newblock {\em Biometrika}, 103(4):985--991.

\bibitem[Bhattacharya et~al., 2015]{bhattacharya_dirichletlaplace_2015}
Bhattacharya, A., Pati, D., Pillai, N.~S., and Dunson, D.~B. (2015).
\newblock Dirichlet–{L}aplace priors for optimal shrinkage.
\newblock {\em Journal of the American Statistical Association},
  110(512):1479--1490.

\bibitem[Bitto and Frühwirth-Schnatter, 2019]{bitto2019achieving}
Bitto, A. and Frühwirth-Schnatter, S. (2019).
\newblock Achieving shrinkage in a time-varying parameter model framework.
\newblock {\em Journal of Econometrics}, 210(1):75--97.

\bibitem[Brown and Griffin, 2010]{griffin2010}
Brown, P.~J. and Griffin, J.~E. (2010).
\newblock {Inference with normal-gamma prior distributions in regression
  problems}.
\newblock {\em Bayesian Analysis}, 5(1):171--188.

\bibitem[Carriero et~al., 2022]{carriero_corrigendum_2021}
Carriero, A., Chan, J., Clark, T.~E., and Marcellino, M. (2022).
\newblock Corrigendum to “{Large} {Bayesian} vector autoregressions with
  stochastic volatility and non-conjugate priors” [{J}. {E}conometrics 212
  (1) (2019) 137–154].
\newblock {\em Journal of Econometrics}, 227(2):506--512.

\bibitem[Carriero et~al., 2019]{carriero_large_2019}
Carriero, A., Clark, T.~E., and Marcellino, M. (2019).
\newblock Large {Bayesian} vector autoregressions with stochastic volatility
  and non-conjugate priors.
\newblock {\em Journal of Econometrics}, 212(1):137--154.

\bibitem[Carvalho et~al., 2010]{carvalhi2010}
Carvalho, C.~M., Polson, N.~G., and Scott, J.~G. (2010).
\newblock The horseshoe estimator for sparse signals.
\newblock {\em Biometrika}, 97(2):465--480.

\bibitem[Chan, 2021a]{chan_VAR_SV}
Chan, J. C.~C. (2021a).
\newblock Comparing stochastic volatility specifications for large {Bayesian}
  {VARs}.
\newblock https://joshuachan.org/papers/ML-LargeVARSV.pdf ; accessed:
  2021-05-19.

\bibitem[Chan, 2021b]{chan_minnesota-type_2021}
Chan, J. C.~C. (2021b).
\newblock Minnesota-type adaptive hierarchical priors for large {Bayesian}
  {VARs}.
\newblock {\em International Journal of Forecasting}, 37(3):1212--1226.

\bibitem[Chan, 2022]{chan2022}
Chan, J. C.~C. (2022).
\newblock Asymmetric conjugate priors for large {Bayesian} {VARs}.
\newblock {\em Quantitative Economics}, 13(3):1145--1169.

\bibitem[Chan, 2023]{chanComparingStochasticVolatility2023}
Chan, J. C.~C. (2023).
\newblock Comparing stochastic volatility specifications for large {Bayesian}
  {VARs}.
\newblock {\em Journal of Econometrics}, 235(2):1419--1446.

\bibitem[Chan et~al., 2024]{chan_large_2021}
Chan, J. C.~C., Koop, G., and Yu, X. (2024).
\newblock Large order-invariant {Bayesian} {VARs} with stochastic volatility.
\newblock {\em Journal of Business \& Economic Statistics}, 42(2):825--837.

\bibitem[Clark and Ravazzolo, 2015]{clark2015}
Clark, T.~E. and Ravazzolo, F. (2015).
\newblock Macroeconomic forecasting performance under alternative
  specifications of time-varying volatility.
\newblock {\em Journal of Applied Econometrics}, 30(4):551--575.

\bibitem[Cogley and Sargent, 2005]{cogley_drifts_2005}
Cogley, T. and Sargent, T.~J. (2005).
\newblock Drifts and volatilities: monetary policies and outcomes in the post
  {WWII} {US}.
\newblock {\em Review of Economic Dynamics}, 8(2):262--302.

\bibitem[Cross et~al., 2020]{cross_macroeconomic_2020}
Cross, J.~L., Hou, C., and Poon, A. (2020).
\newblock Macroeconomic forecasting with large {Bayesian} {VARs}: Global-local
  priors and the illusion of sparsity.
\newblock {\em International Journal of Forecasting}, 36(3):899--915.

\bibitem[Eddelbuettel and Fran\c{c}ois, 2011]{rcpp}
Eddelbuettel, D. and Fran\c{c}ois, R. (2011).
\newblock {Rcpp}: Seamless {R} and {C++} integration.
\newblock {\em Journal of Statistical Software}, 40(8):1--18.

\bibitem[Eddelbuettel and Sanderson, 2014]{rcpparmadillo}
Eddelbuettel, D. and Sanderson, C. (2014).
\newblock Rcpparmadillo: Accelerating {R} with high-performance {C}++ linear
  algebra.
\newblock {\em Computational Statistics and Data Analysis}, 71:1054--1063.

\bibitem[Fava and Lopes, 2021]{fava_illusion_2020}
Fava, B. and Lopes, H.~F. (2021).
\newblock {The illusion of the illusion of sparsity: {An} exercise in prior
  sensitivity}.
\newblock {\em Brazilian Journal of Probability and Statistics},
  35(4):699--720.

\bibitem[Feldkircher et~al., 2024]{feldkircherSophisticatedSmallSimple2024}
Feldkircher, M., Gruber, L., Huber, F., and Kastner, G. (2024).
\newblock Sophisticated and small versus simple and sizeable: When does it pay
  off to introduce drifting coefficients in {Bayesian} vector autoregressions?
\newblock {\em Journal of Forecasting}, 43(6):2126--2145.

\bibitem[Follett and Yu, 2019]{follett_achieving_2019}
Follett, L. and Yu, C. (2019).
\newblock Achieving parsimony in {Bayesian} vector autoregressions with the
  horseshoe prior.
\newblock {\em Econometrics and Statistics}, 11:130--144.

\bibitem[Fr\"{u}hwirth-Schnatter-Schnatter and Knaus, 2021]{knaussfs}
Fr\"{u}hwirth-Schnatter-Schnatter, S. and Knaus, P. (2021).
\newblock Sparse {B}ayesian state-space and time-varying parameter models.
\newblock In Tadesse, M.~G. and Vannucci, M., editors, {\em Handbook of
  Bayesian Variable Selection}, pages 297--326. Chapman \& Hall.

\bibitem[George et~al., 2008]{george_bayesian_2008}
George, E.~I., Sun, D., and Ni, S. (2008).
\newblock Bayesian stochastic search for {VAR} model restrictions.
\newblock {\em Journal of Econometrics}, 142(1):553--580.

\bibitem[Geweke and Amisano, 2010]{geweke_comparing_2010}
Geweke, J. and Amisano, G. (2010).
\newblock Comparing and evaluating {Bayesian} predictive distributions of asset
  returns.
\newblock {\em International Journal of Forecasting}, 26(2):216--230.

\bibitem[Giannone et~al., 2015]{giannone_prior_2015}
Giannone, D., Lenza, M., and Primiceri, G.~E. (2015).
\newblock Prior selection for vector autoregressions.
\newblock {\em The Review of Economics and Statistics}, 97(2):436--451.

\bibitem[Giannone et~al., 2019]{giannonePriorsLongRun2019}
Giannone, D., Lenza, M., and Primiceri, G.~E. (2019).
\newblock Priors for the long run.
\newblock {\em Journal of the American Statistical Association},
  114(526):565--580.

\bibitem[Giannone et~al., 2021]{giannone_economic_2021}
Giannone, D., Lenza, M., and Primiceri, G.~E. (2021).
\newblock Economic predictions with big data: {The} illusion of sparsity.
\newblock {\em Econometrica}, 89(5):2409--2437.

\bibitem[Gneiting and Ranjan, 2011]{gneiting2011}
Gneiting, T. and Ranjan, R. (2011).
\newblock Comparing density forecasts using threshold- and quantile-weighted
  scoring rules.
\newblock {\em Journal of Business \& Economic Statistics}, 29(3):411--422.

\bibitem[Hosszejni and Kastner, 2021]{hosszejni_modeling_2021}
Hosszejni, D. and Kastner, G. (2021).
\newblock Modeling univariate and multivariate stochastic volatility in {R}
  with stochvol and factorstochvol.
\newblock {\em Journal of Statistical Software}, 100:1--34.

\bibitem[Hoyer, 2004]{hoyer_non-negative_2004}
Hoyer, P.~O. (2004).
\newblock Non-negative matrix factorization with sparseness constraints.
\newblock {\em The Journal of Machine Learning Research}, 5:1457--1469.

\bibitem[Huber and Feldkircher, 2019]{huber_adaptive_2019}
Huber, F. and Feldkircher, M. (2019).
\newblock Adaptive shrinkage in {Bayesian} vector autoregressive models.
\newblock {\em Journal of Business \& Economic Statistics}, 37(1):27--39.

\bibitem[Huber et~al., 2019]{huber2019should}
Huber, F., Kastner, G., and Feldkircher, M. (2019).
\newblock Should {I} stay or should {I} go? {A} latent threshold approach to
  large-scale mixture innovation models.
\newblock {\em Journal of Applied Econometrics}, 34(5):621--640.

\bibitem[Iacopini et~al., 2023]{iacopiniProperScoringRules2023}
Iacopini, M., Ravazzolo, F., and Rossini, L. (2023).
\newblock Proper scoring rules for evaluating density forecasts with asymmetric
  loss functions.
\newblock {\em Journal of Business \& Economic Statistics}, 41(2):482--496.

\bibitem[Jacquier et~al., 1994]{jacquier_bayesian_1994}
Jacquier, E., Polson, N.~G., and Rossi, P.~E. (1994).
\newblock Bayesian analysis of stochastic volatility models.
\newblock {\em Journal of Business \& Economic Statistics}, 12(4):371--389.

\bibitem[Jarociński and Maćkowiak, 2017]{jarocinski2017}
Jarociński, M. and Maćkowiak, B. (2017).
\newblock {Granger Causal Priority and Choice of Variables in Vector
  Autoregressions}.
\newblock {\em The Review of Economics and Statistics}, 99(2):319--329.

\bibitem[Kastner, 2019]{kastner_sparse_2019}
Kastner, G. (2019).
\newblock Sparse {Bayesian} time-varying covariance estimation in many
  dimensions.
\newblock {\em Journal of Econometrics}, 210(1):98--115.

\bibitem[Kastner and Frühwirth-Schnatter,
  2014]{kastner_ancillarity-sufficiency_2014}
Kastner, G. and Frühwirth-Schnatter, S. (2014).
\newblock Ancillarity-sufficiency interweaving strategy ({ASIS}) for boosting
  {MCMC} estimation of stochastic volatility models.
\newblock {\em Computational Statistics \& Data Analysis}, 76:408--423.

\bibitem[Kastner and Huber, 2020]{kastner_sparse_2020}
Kastner, G. and Huber, F. (2020).
\newblock Sparse {Bayesian} vector autoregressions in huge dimensions.
\newblock {\em Journal of Forecasting}, 39(7):1142--1165.

\bibitem[Kim et~al., 1998]{kim_stochastic_1998}
Kim, S., Shephard, N., and Chib, S. (1998).
\newblock Stochastic volatility: Likelihood inference and comparison with
  {ARCH} models.
\newblock {\em The Review of Economic Studies}, 65(3):361--393.

\bibitem[Koop and Korobilis, 2010]{koop_bayesian_2010}
Koop, G. and Korobilis, D. (2010).
\newblock Bayesian multivariate time series methods for empirical
  macroeconomics.
\newblock {\em Foundations and Trends® in Econometrics}, 3(4):267--358.

\bibitem[Koop and Korobilis, 2012]{koop_forecasting_2012}
Koop, G. and Korobilis, D. (2012).
\newblock Forecasting inflation using dynamic model averaging.
\newblock {\em International Economic Review}, 53(3):867--886.

\bibitem[Koop and Korobilis, 2013]{koopLargeTimevaryingParameter2013}
Koop, G. and Korobilis, D. (2013).
\newblock Large time-varying parameter {VARs}.
\newblock {\em Journal of Econometrics}, 177(2):185--198.

\bibitem[Koop, 2013]{koop_forecasting_2013}
Koop, G.~M. (2013).
\newblock Forecasting with medium and large {Bayesian} {VARs}.
\newblock {\em Journal of Applied Econometrics}, 28(2):177--203.

\bibitem[Leydold and H\"{o}rmann, 2022]{leydold_gigrvg_2017}
Leydold, J. and H\"{o}rmann, W. (2022).
\newblock {\em GIGrvg: Random Variate Generator for the GIG Distribution}.
\newblock R package version 0.7.

\bibitem[Litterman, 1986]{litterman_forecasting_1986}
Litterman, R.~B. (1986).
\newblock Forecasting with {Bayesian} vector autoregressions: Five years of
  experience.
\newblock {\em Journal of Business \& Economic Statistics}, 4(1):25--38.

\bibitem[Luo and Griffin, 2022]{griffin_luo_2022}
Luo, Y. and Griffin, J.~E. (2022).
\newblock Bayesian methods of vector autoregressions with tensor
  decompositions.
\newblock arXiv:2211.01727 [stat.ME].

\bibitem[Lütkepohl, 2005]{lutkepohl_new_2005}
Lütkepohl, H. (2005).
\newblock {\em New Introduction to Multiple Time Series Analysis}.
\newblock Springer Berlin Heidelberg, Berlin, Heidelberg.

\bibitem[Makalic and Schmidt, 2016]{makalic2016}
Makalic, E. and Schmidt, D.~F. (2016).
\newblock A simple sampler for the horseshoe estimator.
\newblock {\em IEEE Signal Processing Letters}, 23(1):179--182.

\bibitem[Mathai et~al., 2009]{mathai2009h}
Mathai, A.~M., Saxena, R.~K., and Haubold, H.~J. (2009).
\newblock {\em The H-function: theory and applications}.
\newblock Springer Science \& Business Media.

\bibitem[Mavroeidis, 2021]{mavroeidisIdentificationZeroLower2021}
Mavroeidis, S. (2021).
\newblock Identification at the zero lower bound.
\newblock {\em Econometrica}, 89(6):2855--2885.

\bibitem[McCracken and Ng, 2021]{mccracken_fred-qd_2020}
McCracken, M.~W. and Ng, S. (2021).
\newblock {FRED}-{QD}: A quarterly database for macroeconomic research.
\newblock {\em Federal Reserve Bank of St. Louis Review}, 103(1):1--44.

\bibitem[Onorante and Raftery, 2016]{onorante_dynamic_2016}
Onorante, L. and Raftery, A.~E. (2016).
\newblock Dynamic model averaging in large model spaces using dynamic {Occam}'s
  window.
\newblock {\em European Economic Review}, 81:2--14.

\bibitem[Polson and Scott, 2011]{polson_shrink_2010}
Polson, N.~G. and Scott, J.~G. (2011).
\newblock Shrink globally, act locally: Sparse {Bayesian} regularization and
  prediction.
\newblock In Bernardo, J.~M., Bayarri, M.~J., Berger, J.~O., Dawid, A.~P.,
  Heckerman, D., Smith, A. F.~M., and West, M., editors, {\em Bayesian
  Statistics 9: Proceedings of the Ninth Valencia International Meeting}, pages
  501--538, Oxford, UK. Oxford University Press.

\bibitem[{R Core Team}, 2023]{R_core}
{R Core Team} (2023).
\newblock {\em R: A Language and Environment for Statistical Computing}.
\newblock R Foundation for Statistical Computing, Vienna, Austria.

\bibitem[Raftery et~al., 2010]{raftery_online_2010}
Raftery, A.~E., Kárný, M., and Ettler, P. (2010).
\newblock Online prediction under model uncertainty via dynamic model
  averaging: Application to a cold rolling mill.
\newblock {\em Technometrics}, 52(1):52--66.

\bibitem[Rossini et~al., forthcoming]{rossiniLossbasedPriorDegrees2024}
Rossini, L., Villa, C., Prevenas, S., and McCrea, R. (forthcoming).
\newblock Loss-based prior for the degrees of freedom of the wishart
  distribution.
\newblock {\em Econometrics and Statistics}.

\bibitem[Sims, 1980]{simsMacroeconomicsReality1980}
Sims, C.~A. (1980).
\newblock Macroeconomics and reality.
\newblock {\em Econometrica}, 48(1):1--48.

\bibitem[Sims and Zha, 1998]{sims_bayesian_1998}
Sims, C.~A. and Zha, T. (1998).
\newblock Bayesian methods for dynamic multivariate models.
\newblock {\em International Economic Review}, 39(4):949--968.

\bibitem[Stock and Watson, 2012]{stock_watson_2012}
Stock, J.~H. and Watson, M.~W. (2012).
\newblock Disentangling the {Channels} of the 2007–09 {Recession}.
\newblock {\em Brookings Papers on Economic Activity}, pages 81--156.

\bibitem[Villani, 2009]{villaniSteadystatePriorsVector2009}
Villani, M. (2009).
\newblock Steady-state priors for vector autoregressions.
\newblock {\em Journal of Applied Econometrics}, 24(4):630--650.

\bibitem[West, 2020]{west_bayesian_2020}
West, M. (2020).
\newblock Bayesian forecasting of multivariate time series: scalability,
  structure uncertainty and decisions.
\newblock {\em Annals of the Institute of Statistical Mathematics}, 72(1).

\bibitem[Wu and Koop, 2022]{WuKoop2022}
Wu, P. and Koop, G. (2022).
\newblock Fast, order-invariant {Bayesian} inference in {VARs} using the
  eigendecomposition of the error covariance matrix.
\newblock Working Papers 2310, University of Strathclyde Business School,
  Department of Economics.

\bibitem[Yu and Meng, 2011]{yumeng}
Yu, Y. and Meng, X.-L. (2011).
\newblock To center or not to center: That is not the question---an
  ancillarity–sufficiency interweaving strategy ({ASIS}) for boosting {MCMC}
  efficiency.
\newblock {\em Journal of Computational and Graphical Statistics},
  20(3):531--570.

\bibitem[Zhang et~al., 2022]{zhang_bayesian_2020}
Zhang, Y.~D., Naughton, B.~P., Bondell, H.~D., and Reich, B.~J. (2022).
\newblock Bayesian regression using a prior on the model fit: The {R2-D2}
  shrinkage prior.
\newblock {\em Journal of the American Statistical Association},
  117(538):862--874.

\bibitem[Zhou and Carin, 2012]{zhou_carin_2012}
Zhou, M. and Carin, L. (2012).
\newblock Negative binomial process count and mixture modeling.
\newblock arXiv:1209.3442v1 [stat.ME].

\bibitem[Zwillinger and Jeffrey, 2007]{zwillinger2007table}
Zwillinger, D. and Jeffrey, A. (2007).
\newblock {\em Table of integrals, series, and products}.
\newblock Elsevier.

\end{thebibliography}

\newpage
\appendix
\section{Proof of Theorem \ref{thm:lim}.}\label{ap:thm}

In order to prove Theorem \ref{thm:lim}, we first have to establish the densities of the priors marginalized over the local scale $\vartheta_i$.

\begin{proposition}\label{prop:dl}
Given the DL prior \eqref{eq:dle}, the density of $\phi_i$ for any $i\in \mathcal{A}_j$, $j=1,\dots,k$, marginalized over the local scale $\vartheta_i$, is given by
\begin{align}\label{eq:dl_cond}
    p_{DL}(\phi_i|a_j^\gamma)= \frac{|\phi_i|^{(a_j^\gamma-1)/2} K_{1-a_j^\gamma}(\sqrt{2|\phi_i|})}{2^{(1+a_j^\gamma)/2}\Gamma(a_j^\gamma)},
\end{align}
where $K_\nu(x)$ is the modified Bessel function of the second kind.\footnote{Modified Bessel function of the second kind: $K_{\nu}(x)=\frac{\Gamma(\nu+1)(2x)^\nu}{\sqrt{2\pi}}\int_0^\infty \frac{\cos t}{(t^2+x^2)^{\nu+1/2}}dt$.}
\end{proposition}
\begin{proof}
    Cf.\ \citet{bhattacharya_dirichletlaplace_2015}.
\end{proof}
\begin{proposition}\label{prop:ng}
    Given the NG prior \eqref{eq:ng}, the density of $\phi_i$ for any $i\in \mathcal{A}_j$, $j=1,\dots,k$, marginalized over the local scale $\vartheta_i$, is given by
    \begin{align}\label{eq:ng_cond}
    p_{NG}(\phi_i|\zeta_j a_j^\delta) = \frac{2^{1/2-a_j^\delta} \left(\frac{a_j^\delta}{\zeta_j}\right)^{a_j^\delta/2 + 1/4}}{\sqrt{\pi} \Gamma(a_j^\delta)} K_{(a_j^\delta-1/2)}\left(\sqrt{\frac{a_j^\delta}{\zeta_j}} |\phi_i|\right) |\phi_i|^{a_j^\delta-1/2}
\end{align}
\end{proposition}
\begin{proof}
    \begin{align*}
    p_{NG}(\phi_i|\zeta_j a_j^\delta) =& \int_0^\infty \frac{1}{\sqrt{2 \pi \vartheta_i \zeta_j}} \exp\left\lbrace -\frac{\phi_i^2}{2 \vartheta_i \zeta_j} \right\rbrace \frac{\left(\frac{a_j^\delta}{2}\right)^{a^\delta_j}}{\Gamma(a_j^\delta)} \vartheta_i^{a_j^\delta-1} \exp\left\lbrace -\frac{a_j^\delta \vartheta_i}{2}\right\rbrace d\vartheta_i  \\
    =& \frac{1}{\sqrt{2 \pi \zeta_j}} \frac{\left(\frac{a_j^\delta}{2}\right)^{a^\delta_j}}{\Gamma(a_j^\delta)} \frac{2 K_{(a_j^\delta-1/2)}\left(\sqrt{\frac{a_j^\delta}{\zeta_j}} |\phi_i|\right)}{\left(\frac{a_j^\delta\zeta_j}{\phi_i^2}\right)^{\frac{a_j^\delta-1/2}{2}}} \times  \\
    & \int_0^\infty \overbrace{\frac{\left(\frac{a_j^\delta\zeta_j}{\phi_i^2}\right)^{\frac{a_j^\delta-1/2}{2}}}{2 K_{(a_j^\delta-1/2)}\left(\sqrt{\frac{a_j^\delta}{\zeta_j}} |\phi_i|\right)} \vartheta_i^{a_j^\delta-1/2-1} \exp \left\lbrace -\frac{1}{2} \left(a_j^\delta \vartheta_i + \frac{\phi_i^2}{\vartheta_i\zeta_j}\right) \right\rbrace}^{\text{density of GIG$\left(a_j^\delta-1/2,a_j^\delta,\frac{\phi_i^2}{\vartheta_i\zeta_j}\right)$}} d\vartheta_i  \\
    =& \frac{2^{1/2-a_j^\delta} \left(\frac{a_j^\delta}{\zeta_j}\right)^{a_j^\delta/2 + 1/4}}{\sqrt{\pi} \Gamma(a_j^\delta)} K_{(a_j^\delta-1/2)}\left(\sqrt{\frac{a_j^\delta}{\zeta_j}} |\phi_i|\right) |\phi_i|^{a_j^\delta-1/2}.
\end{align*}
\end{proof}
\begin{proposition}
    Given the R2D2 prior (\ref{eq:r2d2}), the density of $\phi_i$ for any $i\in \mathcal{A}_j$, $j=1,\dots,k$, marginalized over the local scales $\vartheta_i$, is given by
    \begin{align}\label{eq:r2d2_cond}
p_{R2D2}(\phi_i|\zeta_j a^{\pi}_j) &=
    \frac{\sqrt{a^{\pi}_j}}{2\sqrt{\pi \zeta_j}\Gamma(a^{\pi}_j)} 
  G^{3,0}_{0,3}\left(\begin{array}{l}
 			-\\
 			\frac{1}{2},0, a^{\pi}_j-\frac{1}{2}
 		\end{array}\middle\vert
 	\frac{\phi_i^2 a^{\pi}_j  }{4\zeta_j}\right),
\end{align}
where $G^{m,n}_{p,q}\left(\begin{array}{l}
 			b_1,\dots, b_q\\
 			a_1,\dots,a_p
 		\end{array}\middle\vert
 	z\right)$ is the Meijer-G function.\footnote{A detailed definition of the Meijer-G function can be found in \citet{bateman1953higher} and in the Appendix of \citet{zhang_bayesian_2020}.}
\end{proposition}
\begin{proof}
    \begin{align*}
    p(\phi_i|\zeta_j a^{\pi}_j) &=\int_0^\infty \frac{1}{2\sqrt{\vartheta_i \zeta_{j}/2}} \exp\left\lbrace - \frac{|\phi_i|}{\sqrt{\vartheta_i \zeta_{j}/2}} \right\rbrace \frac{(a^{\pi}_j/2)^{a^{\pi}_j}}{\Gamma(a^{\pi}_j)} \vartheta_i^{(a^{\pi}_j-1)} \exp\left\lbrace - \frac{a^{\pi}_j}{2} \vartheta_i \right\rbrace d \vartheta_i.
\end{align*}
The Meijer-G function nests many functions as special cases, e.g.\ $G^{1,0}_{0,1}\left(\begin{array}{l}
 			-\\
 			0
 		\end{array}\middle\vert
 	x \right) = G^{0,1}_{1,0}\left(\begin{array}{l}
 			1\\
 			-
 		\end{array}\middle\vert
 	\frac{1}{x} \right)=\exp\{-x\}$.
Hence, 
\begin{align*}
    p(\phi_i|\zeta_j a^{\pi}_j) &= \frac{(a^{\pi}_j/2)^{a^{\pi}_j}}{\sqrt{2 \zeta_j}\Gamma(a^{\pi}_j)} \int_0^\infty \vartheta_i^{(a^\pi_j-3/2)} \exp\left\lbrace - \frac{|\phi_i|}{\sqrt{\vartheta_i \zeta_{j}/2}} \right\rbrace G^{0,1}_{1,0}\left(\begin{array}{l}
 			1\\
 			-
 		\end{array}\middle\vert
 	\frac{2}{a^{\pi}_j\vartheta_i} \right) d\vartheta_i \\
  &= \frac{ a_j^{\pi^{(3/2)}}}{4\sqrt{\zeta_j}\Gamma(a^{\pi}_j)} \int_0^\infty  
  \exp\left\lbrace - \frac{|\phi_i|}{\sqrt{\vartheta_i \zeta_{j}/2}} \right\rbrace 
  \left(\frac{2}{a^{\pi}_j\vartheta_i}\right)^{(3/2-a^{\pi}_j)}
  G^{0,1}_{1,0}\left(\begin{array}{l}
 			1\\
 			-
 		\end{array}\middle\vert
 	\frac{2}{a^{\pi}_j\vartheta_i} \right) d\vartheta_i \\
  &= \frac{ a_j^{\pi^{(3/2)}}}{4\sqrt{\zeta_j}\Gamma(a^{\pi}_j)} \int_0^\infty
  \exp\left\lbrace - \frac{|\phi_i|}{\sqrt{\vartheta_i \zeta_{j}/2}} \right\rbrace 
  G^{0,1}_{1,0}\left(\begin{array}{l}
 			\frac{5}{2}-a^{\pi}_j\\
 			-
 		\end{array}\middle\vert
 	\frac{2}{a^{\pi}_j\vartheta_i} \right) d\vartheta_i.
\end{align*}
The last equality follows from 9.31.(5) in \citet{zwillinger2007table}.
Now, let $\theta_i = \frac{1}{\sqrt{\vartheta_j}}$. Then,
\begin{align*}
    p(\phi_i|\zeta_j a^{\pi}_j) & = \frac{ a_j^{\pi^{(3/2)}}}{4\sqrt{\zeta_j}\Gamma(a^{\pi}_j)} \int_0^\infty
  \exp\left\lbrace - \frac{|\phi_i| \theta_i}{\sqrt{\zeta_{j}/2}} \right\rbrace 
  G^{0,1}_{1,0}\left(\begin{array}{l}
 			\frac{5}{2}-a^{\pi}_j\\
 			-
 		\end{array}\middle\vert
 	\frac{2\theta_i^2}{a^{\pi}_j} \right) \frac{2}{\theta^3} d\theta_i \\
  & = \frac{\sqrt{2}}{\sqrt{\zeta_j}\Gamma(a^{\pi}_j)} \int_0^\infty
  \exp\left\lbrace - \frac{|\phi_i| \theta_i}{\sqrt{\zeta_{j}/2}} \right\rbrace 
  G^{0,1}_{1,0}\left(\begin{array}{l}
 			\frac{5}{2}-a^{\pi}_j\\
 			-
 		\end{array}\middle\vert
 	\frac{2\theta_i^2}{a^{\pi}_j} \right) \left(\frac{2\theta_i^2}{a^{\pi}_j}\right)^{-3/2} d\theta_i\\
  & = \frac{\sqrt{2}}{\sqrt{\zeta_j}\Gamma(a^{\pi}_j)} \int_0^\infty
  \exp\left\lbrace - \frac{|\phi_i| \theta_i}{\sqrt{\zeta_{j}/2}} \right\rbrace 
  G^{0,1}_{1,0}\left(\begin{array}{l}
 			1-a^{\pi}_j\\
 			-
 		\end{array}\middle\vert
 	\frac{2\theta_i^2}{a^{\pi}_j} \right)  d\theta_i \\
  &=\frac{\sqrt{2}}{\sqrt{\zeta_j}\Gamma(a^{\pi}_j)} 
  \frac{1}{\sqrt{\pi}\frac{|\phi_i|}{\sqrt{\zeta_{j}/2}}}
  G^{0,3}_{3,0}\left(\begin{array}{l}
 			0,\frac{1}{2}, 1 - a^{\pi}_j\\
 			-
 		\end{array}\middle\vert
 	\frac{4\zeta_j}{a^{\pi}_j \phi_i^2 } \right).
\end{align*}
The last equality follows from 7.813.(2) in \citet{zwillinger2007table}.
\begin{align*}
    p(\phi_i|\zeta_j a^{\pi}_j) & =\frac{a_j^{\pi^{\frac{1}{2}}}}{2\sqrt{\zeta_j}\Gamma(a^{\pi}_j)} 
  \frac{1}{\sqrt{\pi}}
  \left( \frac{4\zeta_j}{a^{\pi}_j \phi_i^2 } \right)^{\frac{1}{2}}
  G^{0,3}_{3,0}\left(\begin{array}{l}
 			0,\frac{1}{2}, 1 - a^{\pi}_j\\
 			-
 		\end{array}\middle\vert
 	\frac{4\zeta_j}{a^{\pi}_j \phi_i^2 } \right) \\
  & =\frac{a_j^{\pi^{\frac{1}{2}}}}{2\sqrt{\zeta_j}\sqrt{\pi}\Gamma(a^{\pi}_j)} 
  G^{0,3}_{3,0}\left(\begin{array}{l}
 			\frac{1}{2},1, \frac{3}{2} - a^{\pi}_j\\
 			-
 		\end{array}\middle\vert
 	\frac{4\zeta_j}{a^{\pi}_j \phi_i^2 } \right) \\
  & =\frac{\sqrt{a^{\pi}_j}}{2\sqrt{\zeta_j}\sqrt{\pi}\Gamma(a^{\pi}_j)} 
  G^{3,0}_{0,3}\left(\begin{array}{l}
 			-\\
 			\frac{1}{2},0, a^{\pi}_j-\frac{1}{2}
 		\end{array}\middle\vert
 	\frac{\phi_i^2 a^{\pi}_j  }{4\zeta_j}\right).
\end{align*}
The last equality follows from 9.31.(2) in \citet{zwillinger2007table}.
\end{proof}
\begin{proposition}
    Given the HS prior \eqref{eq:hs}, the density of $\phi_i$ for any $i\in \mathcal{A}_j$, $j=1,\dots,k$, marginalized over the local scales $\vartheta_i$, is given by
    \begin{align}\label{eq:hs_cond}
    p(\phi_i|\zeta_j ) = \frac{1}{\sqrt{2}\pi^{3/2}} \exp\left\lbrace \frac{\phi_i^2}{2\zeta_j} \right\rbrace E_1\left(\frac{\phi_i^2}{2\zeta_j}\right),
\end{align}
where $E_1(\cdot)$ is the exponential integral function.
\end{proposition}
\begin{proof}
    Cf. \citet{carvalhi2010}.
\end{proof}

\paragraph{Proof of Theorem \ref{thm:lim}\ref{thm:lim_tails}.}

For \eqref{thm:dl_tail} cf.\ \citet{zhang_bayesian_2020}, for \eqref{thm:hs_tail} cf.\ \citet{carvalhi2010}. \eqref{thm:shm_tail} and \eqref{thm:ssvs_tail} are standard and hence not derived.

\eqref{thm:ng_tail}: According to 10.25.3 in \url{https://dlmf.nist.gov/10.25}, when both $\nu$ and $z$ are real, if $z \rightarrow \infty$, then $K_\nu(z) \approx \pi^{1/2} (2z)^{-1/2}\exp\left\lbrace - z \right\rbrace$. Then as $|\phi| \rightarrow \infty$, using Proposition \ref{prop:ng},
	\begin{align*}
		p_{NG}^\infty(\phi_i|\zeta_j a_j^\delta) &=	 \frac{2^{1/2-a_j^\delta} \left(\frac{a_j^\delta}{\zeta_j}\right)^{a_j^\delta/2 + 1/4}}{\sqrt{\pi} \Gamma(a_j^\delta)} \pi^{1/2} \left(2 \sqrt{\frac{a_j^\delta}{\zeta_j}} |\phi_i|\right)^{-1/2} \exp\left\lbrace -\sqrt{\frac{a_j^\delta}{\zeta_j}} |\phi_i| \right\rbrace |\phi_i|^{a_j^\delta-1/2} \\
  & \propto |\phi_i|^{a_j^\delta-1} \exp\left\lbrace -\sqrt{\frac{a_j^\delta}{\zeta_j}} |\phi_i| \right\rbrace.
	\end{align*}

 \eqref{thm:r2d2_tail}: Follows from Theorem 1.3.(i) in \citet{mathai2009h}.

\paragraph{Proof of Theorem \ref{thm:lim}\ref{thm:lim_zero}.}
For \eqref{thm:hs_zero} cf. \citet{carvalhi2010}.

\eqref{thm:dl_zero}: Case $0 < a_j^\gamma < 1$: Cf. \cite{zhang_bayesian_2020}.

Case $a_j^\gamma=1$: According to 10.30.3 in \url{https://dlmf.nist.gov/10.30}, when $\nu = 0 $, $z \rightarrow 0$, $K_\nu(z) \approx - \ln z$. Given $a_j^\gamma=1$, $|\phi|\rightarrow 0$, using Proposition \ref{prop:dl}, 
\begin{align*}
%
%
    p_{DL}^0(\phi_i| a_j^\gamma) &\approx \frac{\ln\left((2|\phi_i|)^{-1/2}\right)}{2^{(1+a_j^\gamma)/2}\Gamma(a_j^\gamma)} \propto \ln\left(\frac{1}{|\phi_i|}\right).
\end{align*}

\eqref{thm:ng_zero}:
Case $0<a_j^\delta<0.5$: According to 10.27.3 in \url{https://dlmf.nist.gov/10.27} $K_\nu(z) = K_{-\nu}(z)$. Moreover, according to 10.30.2 in \url{https://dlmf.nist.gov/10.30}, when $\nu > 0 $, $z \rightarrow 0$ and $z$ is real, $K_\nu(z) \approx \Gamma(\nu)(z/2)^{-\nu}/2$. Given $0<a_j^\delta<0.5$, $|\phi|\rightarrow 0$, using Proposition \ref{prop:ng},
	\begin{align*}
		p_{NG}^0(\phi_i|\zeta_j a_j^\delta) &\approx \frac{2^{1/2-a_j^\delta} \left(\frac{a_j^\delta}{\zeta_j}\right)^{a_j^\delta/2 + 1/4}}{\sqrt{\pi} \Gamma(a_j^\delta)} \frac{1}{2} \Gamma(1/2-a_j^\delta) \left(\frac{\sqrt{\frac{a_j^\delta}{\zeta_j}} |\phi_i|}{2}\right)^{a_j^\delta-1/2} |\phi_i|^{a_j^\delta-1/2}\\
  & \propto |\phi_i|^{2a_j^\delta-1}.
	\end{align*}

Case $a_j^\delta=\frac{1}{2}$: According 10.30.3 in \url{https://dlmf.nist.gov/10.30}, when $\nu = 0 $, $z \rightarrow 0$, $K_\nu(z) \approx - \ln z$. Given $a_j^\delta=\frac{1}{2}$, $|\phi|\rightarrow 0$, using Proposition \ref{prop:ng}, 
\begin{align*}
    p_{NG}^0(\phi_i|\zeta_j a_j^\delta) &\approx \frac{2^{1/2-a_j^\delta} \left(\frac{a_j^\delta}{\zeta_j}\right)^{a_j^\delta/2 + 1/4}}{\sqrt{\pi} \Gamma(a_j^\delta)} \ln \left(\left(\sqrt{\frac{a_j^\delta}{\zeta_j}} |\phi_i|\right)^{-1}\right) \propto \ln\left(\frac{1}{|\phi_i|}\right).
\end{align*}

\eqref{thm:r2d2_zero}:
Case $0<a_j^\pi<0.5$: Follows from Theorem 1.2.(i) in \citet{mathai2009h}.

\section{A note on posterior estimation}\label{sec:fcp}
Efficient computer implementations of all posterior samplers discussed in this paper are conveniently bundled into 
the R package \texttt{bayesianVARs}, 
available from the Comprehensive R Archive Network (CRAN) at \url{https://cran.r-project.org/package=bayesianVARs}.
In what is to follow, we provide the necessary conditionals used for Gibbs sampling.

The main challenge in terms of computational complexity is to sample from the conditional posterior distribution of $\bm{\phi}$, which is multivariate normal: $\bm{\phi} | \bullet \sim \bm{N}(\bm{\bar{\mu}}, \bm{\bar{V}})$, where $\bm{\bar{V}} = \left(\bm{\underline{V}}^{-1} + \sum_{t=1}^T (\bm{\Sigma}_t^{-1} \otimes \bm{x}_t\bm{x}_t^\prime)\right)^{-1}$ and $\bm{\bar{\mu}} = \bm{\bar{V}}\left(\sum_{t=1}^T(\bm{I}_M \otimes \bm{x}_t) \bm{\Sigma}_t^{-1} \bm{y}_t\right)$. The complexity stems from the required manipulations of the $n \times n$ matrix variance-covariance matrix of the vector of coefficients. To render computation feasible, we use the corrected triangular algorithm as in \citet{carriero_corrigendum_2021}. 

The factorization of the variance-covariance matrix in (\ref{eq:usv}) makes sampling of the free off-diagonal elements in $\bm{U}$ straightforward: These can be sampled ``equation per equation'' using standard Bayesian linear regression results \citep{cogley_drifts_2005}. Let $\bm{u}^{(j)}=(u_{1j},\dots, u_{(j-1),j})^\prime$ denote the $(j-1)$-dimensional column vector that collects the free off-diagonal elements of the $j$th column of $\bm U$ for $j=2,\dots,M$. Assuming a normal prior $\bm u^{(j)} \sim N(\bm 0, \underline{\bm{V}}_u^{(j)})$, the full conditional posterior is also normal: $\bm u^{(j)} | \bullet \sim N(\bm \mu_u^{(j)}, \bar{\bm V}_u^{(j)})$, where $\bar{\bm V}_u^{(j)} = \left( \Tilde{\bm X}^{(j)^{\prime}}\Tilde{\bm X}^{(j)} +  \underline{\bm{V}}_u^{(j)^{-1}}\right)^{-1}$ and $\bm \mu_u^{(j)} = \bar{\bm V}_u^{(j)} \Tilde{\bm X}^{(j)^{\prime}} \Tilde{\bm y}^{(j)}$. Here, $\Tilde{\bm y}^{(j)} = \left(\varepsilon_{j1} e^{-h_{j1}/2}, \dots,  \varepsilon_{jT} e^{-h_{jT}/2}\right)^\prime$ denotes the normalized observation vector and $$\Tilde{\bm X}^{(j)} = \left( \left(\varepsilon_{11} e^{-h_{j1}/2},\dots, \varepsilon_{1T} e^{-h_{jT}/2}\right)^\prime, \dots , \left(\varepsilon_{(j-1)1} e^{-h_{j1}/2},\dots, \varepsilon_{(j-1)T} e^{-h_{jT}/2}\right)^\prime\right)$$ is the $T \times (j-1)$ design matrix.

In the following paragraphs, we provide the conditionals for the latent variables introduced by the hierarchical shrinkage priors. We abstain from providing the conditionals if a hierarchical prior is placed on $\bm u$, because it is basically the same as for $\bm \phi$.
\paragraph{DL}
\citet{bhattacharya_dirichletlaplace_2015} propose an efficient algorithm how to sample the hyperparameters of the DL prior, shown in Eq.~\ref{eq:dl0}. An important feature of this algorithm is the joint update of $(\bm{\varrho}, \bm{\omega}, \bm{\psi})$, as one-at-a-time updates of Dirichlet vectors usually lead to slow mixing. We, however, propose to sample the hyperparameters of the DL prior shown in Eq.~\ref{eq:dle}. Let $\tilde{\vartheta}_i=\sqrt{\vartheta_i/2}$. Then, one important aspect of our algorithm is the joint update of $p(\bm{\tilde{\vartheta}}, \bm{\psi}|\bm{\phi}) = p(\bm{\tilde{\vartheta}}|\bm{\phi})p(\bm{\psi}|\bm{\phi},\bm{\tilde{\vartheta}})$: A draw from the joint conditional posterior distribution is generated by first sampling $
	\tilde{\vartheta}_i|\phi_i \sim GIG\left(a_j^\gamma-1,1,2|\phi_i|\right), i \in \mathcal{A}_j, j=1,\dots,k
$, where $GIG$ denotes the generalized inverse Gaussian distribution\footnote{The density of $GIG(\theta, \psi, \chi)$ is proportional to $x^{\theta -1} \exp \left\lbrace -\frac{1}{2} \left( \psi x + \frac{\chi}{x} \right) \right\rbrace$. An efficient algorithm to generate draws from GIG is implemented in the \texttt{R} package \texttt{GIGrvg} \citep{leydold_gigrvg_2017}.} and then $
 \psi_{i} | \phi_i, \tilde{\vartheta_i} \sim GIG \left(\frac{1}{2},1, \left(\frac{\phi_i}{\tilde{\vartheta_i}}\right)^2\right), i \in \mathcal{A}_j, j=1,\dots,k
$. What is left, is the full conditional posterior of $a_j^\gamma$ which is discrete with the same support points as the prior:
$
Pr(a_j^\gamma=\tilde{a}^\gamma_r|\bullet) = \frac{p^\gamma_r \prod_{i \in \mathcal{A}_j}f_{G}(\kappa_i;\tilde{a}^\gamma_r,\frac{1}{2})}{\sum_{r=1}^R p^\gamma_r \prod_{i \in \mathcal{A}_j}f_{G}(\kappa_i;\tilde{a}^\gamma_r,\frac{1}{2})}, i \in \mathcal{A}_j, j=1,\dots,k,
$
where $f_G(x;\alpha,\beta)$ denotes the density of the gamma distribution with shape $\alpha$ and rate $\beta$ evaluated at $x$.

Overall, our algorithm consists of one step less than the algorithm in \citet{bhattacharya_dirichletlaplace_2015}, because we directly update the product $\tilde{\vartheta}_i = \varrho_i\omega_j$. It has to be noted, if there is interest in the posterior of $\omega_j$ (Eq.~\ref{eq:dl0}), draws of this distribution can be easily generated by transforming the posterior draws of the local scales: According to the prior in Eq.~\ref{eq:dl0}, it holds that $\omega_j=\sum_{i \in \mathcal{A}_j} \varrho_i \omega_j$, because a Dirichlet distributed vector always sums up to one. Earlier, we defined $\sqrt{\vartheta_i/2}=\varrho_i\omega_j$. Hence, we get $\omega_j=\sum_{i \in \mathcal{A}_j} \sqrt{\vartheta_i/2}=\sum_{i \in \mathcal{A}_j} \tilde{\vartheta}_i$. 
\paragraph{NG}
In order to draw from the posterior distribution under the NG prior shown in Eq.~\ref{eq:ng}, we use the following equivalent representation:
\begin{align}
    \phi_i|\varpi_i \sim N(0,\varpi_i), \quad \varpi_i|\xi_j\sim G(a^\delta_j, \xi_j), \quad \xi_j \sim G\left(b^\delta,\frac{2c}{a^\delta_j}\right).
\end{align}
where $\varpi_i=\vartheta_i\zeta_j$ and $\xi_j=\frac{a^\delta_j}{2\zeta_j}$. The full conditional posterior distribution of $\varpi_i$ is generalized-inverse-Gaussian: $\varpi_i|\bullet \sim GIG(a^\delta_j - \frac{1}{2}, 2\xi_j, \phi_i^2)$. The full conditional posterior of $\xi_j$ follows a gamma distribution: $\xi_j|\bullet \sim G(n_j a^\delta_j + b^\delta, \frac{2c}{a^\delta_j} + \sum_{i \in \mathcal{A}_j} \varpi_i)$. The full conditional posterior of $a_j^\delta$ is discrete, with the same support points as the prior:
$
Pr(a_j^\delta=\tilde{a}^\delta_r|\bullet) = \frac{p^\delta_r \prod_{i \in \mathcal{A}_j}f_{G}(\varpi_i;\tilde{a}^\delta_r,\xi_j)}{\sum_{r=1}^R p^\delta_r \prod_{i \in \mathcal{A}_j}f_{G}(\varpi_i;\tilde{a}^\delta_r,\xi_j)}, i \in \mathcal{A}_j, j=1,\dots,k.
$
\paragraph{R2D2} \citet{zhang_bayesian_2020} propose an algorithm to sample the hyperparameters of the R2D2 prior shown in Eq.~\ref{eq:r2d20}, which relies on the joint update of the Dirichlet vector proposed by \citet{bhattacharya_dirichletlaplace_2015}. With the same arguments as before, we prefer to sample the hyperparameters of the R2D2 prior using an equivalent representation, which is similar in spirit to Eq.~\ref{eq:r2d2}:
\begin{align}
    \phi_i|\psi_i,\varpi_i \sim N\left(0,\psi_i\varpi_i/2\right), \quad \psi_i \sim Exp(1/2),\quad \varpi_i|\xi_j \sim G(a^{\pi}_j,\xi_j), \quad \xi_j \sim G(b^\pi,1).
\end{align}
The full conditional posteriors of $\psi_i$ and $\varpi_i$ both are generalized inverse Gaussian: $\psi_i|\bullet GIG(\frac{1}{2}, 1, \frac{2 \phi_i^2}{\varpi_i})$ and $\varpi_i|\bullet \sim GIG(a^\pi_j-\frac{1}{2}, 2\xi_j, \frac{2\phi_i^2}{\psi_i})$, for $i \in \mathcal{A}_j, j=1,\dots,k$. The conditional posterior of $\xi_j$ follows a gamma distribution: $\xi_j|\bullet \sim G(n_j a^\pi_j + b^\pi, 1 + \sum_{i \in \mathcal{A}_j}\varpi_i), j=1,\dots,k$. The full conditional posterior of $a_j^\pi$ is discrete, with the same support points as the prior:
$
Pr(a_j^\pi=\tilde{a}^\pi_r|\bullet) = \frac{p^\pi_r \prod_{i \in \mathcal{A}_j}f_{G}(\varpi_i;\tilde{a}^\pi_r,\xi_j)}{\sum_{r=1}^R p^\pi_r \prod_{i \in \mathcal{A}_j}f_{G}(\varpi_i;\tilde{a}^\pi_r,\xi_j)}, i \in \mathcal{A}_j, j=1,\dots,k.
$
\paragraph{Horseshoe}
We follow \citet{makalic2016} in sampling the hyperparameters of the HS prior shown in Eq.~\ref{eq:hs}. They propose the following scale-mixture representation of the prior:
\begin{align}
    \phi_i | \vartheta_i\zeta_j \sim N(0,\vartheta_i \zeta_j), \quad \vartheta_i|\nu_i \sim G^{-1}(1/2,1/\nu_i), \quad \zeta_j|\tau_j \sim G^{-1}(1/2,1/\tau_j), \nonumber\\ \nu_i \sim G^{-1}(1/2,1), \quad \tau_j \sim G^{-1}(1/2,1),
\end{align}
where $G^{-1}(\alpha, \beta)$ denotes the inverse-Gamma distribution with shape $\alpha$ and scale $\beta$. The scale-mixture representation has the advantage, that the full conditional posterior distributions are of well-known forms. The full conditional posteriors of both the local parameters $\vartheta_i$ and the (semi-)global parameters $\zeta_j$ are inverse-gamma: $\vartheta_i|\bullet \sim G^{-1}(1, \frac{1}{\nu_i} + \frac{\phi_i^2}{2 \tau_j}), \quad \zeta_j|\bullet \sim G^{-1}(\frac{n_j+1}{2}, \frac{1}{\tau_j} + \frac{1}{2}\sum_{i=1}^{\mathcal{A}_j} \frac{\phi_i^2}{\vartheta_i})$. Also, the full conditional posteriors of both the local and (semi-)global auxiliary variables $\nu_i$ and $\tau_j$ are inverse-gamma: $\nu_i|\bullet \sim G^{-1}(1,1+\frac{1}{\vartheta_i}), \quad
\tau_j|\bullet \sim G^{-1}(1,1+\frac{1}{\zeta_j})$.
\paragraph{SHM}
\citet{huber_adaptive_2019} use a (log-)random walk Metropolis step to sample the shrinkage parameters of SHM. The gamma distribution, however, arises as a special case of the generalized inverse Gaussian (GIG) distribution. It is well known that GIG is a conjugate prior for the variance parameter of a normal distribution. Hence, the shrinkage parameters  $\lambda_j$ for $j=1,2$ can be sampled from their full conditional posterior distributions. Let $\mathcal{M}_1$ and $\mathcal{M}_2$ be the index sets that collect own-lag and cross-lag coefficients, respectively. Then,
$
	\lambda_j |\bullet \sim GIG \left( \left(c_j - \frac{|\mathcal{M}_j|}{2} \right), 2d_j, \sum_{i \in \mathcal{M}_j} \frac{\phi_i^2}{\tilde{r}_i} \right),
$
where $\tilde{r}_i$ is the constant term determined in (\ref{eq:HMv}). This has the benefit that no tuning of proposal distributions is required. 

\paragraph{SSVS}
 The full conditional posterior distribution of $\gamma_i$ follows a Bernoulli distribution:
$
	\gamma_i|\bullet \sim Bernoulli(\bar{p}_i)$,  $i=\mathcal{A}_j$, $j=1,\dots,k$,
where
\begin{align*}
    \bar{p}_i = Pr(\gamma_i=1|\phi_i) = \frac{\frac{1}{\tau_{1i}} \exp\left(- \frac{\phi_i^2}{2 \tau_{1i}^2} \right) \underline{p}_j}{\frac{1}{\tau_{1i}} \exp\left(- \frac{\phi_i^2}{2 \tau_{1i}^2} \right) \underline{p}_j + \frac{1}{\tau_{0i}} \exp\left(- \frac{\phi_i^2}{2 \tau_{0i}^2} \right) (1 - \underline{p}_j)}.
\end{align*}
The full conditional posterior distribution of $\underline{p}_j$ is beta:
$
	\underline{p}_j|\bullet \sim Beta(s_1 + \sum_{i \in \mathcal{A}_j} \gamma_i, s_2 + |\mathcal{A}_j| - \sum_{i \in \mathcal{A}_j} \gamma_i).
$ 

\paragraph{Latent volatilities}
We sample all the parameter of (\ref{eq:usv}) using the \textit{ancillarity-sufficiency interweaving strategy} \citep{yumeng} as in \citet{kastner_ancillarity-sufficiency_2014} via the \texttt{R} package \texttt{stochvol} \citep{hosszejni_modeling_2021}.

\section{Data}
\label{sec:Data}

\begin{table}[ht]
\centering
\resizebox*{\textwidth}{!}{
\begin{tabular}{l  l l}
\hline\noalign{\smallskip}
Name & FRED MNEMONIC & Transformation \\
\hline\noalign{\smallskip}
\multicolumn{3}{c}{\textit{National Income and Product Accounts}} \\
Real Gross Domestic Product & GDPC1 & $\Delta \log(x_t)$ \\
Real Personal Consumption Expenditures & PCECC96 & $\Delta \log(x_t)$ \\
Real Gross Private Domestic Investment & GPDIC1 & $\Delta \log(x_t)$ \\
Real private fixed investment: Residential & PRFIx & $\Delta \log(x_t)$ \\
\hline\noalign{\smallskip}
\multicolumn{3}{c}{\textit{Industrial Production}} \\
Industrial Production Index & INDPRO & $\Delta \log(x_t)$ \\
Capacity Utilization: Manufacturing & CUMFNS & $\Delta \log(x_t)$ \\
\hline\noalign{\smallskip}
\multicolumn{3}{c}{\textit{Employment}} \\
All employees: Service-Providing Industries & SRVPRD & $\Delta \log(x_t)$ \\
Civilian Employment & CE16OV & $\Delta \log(x_t)$ \\
Average Weekly Hours of Production and Nonsupervisory Employees: & AWHMAN & $\Delta \log(x_t)$ \\
Manufacturing && \\
\hline\noalign{\smallskip}
\multicolumn{3}{c}{Prices} \\
Personal Consumption Expenditures: Chain-type Price Index & PCECTPI & $\Delta \log (x_t)$ \\
Gross Domestic Product: Chain-type Price Index & GDPCTPI & $\Delta \log (x_t)$ \\
Gross Private Domestic Investment: Chain-type Price Index & GPDICTPI & $\Delta \log (x_t)$ \\
Consumer Price Index for All Urban Consumers: All Items & CPIAUCSL & $\Delta \log (x_t)$ \\ 
\hline\noalign{\smallskip}
\multicolumn{3}{c}{\textit{Earnings and Productivity}} \\
Real Average Hourly Earnings of Production and Nonsupervisory Employees: & CES2000000008x & $\Delta \log(x_t)$ \\
Construction && \\
\hline\noalign{\smallskip}
\multicolumn{3}{c}{\textit{Interest Rates}} \\
Effective Federal Funds Rate & FEDFUNDS & no transformation \\
1-Year Treasury Constant Maturity Rate & GS1 & no transformation \\
10-Year Treasury Constant Maturity Rate & GS10 & no transformation \\
\hline\noalign{\smallskip}
\multicolumn{3}{c}{\textit{Money}} \\
Real M2 Money Stock & M2REAL & $\Delta \log(x_t)$ \\
\hline\noalign{\smallskip}
\multicolumn{3}{c}{\textit{Exchange Rates}} \\
U.S. / U.K. Foreign Exchange Rate & EXUSUKx & $\Delta \log(x_t)$ \\
\hline\noalign{\smallskip}
\multicolumn{3}{c}{\textit{Other}} \\
University of Michigan: Consumer Sentiment & UMCSENTx & $\Delta \log(x_t)$ \\
\hline\noalign{\smallskip}
\multicolumn{3}{c}{\textit{Stock Markets}} \\
S{\&}P’s Common Stock Price Index: Composite & S{\&}P 500 & $\Delta \log(x_t)$\\
\end{tabular}
}
\caption{Variables used in the empirical application.}
\label{tab:data}
\end{table}

\section{A note on priors for VAR coefficients}
\label{sec:reducedstructural}
Many of the priors used in the Bayesian VAR literature, amongst them the Minnesota prior as in \citet{litterman_forecasting_1986}, the stochastic search variable selection (SSVS) prior as in \citet{george_bayesian_2008}, the Dirichlet-Laplace (DL) prior as in \citet{kastner_sparse_2020}, the adaptive normal-gamma prior as in \citet{huber_adaptive_2019}, or the Horseshoe prior as in \citet{follett_achieving_2019}, were proposed using the reduced form VAR in (\ref{eq:VAR}). Nowadays, as it has become a trend using large datasets, many authors, however, opt for working with the structural VAR formulation. The independence of the structural VAR equations allows for more efficient computations \citep{chan_minnesota-type_2021, cross_macroeconomic_2020, chan2022}. The structural VAR is
\begin{align}
	\bm{y}_t^\prime \bm{U} &= \bm{x}_t^\prime \bm{B} + \bm{\xi}_t^\prime \nonumber \\
	\bm{y}_t^\prime &= \bm{x}_t^\prime \bm{B} + \bm{y}_t^\prime \tilde{\bm{U}} + \bm{\xi}_t^\prime, 
\end{align}
where $\bm{B} = \bm{\Phi U}$ are the structural VAR coefficients, $\tilde{\bm{U}} = \bm{I} - \bm{U}$ the contemporaneous coefficients and $\bm{\xi}_t = \bm{\epsilon}_t \bm{U}$ the orthogonalized errors. It is easy to see that, given $\bm{B}$ and $\bm{U}$, the reduced form coefficients can be recovered by computing $\bm{\Phi} = \bm{BU}^{-1}$.

However, we want to point out that priors in general are not translation-invariant: Changing the form without adjusting the prior accordingly will result in two different models. Simply put, imposing a specific prior on the structural coefficients $\bm{B}$ is not the same as imposing it on the reduced form coefficients $\bm{\Phi}$. Denote $b_{ij}$ the $ij$th element of $\bm{B}$ and $u^{-1}_{ij}$ the $ij$th element in $\bm{U}^{-1}$. Then as ``recovered'' matrix $\bm{\Phi} = \bm{BU}^{-1}$ we have:
\begin{equation}\label{eq:phi_induced}
	\bm{\Phi} =
	\begin{pmatrix}
		b_{11} & b_{11} u^{-1}_{12} + b_{12} & \cdots &  b_{11} u^{-1}_{1M} + \dots + b_{1(M-1)}u^{-1}_{(M-1)M} + b_{1M}\\
		b_{21} & b_{21} u^{-1}_{12} + b_{22} & \cdots &  b_{21} u^{-1}_{1M} + \dots + b_{2(M-1)}u^{-1}_{(M-1)M} + b_{2M}\\
		\vdots & \vdots & \ddots & \vdots \\
		b_{K1} & b_{K1} u^{-1}_{12} + b_{K2} & \cdots &  b_{K1} u^{-1}_{1M} + \dots + b_{K(M-1)}u^{-1}_{(M-1)M} + b_{KM}\\
	\end{pmatrix}.
\end{equation}
It is clear that the induced prior on the reduced form coefficients now depends on $\bm{U}$. Moreover, a potential drawback of using the parametrization in form of $\bm{B}=\bm{\Phi U}$ is that the posterior is not permutation invariant: The triangular form of $\bm{U}$ causes results to depend on the order of the vector $\bm{y}_t$. Especially in high-dimensional settings, justifying a specific order of the endogenous variables is implausible.\footnote{Our reduced form results also depend on the ordering, because of the factorization of $\bm{\Sigma}_t$ in Eq.\ \ref{eq:cogley}. Yet, the prior imposed on $\bm{\Phi}$ remains order-invariant. Using the structural form, the ordering affects both $\bm{U}$ and $\bm{\Phi}$, which might aggravate the problem. We refer to \citet{kastner_sparse_2020} and \citet{chan_large_2021} for methods that are completely order-invariant.} 

\begin{figure}[t]
	\includegraphics[width=\textwidth]{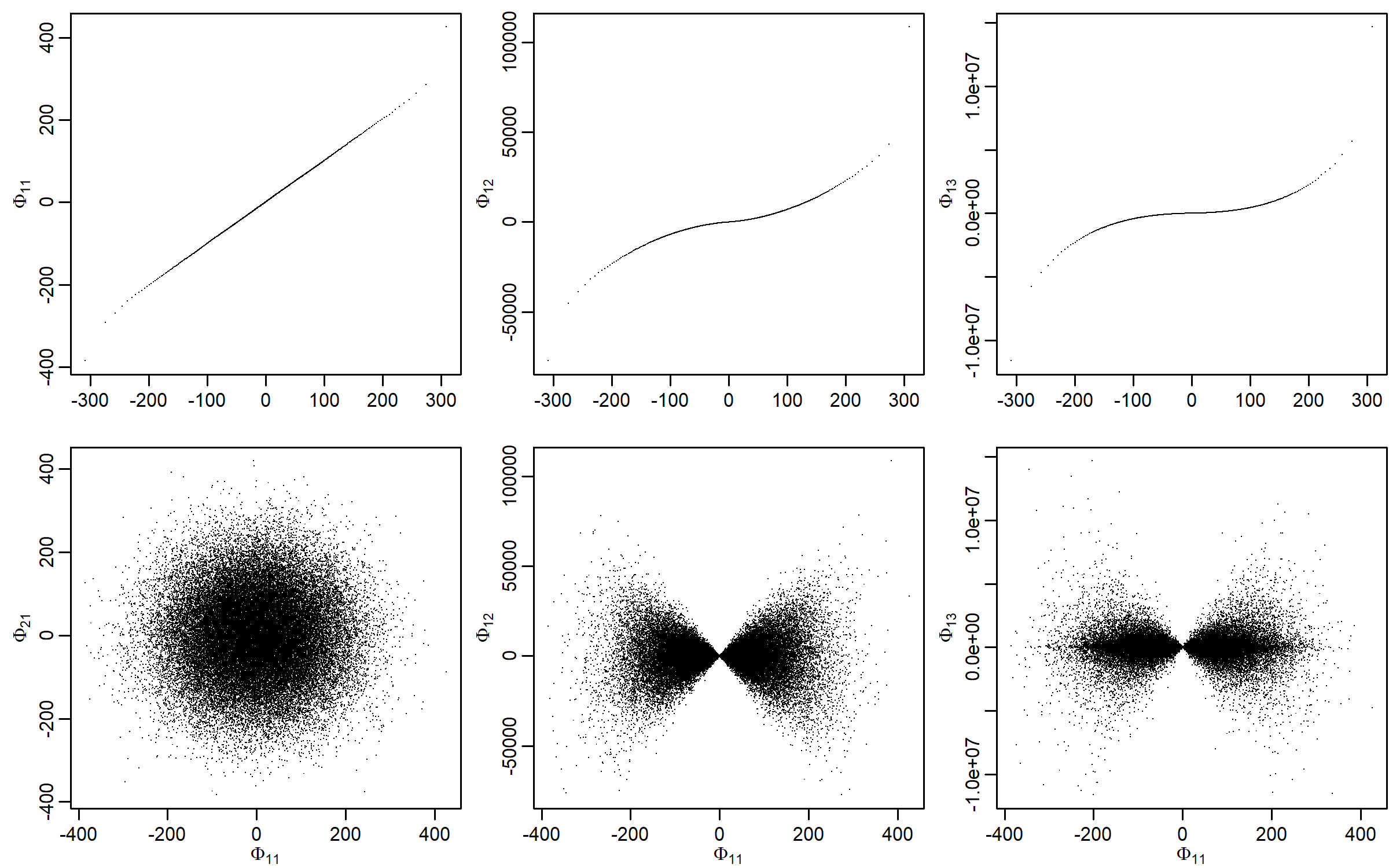}
	\caption{Prior simulations: Induced prior distributions for $\bm{\Phi}^{3\times 3}$ resulting from i.i.d. normal priors (zero mean and variance 10) for the elements in $\bm{B}^{3\times 3}$ and $\bm{u}$. Top panel: quantile-quantile plots for $\Phi_{11}, \Phi_{12}$ and $\Phi_{13}$ against theoretical quantiles of N(0,10). Bottom panel: Scatter plots of the pairs $(\Phi_{11}, \Phi_{21}), (\Phi_{11}, \Phi_{12})$, and $(\Phi_{11}, \Phi_{13})$.}
	\label{fig:str_red}
\end{figure}

To get a rough feeling of the induced prior distributions on $\bm{\Phi}$ we simulate from i.i.d. Gaussian priors for the elements in $\bm{B}$ and $\bm{u}$. Figure \ref{fig:str_red} depicts QQ plots and scatter plots of the recovered reduced-form coefficients. Except for the coefficients of the first columns (where $\bm{\Phi}$ and $\bm{B}$ are identical), the induced prior distributions have much heavier tails. The tails get heavier from column to column. Moreover, the distributions are of no well-known forms. 

Obviously, reparametrization can heavily influence results in the Bayesian paradigm; one must consider that a good prior for $\bm{\Phi}$ is not necessarily a good prior for $\bm{B}$ and vice versa. Unfortunately, this issue is often ignored. E.g., \citet{cross_macroeconomic_2020} find that their implementation of a hierarchical Minnesota prior (SHM) generates better forecasts of macroeconomic data than various global-local priors (GL), which seems contradictory to the results in \citet{huber_adaptive_2019}. Consequently, they claim that their hierarchical setup of SHM is better, disregarding that they use different parametrizations: Whereas \citet{huber_adaptive_2019} place the priors on $\bm{\Phi}$, \citet{cross_macroeconomic_2020} place them on $\bm{B}$. Since neither the same data (same time span, same transformations, etc.) nor exactly the same evaluation metrics are used, it is unclear which full model specification (i.e., a specific prior on a specific form of the VAR) yields better forecasts. It seems plausible that the reparametrization of the VAR makes a bigger difference than some subtleties in the prior configuration. Indeed, in an empirical application predicting stock returns, \citet{bernardi2022} find that various shrinkage priors lead to better out-of sample forecasts when they are imposed on the reduced-form coefficients instead of the structural-form coefficients. Another example can be found in \citet{chan_VAR_SV} where different SV specifications of the errors are examined. For one SV specification, the prior for the coefficients is placed on $\bm{\Phi}$, whereas for another SV specification, the same prior is placed on $\bm{B}$. The question arises to which degree the results are driven by different SV specifications, and to which degree by different parametrizations of the VAR coefficients.
\end{document}